%% file: main.tex
\DocumentMetadata{} 
\documentclass[acmtog]{acmart}

\AtBeginDocument{%
  \providecommand\BibTeX{{%
    \normalfont B\kern-0.5em{\scshape i\kern-0.25em b}\kern-0.8em\TeX}}}

\setcopyright{acmlicensed}
\acmJournal{TOG}
\acmYear{2024} \acmVolume{43} \acmNumber{6} \acmArticle{218} \acmMonth{12}\acmDOI{10.1145/3687763}

\usepackage[boxed,ruled,linesnumbered]{algorithm2e} 

\SetCommentSty{mycommfont}
\let\oldnl\nl
\newcommand{\nonl}{\renewcommand{\nl}{\let\nl\oldnl}}

\SetAlFnt{\small}
\SetAlCapFnt{\small}
\SetAlCapNameFnt{\small}
\SetAlCapHSkip{0pt}
\IncMargin{-\parindent}

\newlength\savedwidth

\usepackage{booktabs} 
\usepackage{enumitem}
\usepackage{amsmath}
\usepackage{color, colortbl}
\definecolor{lightgray}{gray}{0.9}
\usepackage{xcolor}
\usepackage{wrapfig}
\usepackage[normalem]{ulem}
\usepackage[export]{adjustbox}
\usepackage[most]{tcolorbox}
\newtcolorbox{mybox}{colback=white,colframe=black,left=7 pt,top=5 pt,
bottom=0 pt,right=5 pt,boxsep=2 pt,toprule=0.5 pt,leftrule=0.5 pt, bottomrule=0.5 pt, rightrule=0.5 pt,breakable, enhanced, arc=0pt,outer arc=0pt}

\usepackage{romannum}
\usepackage{soul}
\usepackage{overpic}
\usepackage{multirow}
\usepackage{tabulary}
\input{newcommands}

\usepackage{bbding}
\usepackage{tikz}
\newcommand{\ExternalLink}{%
\tikz[x=1.2ex, y=1.2ex, baseline=-0.05ex]{%
    \begin{scope}[x=1ex, y=1ex]
        \clip (-0.1,-0.1) 
            --++ (-0, 1.2) 
            --++ (0.6, 0) 
            --++ (0, -0.6) 
            --++ (0.6, 0) 
            --++ (0, -1);
        \path[draw, 
            line width = 0.5, 
            rounded corners=0.5] 
            (0,0) rectangle (1,1);
    \end{scope}
    \path[draw, line width = 0.5] (0.5, 0.5) 
        -- (1, 1);
    \path[draw, line width = 0.5] (0.6, 1) 
        -- (1, 1) -- (1, 0.6);
    }
}

\begin{document}
\title{MATTopo: Topology-preserving Medial Axis Transform with Restricted Power Diagram}

\author{Ningna Wang}
\affiliation{%
  \institution{University of Texas at Dallas}
  \state{Texas}
  \country{USA}
}
\email{ningna.wang@utdallas.edu}

\author{Hui Huang}
\affiliation{%
  \institution{Shenzhen University}
  \city{Shenzhen}
  \country{China}
}
\email{hhzhiyan@gmail.com}

\author{Shibo Song}
\affiliation{%
  \institution{Independent Researcher}
  \city{Shanghai}
  \country{China}
}
\email{longmaythess@outlook.com}

\author{Bin Wang}
\affiliation{%
  \institution{Tsinghua University}
  \city{Beijing}
  \country{China}
}
\email{wangbins@tsinghua.edu.cn}

\author{Wenping Wang}
\affiliation{%
  \institution{Texas A\&M University}
  \state{Texas}
  \country{USA}
}
\email{wenping@tamu.edu}

\author{Xiaohu Guo}\authornote{Corresponding author}
\affiliation{
  \institution{University of Texas at Dallas}
  \state{Texas}
  \country{USA}
}
\email{xguo@utdallas.edu}

\renewcommand\shortauthors{Wang et. al}

\begin{abstract}
We present a novel topology-preserving 3D medial axis computation framework based on volumetric restricted power diagram (RPD), while preserving the medial features and geometric convergence simultaneously, for both 3D CAD and organic shapes.
The volumetric RPD discretizes the input 3D volume into sub-regions given a set of medial spheres. With this intermediate structure, we convert the homotopy equivalency between the generated medial mesh and the input 3D shape into a localized contractibility checking for each restricted element (power cell, power face, power edge), by checking their connected components and Euler characteristics. We further propose a fractional Euler characteristic algorithm for efficient GPU-based computation of Euler characteristic for each restricted element on the fly while computing the volumetric RPD.
Compared with existing voxel-based or point-cloud-based methods, our approach is the first to adaptively and directly revise the medial mesh without globally modifying the dependent structure, such as voxel size or sampling density, while preserving its topology and medial features. In comparison with the feature preservation method MATFP \cite{2022MATFP}, our method provides geometrically comparable results with fewer spheres and more robustly captures the topology of the input 3D shape.
\end{abstract}

%
%



\begin{teaserfigure}
    \centering
     \begin{overpic}[width=\linewidth]{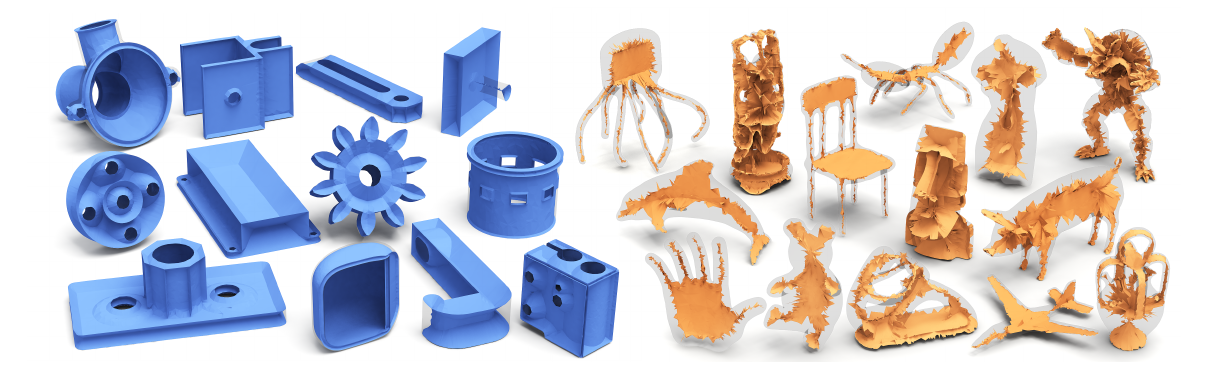}
    \end{overpic}
    \caption{A gallery of our topology-preserving 3D medial axis results. The input shapes are shown in transparency, while the computed medial axis are shown in blue (left) and brown (right) colors. The left side displays the medial axis computed on CAD models, and the right side shows those on organic models.}
    \label{fig:teaser}
\end{teaserfigure}

\maketitle

\input{1_introduction}
\input{2_relatedworks}

\input{3_preliminaries}

\input{4_methods}

\input{5_tech_detail}
\input{6_results}
\input{7_conclusion}
\begin{acks}
The authors would like to thank the anonymous reviewers for their valuable comments and suggestions. Ningna Wang and Xiaohu Guo were partially supported by the National Science Foundation (OAC-2007661). Hui Huang was partially supported by the National Natural Science Foundation of China (U21B2023, U2001206) and Guangdong Basic and Applied Basic Research Foundation (2023B1515120026).
\end{acks}

\bibliographystyle{ACM-Reference-Format}
\bibliography{reference}

\clearpage
\input{8_figureonly}

\clearpage



\end{document}

%% file: newcommands.tex
\theoremstyle{assumption}
\newtheorem*{assumption}{Assumption}
\theoremstyle{definition}
\newtheorem{definition}{Definition}
\newtheorem{theorem}{Theorem}

\newcommand{\red}[1]{\textcolor{red}{#1}}

\newcommand{\violet}[1]{\textcolor{violet}{#1}}
\newcommand{\rspace}{\mathbb{R}}
\newcommand{\model}{\mathcal{S}}
\newcommand{\bmodel}{\partial \mathbf{S}}

\newcommand{\ma}{\mathcal{M}}
\newcommand{\mmesh}{\mathcal{M}_s}
\newcommand{\msphereset}{\Gamma}
\newcommand{\msphere}{\mathbf{m}}
\newcommand{\mcenter}{\boldsymbol \theta}
\newcommand{\anypoint}{\mathbf{x}}

\newcommand{\melement}{\mathbf{s}}

\newcommand{\powerdist}{d_{pow}}
\newcommand{\domain}{\Omega}
\newcommand{\powercell}{\Omega^{pow}}

\newcommand{\euler}{\mathcal{X}}
\newcommand{\rpd}{\mathcal{Q}}
\newcommand{\rpset}{\mathbf{\omega}}
\newcommand{\rpc}{\mathbf{\omega_c}} 
\newcommand{\rpf}{\mathbf{\omega_f}} 
\newcommand{\rpe}{\mathbf{\omega_e}} 
\newcommand{\rpv}{\mathbf{\omega_v}} 
\newcommand{\halfspace}{\Pi^{+}}

\newcommand{\ie}{\textit{i.e., }}
\newcommand{\eg}{\textit{e.g., }}
\newcommand{\hd}{\epsilon}

\newcommand{\geoerror}{\delta^{\epsilon}}

%% file: 1_introduction.tex
\section{Introduction}

As a fundamental geometric structure1, the medial axis ~\cite{blum1967transformation} captures the topological equivalence and geometric protrusions of the input shape. The medial axis $\ma$ of a shape $\model$ is defined as the set of vertices with two or more nearest neighbors on the shape boundary $\bmodel$. The \textit{medial axis transform} (MAT) is a combination of the medial axis and its radius function. The topological and geometric properties of medial axis allow it to become the foundation for other skeletal shape descriptors~\cite{tagliasacchi20163d} and has been used in approximating~\cite{hu2022immat, hu2023s3ds, ge2023point2mm, yang2020p2mat, yang2018dmat, Lan2020MedialElastics, petrov2024gem3d}, simplifying~\cite{li2015qmat, dou2021coverage, yan2016erosion}, and analyzing shapes~\cite{Hu2019MATNet, dou2020top, lin2021point2skeleton, fu2022easyvrmodeling, xu2024cwf, BRepVP24, xu2022rfeps, xu2023globally, noma2024surface}.
Some literature~\cite{tagliasacchi20163d, kustra2013, kustra2015} refers to the term `medial axis' as 2D skeletons and uses `medial surface' for 3D structures. For clarification, we consider `medial axis' as a broader definition that includes both 2D and 3D. However, in this paper, we focus exclusively on the 3D medial axis.

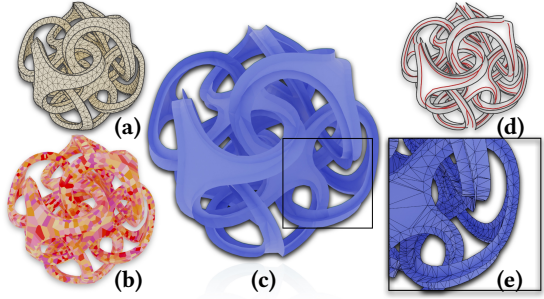
\begin{figure}[!h]
    \centering
    \begin{overpic}[width=0.85\linewidth]{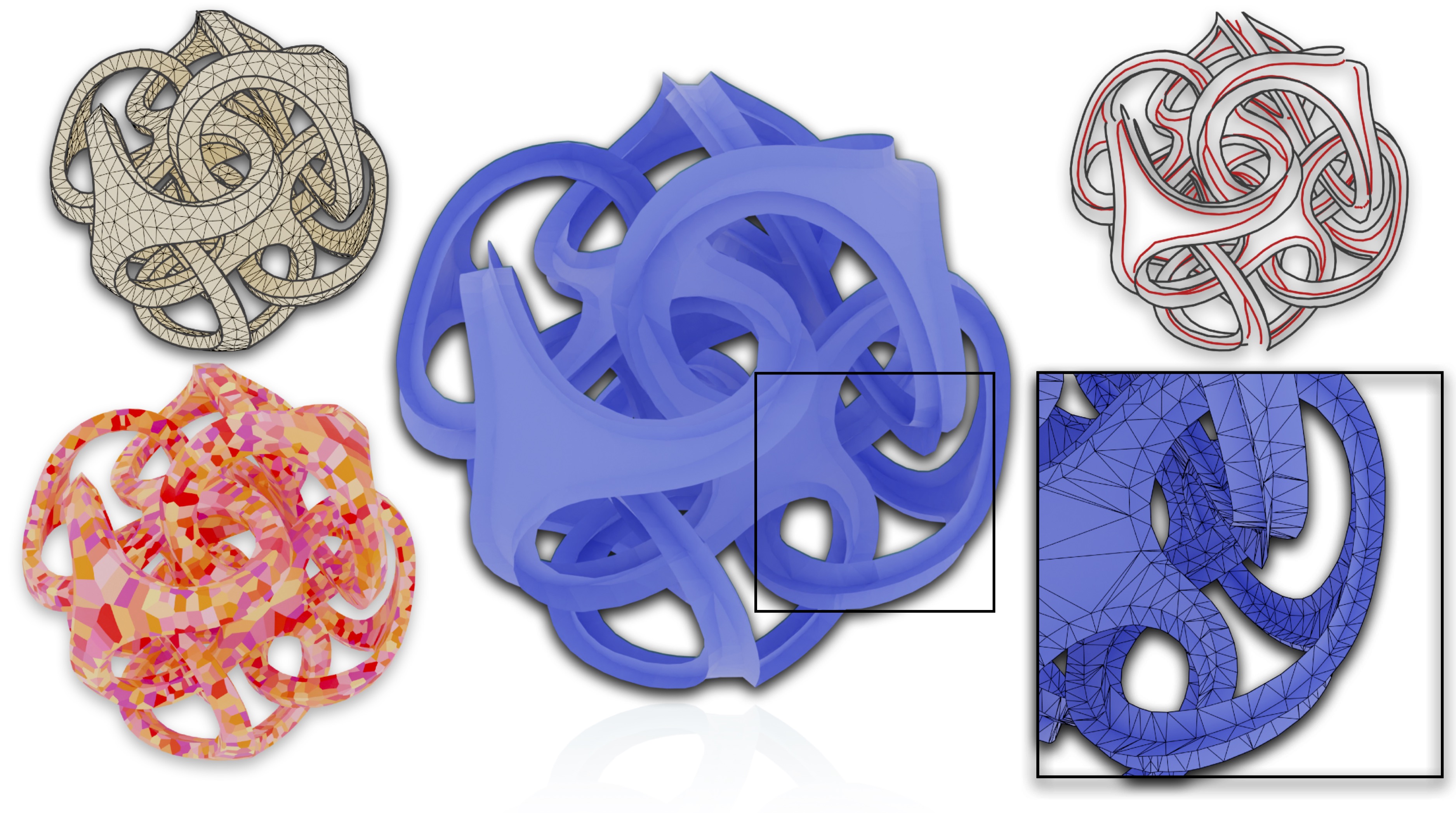}
    \put(21, 31){\textbf{(a)}}
    \put(21, 3){\textbf{(b)}}
    \put(46, 3){\textbf{(c)}}
    \put(91, 31){\textbf{(d)}}
    \put(91, 3){\textbf{(e)}}
    \end{overpic}
    \caption{We propose a novel volumetric RPD-based framework for computing the medial axis while preserving topology, medial features, and geometry. (a) Input tetrahedral mesh with pre-detected surface sharp features; (b) the RPD; (c) the generated medial mesh; (d) the generated external (in black) and internal (in red) features; (e) a zoomed-in view of the generated medial mesh.}
    \label{fig:teaser_intro}
\end{figure}

\begin{figure}[!h]
    \centering
    \begin{overpic}[width=\linewidth]{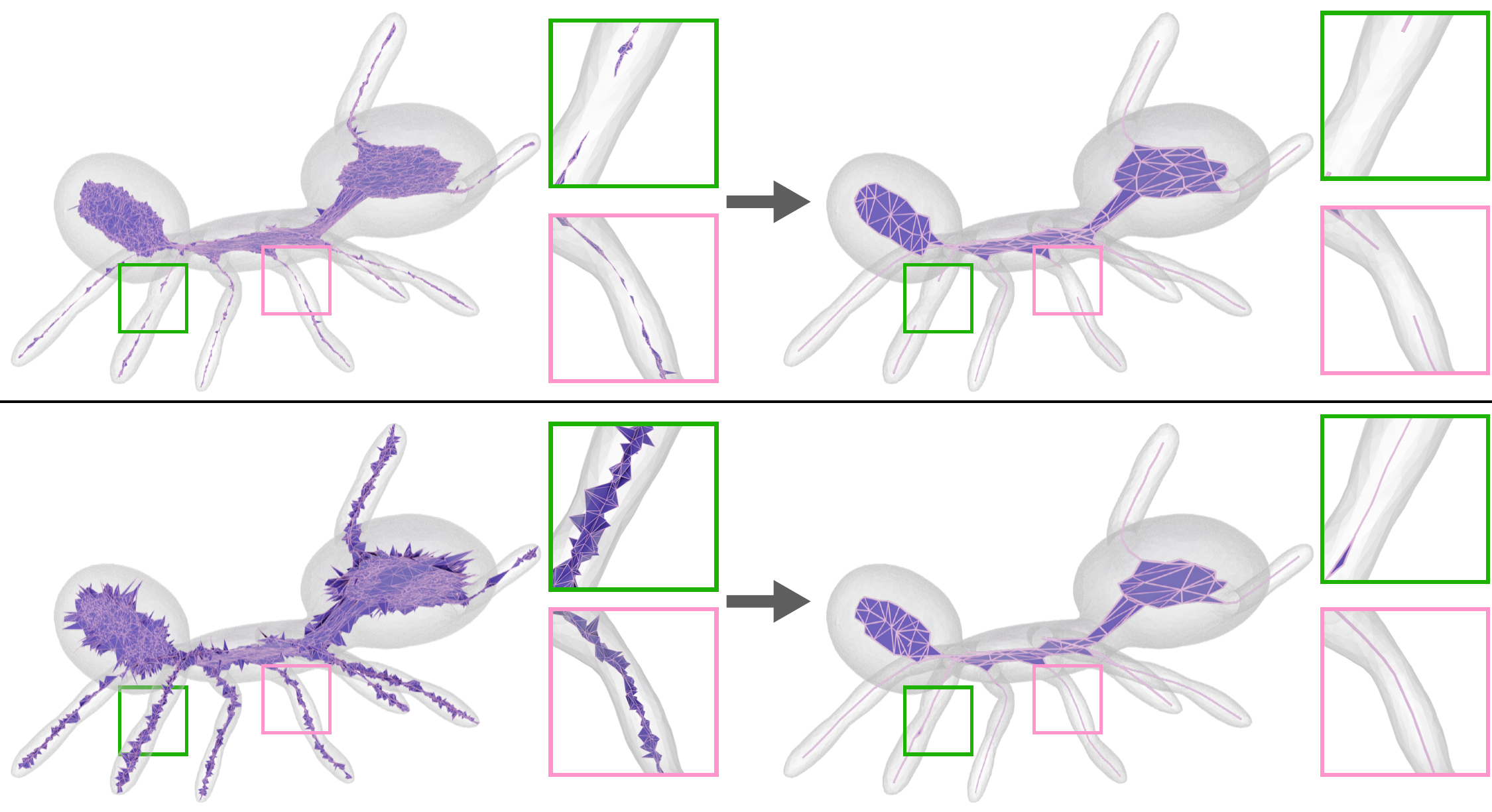}
    \put(50, 50){\textbf{MATFP}}
    \put(2, 50){\small{\red{Euler$=4$}}}
    \put(2, 46){\small{\red{CC$=3$}}}
    \put(50, 23){\textbf{Ours}}
    \put(2, 23){\small{Euler$=1$}}
    \put(2, 19){\small{CC$=1$}}
    \put(15, -2){\textbf{(a)}}
    \put(68, -2){\textbf{(b)}}
    \end{overpic}
    \caption{Illustration of the importance of the medial axis's homotopy equivalence property. (a) Two medial meshes generated by MATFP~\cite{2022MATFP} and our method. Both initial medial meshes contain approximately $5k$ medial spheres. The Euler characteristic of MATFP's mesh is $4$, while our result is $1$ (the ground truth is $1$). The connected component (CC) results are $3$ and $1$, respectively (the ground truth is $1$). (b) Two simplified medial meshes generated using the simplification algorithm Q-MAT~\cite{li2015qmat} with the target number of medial spheres set to $170$. Both the initial and simplified medial meshes from MATFP~\shortcite{2022MATFP} exhibit `broken' legs, whereas our method preserves the structure.}
    \label{fig:intro_topo}
\end{figure}
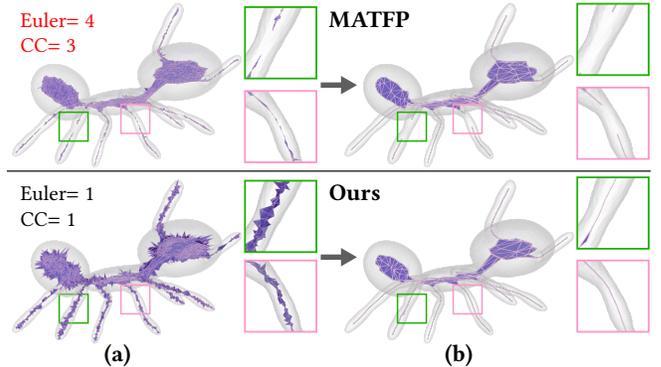

Since an exact 3D medial axis is notoriously difficult to compute, most existing methods resort to computing an approximated MAT. This approximation aims to retain as many properties of the medial axis as possible, both topologically and geometrically, while being capable of handling various inputs, including smooth shapes (\ie organic models) and non-smooth shapes (\ie CAD models). Existing MAT approximation methods often consider these properties as a trade-off.
For instance, the Voxel Cores (VC) method~\cite{yan2018voxel} provides a strong theoretical guarantee, allowing it to approximate the medial axes of $\mathcal{C}^2$ smooth shapes with homotopy equivalence. 
However, it performs poorly in terms of geometric convergence when approximating CAD models with non-smooth external features (\ie convex sharp edges and corners). Recent progress for computing MAT has demonstrated advantages in preserving \textit{medial features} (see Fig.~\ref{fig:teaser_intro} (d) and Sec.~\ref{sec:pre_ma_features}) as well as geometric convergence for CAD models using the surface restricted power diagram (RPD) based framework~\cite{2022MATFP}. However, it cannot guarantee the topological preservation for its generated medial mesh w.r.t. to the input model, as their experiments show inconsistency of the Euler characteristics. 
The topology preservation property (also called `homotopy equivalence') of the medial axis with respect to the input shape refers to the concept that these two spaces can be continuously deformed into one another without tearing or gluing, ensuring they share the same fundamental shape or structure. Without this property, the computed medial mesh may encounter challenges in various downstream applications. For example, Fig.~\ref{fig:intro_topo} shows two medial meshes generated using MATFP~\cite{2022MATFP} and our method for the Ant model, along with two skeletonization results from the Q-MAT algorithm~\cite{li2015qmat}, using either method as the initial medial mesh for MAT simplification to identify significant and stable parts of the medial axis. Due to the lack of homotopy equivalence in MATFP~\cite{2022MATFP}, both the initial and simplified medial meshes for the Ant model exhibit `broken' legs, whereas our method successfully preserves this property.


In this paper, we present a novel framework for computing a topology-preserving MAT that is homotopy-equivalent to the input 3D shape. 
Our framework can also preserve \textit{medial features} (Sec.~\ref{sec:pre_ma_features}) and ensure the geometric approximation accuracy. We have found that the volumetric restricted power diagram (RPD, see the definition in Sec.~\ref{sec:pre_rpd_mm}) serves as a simple but effective intermediate structure between the input volumetric shape and the generated medial mesh. The medial mesh is generated as the dual structure of the volumetric RPD, followed by a thinning process (Sec.~\ref{sec:tech_thinning}). This structure allows us to localize the topological inconsistencies between the restricted power elements (cells, faces, edges, vertices) and its dual mesh simplices (vertices, edges, triangles, tetrahedrons), based on the Nerve Theorem (Sec.~\ref{sec:pre_nerve}).
Owing to recent progress in GPU-based 3D power diagram computations ~\cite{ray2018meshless, liu2020RVD, basselin2021RPD}, we propose a volumetric RPD-based strategy for topological checking and fixing that amends the medial mesh in a local manner. Additionally, we introduce a novel \textit{fractional Euler characteristic} strategy for efficient GPU-based computation of the Euler characteristic of each restricted element (such as cells, faces, edges) on the fly while computing the volumetric RPD.

The initial medial mesh is generated using a small number of spheres ($\eg$ 50 spheres), which is then revised iteratively by adding new spheres in local regions and updating the corresponding partial RPD.
Each revision only happens locally, involving the checking of each individual restricted power cell (Sec.~\ref{sec:fix_topo}), their relation with neighboring power cells (Sec.~\ref{sec:fix_extf_intf}), and their approximation to the input surface boundary (Sec.~\ref{sec:fix_geo}). Compared with existing point-cloud-based methods~\cite{amenta2001power, miklos2010sat}, voxel-based methods~\cite{yan2018voxel}, and surface-RPD based method~\cite{2022MATFP}, our adaptive strategy can approximate the medial axis of both 3D CAD shapes and organic models with the preservation of homotopy equivalence, medial features, and geometric convergence, requiring a lower number of medial spheres.

The main contributions of this paper include:
\begin{itemize}
    \item We present a complete volumetric RPD-based framework for computing 3D medial mesh of both CAD models and organic models, while preserving the homotopy equivalence, medial features, and geometric convergence.
    \item For input volumetric shapes with no cavities, we propose an adaptive partial-RPD updating strategy for topological checking and amendment. This approach involves localized assessment of the connected components and Euler characteristics of each restricted element.
    \item We propose a novel fractional Euler characteristic algorithm for the efficient GPU-based computation of Euler characteristic for each restricted cell, face, and edge on the fly while computing the volumetric RPD in parallel.
    \item 
    We propose a GPU-based local geometric metric that enables us to adaptively refine the 3D medial mesh until its enveloping volume (see ~\ref{fig:fix_geo}) achieves a desired geometric accuracy relative to the input shape.
\end{itemize}

%% file: 2_relatedworks.tex
\section{Related Works}
\label{sec:related_works}

In this section, we review several representative works for approximating the 3D medial axis. Additionally, since our work relies on the computation of power diagrams in restricted domains, we also review algorithms for computing restricted Voronoi and power diagrams in volumes.

\subsection{Medial Axis Computation}
The approximation of the 3D medial axis mainly falls into three categories. Additionally, there are many regularization methods for pruning noisy branches of the medial axis~\cite{li2015qmat, yan2016erosion, dou2021coverage, wang2024coverage}, which require either an initial approximation of the 3D medial axis or initial candidate inner balls as input.
Please refer to survey materials~\cite{siddiqi2008medial, tagliasacchi20163d} for further discussion on this topic.

\paragraph{Algebraic methods} 
These methods tend to create exact and analytic representation of MAT from a given boundary representation while tracing the features of medial axes from the shape boundary inward ($\eg$ seam and junction)~\cite{culver2004exact, milenkovic1993robust, sherbrooke1996algorithm}. However, these methods are algorithmically challenging and computationally expensive due to the need for solving nonlinear systems, which limits their applicability to a simple class of shapes ($\eg$ polyhedron meshes with at most hundreds of faces).

\paragraph{Voxel-based methods} 
These methods attempt to approximate the 3D medial axis with a piecewise-constant interpolation based on a uniform sampling (\ie voxels)~\cite{sobiecki2014comparison, saha2016survey, yan2018voxel}. Many voxel-based methods are guided by a Euclidean distance field \cite{rumpf2002continuous, hesselink2008euclidean}, a derived gradient field~\cite{siddiqi2002hamilton}, or more global shape information~\cite{jalba2015unified}. 

\paragraph{Point-cloud-based methods} 
These methods place sample points around the boundary of the shape and consider either an interior subset of Voronoi diagram of those surface samples or some derivative structures.
\textit{Sphere-shrinking-based} (SS) methods~\cite{kustra2013, kustra2015} generate maximal inscribed balls (so-called \textit{medial spheres}) from given point clouds with normals using the \textit{sphere-shrinking} algorithm~\cite{ma20123shrink}. 
\textit{Angle-based filtering methods} ~\cite{amenta2001power, brandt1992continuous, dey2002approximate, dey2004approximating} are approaches to filter the Voronoi diagram and select the subset of Voronoi diagram of the boundary samples that meets an angle criterion. 
\textit{$\lambda$-medial axis methods} ~\cite{chazal2005lambda, chazal2008smooth} discard a medial sphere if its radius is smaller than a given filtering threshold $\lambda$. 
The scale axis transform (SAT)~\cite{miklos2010sat} exploits union-of-balls (UoB) and removes spikes while retaining small features by scaling medial spheres.
MATFP~\cite{2022MATFP} uses the inner Voronoi vertices as initial sphere centers, then updates those spheres' centers and radii as close as possible to the ground truth (GT), through an energy optimization framework. 


\begin{table}[h!]
\caption{Summary of properties for five selected 3D medial axis approximation methods: PC~\cite{amenta2001power}, SAT~\cite{miklos2010sat}, $\lambda$MA~\cite{chazal2005lambda, chazal2008smooth}, SS~\cite{kustra2013, kustra2015}, VC~\cite{yan2018voxel}, MATFP~\cite{2022MATFP}, and our method. The \checkmark means the property is fulfilled and $\circledcirc$ represents the property is conditionally satisfied.}
\begin{center}
\begin{tabular}{ |p{2.1cm}||p{0.4cm}|p{0.5cm}|p{0.5cm}|p{0.3cm}|p{0.3cm}|p{0.9cm}|p{0.5cm}|}
 \hline
 Property & PC & $\lambda$MA & SAT & SS & VC & MATFP & Ours\\
 \hline
 Homotopy       
 &$\circledcirc$ &$\circledcirc$ &$\circledcirc$ & &$\circledcirc$ & &\checkmark \\
 Medial Features
 &$\circledcirc$ &$\circledcirc$ &$\circledcirc$ & &$\circledcirc$ & \checkmark &\checkmark \\
 Thinness
 &  &\checkmark  &  &  & \checkmark & \checkmark & \checkmark\\
 Centeredness   
 &$\circledcirc$ &$\circledcirc$ &$\circledcirc$ &\checkmark &$\circledcirc$ &\checkmark & \checkmark \\
 Reconstructibility
 &$\circledcirc$  &$\circledcirc$  &$\circledcirc$ &$\circledcirc$ &$\circledcirc$ &\checkmark &\checkmark \\
 Scalability
 &\checkmark &\checkmark  &\checkmark &\checkmark &  & & \\
 \hline
\end{tabular}
\end{center}
\label{tab:rel_properties}
\end{table}

We select several representative methods mentioned above, and discuss several important properties of 3D medial axis that are addressed by these methods. The summary is given in Table~\ref{tab:rel_properties}.

\paragraph{Homotopy Equivalence} 
While most existing methods maintain the homotopy equivalence between the extracted 3D medial axis with the input shape, such equivalence is often conditioned on the sampling rate of points on the shape boundary. For example, both the PC~\cite{amenta2001power} method and SAT~\cite{miklos2010sat} method rely on the \textit{$r$-sample} condition measured by the \textit{local feature size} (LFS) as the `sufficiently dense' input sample density. The $\lambda$-medial axis~\cite{chazal2005lambda} is homotopy equivalent to the medial axis when $\lambda$ is less than the weak feature size.
Similarly, VC~\cite{yan2018voxel} shows that for any voxel sizes smaller than $(2\sqrt{3}/3)r$, the union of voxels is homeomorphic to the original shape. Our method, on the contrary, has no assumption on the sampling density of the input shape in order to preserve the topology of the generated medial mesh. 
Both the SS method~\cite{kustra2013, kustra2015} and the MATFP method~\cite{2022MATFP} do not provide any guarantee on the topology of their generated medial axis approximation.

\paragraph{Medial Feature Preservation}
Please refer to Sec.~\ref{sec:pre_ma} for a detailed discussion of the medial sphere classification and medial features (\textit{external features} and \textit{internal features}), which are also intensively discussed in seminal works~\cite{tagliasacchi20163d, kustra2015, 2022MATFP}.
To the best of our knowledge, MATFP~\cite{2022MATFP} is the first method that attempts to approximate medial axes with non-smooth regions, such as convex sharp edges and corners (so called external features). Our method uses the same strategy for external feature preservation, by placing zero-radius spheres on non-smooth regions, as described in Sec.~\ref{sec:fix_extf_intf}. All other methods mentioned in Table.~\ref{tab:rel_properties} perform poorly in terms of external features preservation.
Internal features are hidden structure of the shape, which consists of spheres with more than two tangent points on the surface (\ie on seams or junctions). 
The SS methods~\cite{kustra2013, kustra2015} are not able to preserve internal features. This limitation arises because the sphere-shrinking algorithm only provides medial spheres of type $T_2$ (on sheet), as each sphere is only tangent to two surface points. For medial spheres with more than two tangent points on the surface, such as $T_3$ (on seam) or $T_4$ (on junction), these methods cannot generate spheres that lie on the internal features of the 3D medial axis. 
The MATFP~\cite{2022MATFP} and our method are able to detect the ill-posed regions where the internal feature spheres (\eg $T3$ or $T4$) are lacking, and use \textit{sphere-optimization} algorithm to insert new spheres.
All other sampling-based methods (PC, $\lambda$MA, SAT, VC) in Table.~\ref{tab:rel_properties} require dense sampling rate (as discussed above) in order to generate approximated internal feature spheres.


\paragraph{Thinness}
By definition, the 3D medial axis is a thin structure, so the approximated medial mesh should contain no 3-dimensional cells. 
This property does not hold for PC~\cite{amenta2001power} and SAT~\cite{miklos2010sat}, as their results contain large amounts of `closed pockets'. 
The per-manifold connectivity strategy proposed by Kustra et al.~\shortcite{kustra2015} does not consider the thinness property of the 3D medial axis. 
A subset of Voronoi diagram from a sufficiently close and dense noisy sampling, filtered as the $\lambda$MA~\cite{chazal2005lambda,chazal2008smooth} methods, retains the thinness property as no element of this subset is dual to a 0-dimensional vertex. Similarly, the VC~\cite{yan2018voxel} method achieves this thinness because the \textit{voxel core} is a direct consequence of its duality with Delaunay triangulation. 
Our method follows the same thinning process as MATFP~\cite{2022MATFP} (described in Sec.~\ref{sec:tech_thinning}) to preserve this property. 

\paragraph{Centeredness}
The PC~\cite{amenta2001power}, $\lambda$MA~\cite{chazal2005lambda, chazal2008smooth}, SAT~\cite{miklos2010sat}, and VC~\cite{yan2018voxel} methods all require a certain sampling density to ensure that the centeredness of the generated medial spheres is preserved. This requirement arises because these methods use inner Voronoi balls as initial candidate spheres and then select a subset of them. However, Voronoi balls often protrude from the surface as they are naturally circumscribed over the sampling points. 
The MATFP method~\cite{2022MATFP} addresses this by updating and pushing these protruding inner Voronoi balls to be tangential to the surface. Our method inherits this advantage from MATFP and preserves the centeredness property by directly creating medial spheres tangent to at least two points on the surface using the \textit{sphere-shrinking} and \textit{sphere-optimization} algorithms described in Sec.~\ref{sec:tech_sphere_gen}. 
The SS methods~\cite{kustra2013, kustra2015} can also maintain the centeredness property as they use the \textit{sphere-shrinking} algorithm~\cite{ma20123shrink} to generate spheres tangent to exactly two points on the surface.

\paragraph{Reconstructibility}
Since only the MATFP method~\cite{2022MATFP} and our method take external features (also see the discussion of \textit{Medial Features} above) into account, all other methods shown in Table~\ref{tab:rel_properties} would reconstruct rounded shapes for 3D CAD models with convex sharp edges and corners. Our method also proposes a local geometric metric that can adaptively refine the 3D medial mesh until its enveloping volume, as the main component of the reconstruction, reaches a desired accuracy.

\paragraph{Scalability}
The VC method~\cite{yan2018voxel} requires fine vwoxel resolution, hence incurring higher computational cost to achieve a comparable geometric accuracy as point-cloud-based methods such as PC~\cite{amenta2001power}, $\lambda$MA~\cite{chazal2005lambda, chazal2008smooth}, and SAT~\cite{miklos2010sat}. Both RPD-based methods in Table~\ref{tab:rel_properties} (MATFP~\cite{2022MATFP} and Ours) requires multiple runs of surface or volumetric RPD computation through the clipping process, which is inevitably more time-consuming than others. Please see Sec.~\ref{sec:limitations} for more discussions.

\subsection{Restricted Voronoi and Power Diagram in Volumes}

Robust and accurate computation of 3D volumetric RPD is not a trivial task. The classical clipping algorithm~\cite{yan2010efficient} and industrial-quality libraries, such as CGAL~\cite{fabri2009cgal} and Geogram~\cite{levy2015geogram}, are still overly time-consuming to be used in tasks that require frequent iterative RPD computations. Fortunately, recent GPU-based approaches has shown strong ability to compute Voronoi and power diagrams on highly parallel architectures, where the geometry of each Voronoi or power cell can be evaluated independently~\cite{ray2018meshless,liu2020RVD,basselin2021RPD}. Basselin et al.~\shortcite{basselin2021RPD} proposed a method to directly evaluate the integrals over every restricted power cell without computing the combinatorial structure of power diagram explicitly. Since our work requires explicit structure of each power cell, we build upon the `Tet-Cell' strategy proposed by Liu et al.~\shortcite{liu2020RVD}. After discretizing the volume into a tetrahedral mesh~\cite{hu2020ftetwild}, the intersection of a tetrahedron (\textit{tet} for short) with a cell can be calculated in a parallel manner. This parallel GPU-based volumetric RPD implementation reduces the computational cost significantly, $\eg$ from 8s to 0.6s for a model with 10k tets and 10k cells.

%% file: 3_preliminaries.tex
\section{Preliminaries}
\label{sec:pre}

\subsection{Medial Axis and Medial Mesh}
\label{sec:pre_ma}

\begin{figure}[!h]
    \centering
    \begin{overpic}[width=0.8\linewidth]{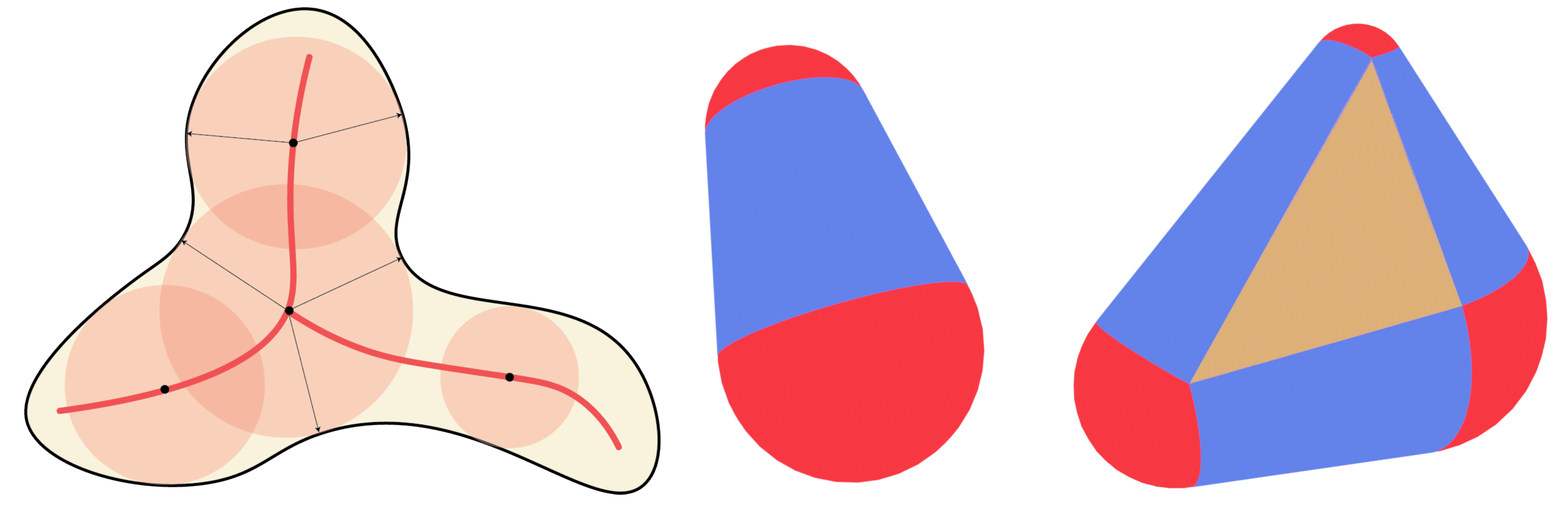}
    \put(20, -3){\textbf{(a)}}
    \put(55, -3){\textbf{(b)}}
    \put(80, -3){\textbf{(c)}}
    \put(5, 16){$\bmodel$}
    \put(25, 10){$\ma$}
    \put(48, 31){$\msphere_i$}
    \put(54, 5){$\msphere_j$}
    \put(84, 32){$\msphere_i$}
    \put(70, 5){$\msphere_j$}
    \put(95, 9){$\msphere_k$}
    \end{overpic}
    \caption{(a) The medial axis $\ma$ of a shape $\model$ in $\rspace^2$. (b) The medial cone as a linear interpolation of two medial spheres $\msphere_i$ and $\msphere_j$. (c) The medial slab as a linear interpolation of three spheres $\msphere_i$, $\msphere_j$, and $\msphere_k$.}
    \label{fig:pre_mm}
\end{figure}

\paragraph{Medial Aixs} 
The medial axis $\ma$ of a closed, oriented, and bounded shape $\model \in \rspace^2 \textbackslash \rspace^3$, as shown in Fig.~\ref{fig:pre_mm} (a), is defined as the locus of centers of spheres that are tangent to at least two boundary points of $\model$, without containing any other boundary points in its interior. The \textit{medial axis transform} (MAT) is formed by the medial axis $\ma$ and its radius function. It has been shown that any bounded open set $\model$ is homotopy equivalent to its medial axis $\ma$~\cite{lieutier2004any, miklos2010sat, lieutier2023hausdorff}.

\paragraph{Medial Features}
\label{sec:pre_ma_features}

\begin{wrapfigure}{r}{2cm}
\vspace{-3.5mm}
  \hspace*{-4mm}
  \centerline{
  \includegraphics[width=25mm]{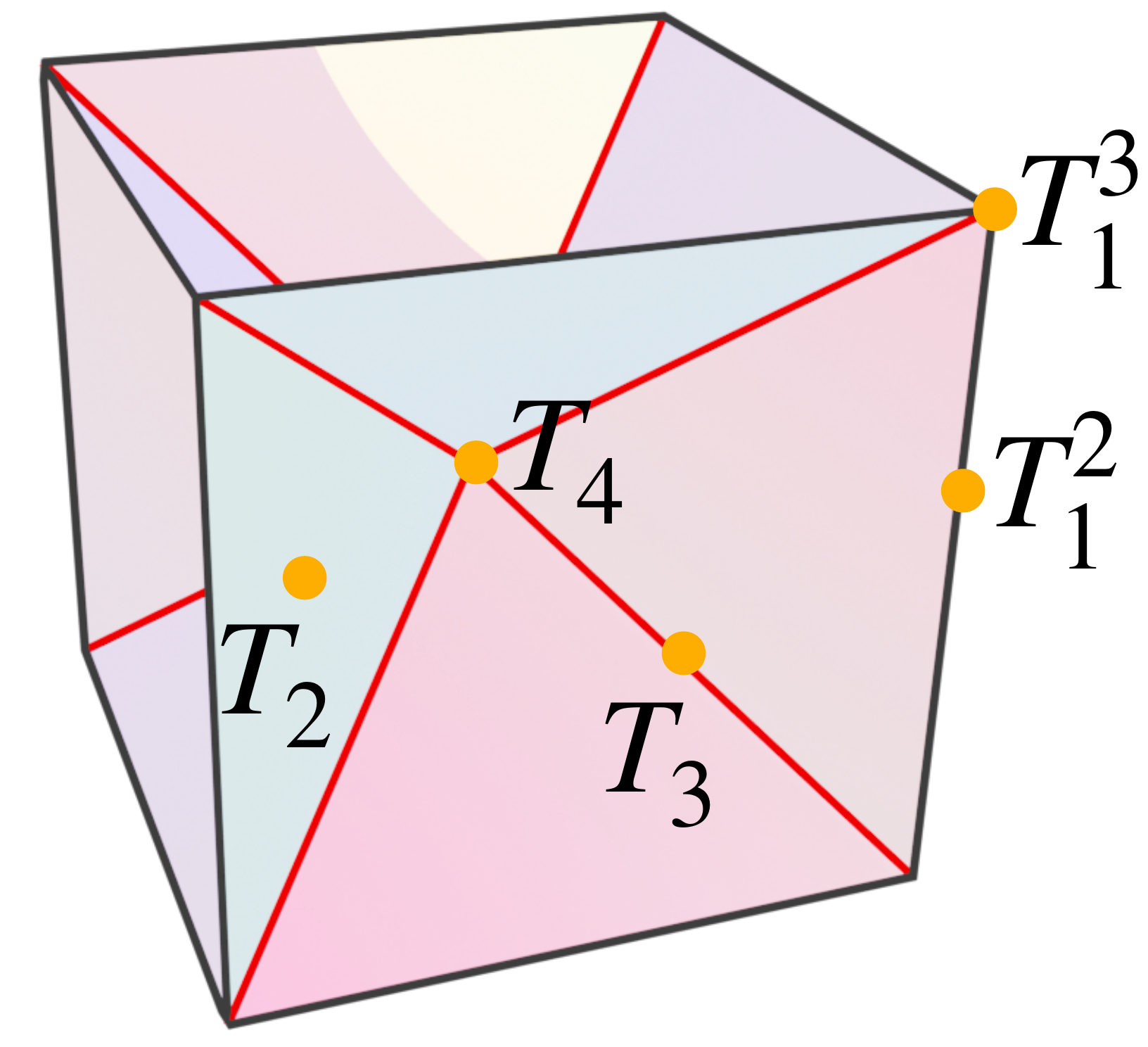}
  }
  \vspace*{-4mm}
\end{wrapfigure}
We follow the same notation for classifying medial spheres as MATFP \cite{2022MATFP}. Taking the medial axis of a cube model as an example. The medial spheres of type $T_2$ are tangent to the boundary of the shape $\bmodel$ at exactly two distinct tangential points, which is the most ordinary case as they lie on 2-manifold \textbf{sheets} of $\ma$. The intersection of three or more sheets forms a \textbf{seam} of $\ma$, consisting of a set of $T_3$ medial spheres tangent to $\bmodel$ at three distinct points. Multiple seams could intersect at a \textbf{junction} sphere of $T_{\geq4}$. 
The \textbf{internal features} of medial axis include those internal spheres located on seams and junctions. 
For a non-smooth model (\ie CAD model) that contains sharp edges and corners, either convex or concave, we use a dihedral angle less than $\pi-\phi$ and greater than $\pi+\phi$ to define the \textit{convex sharp edge} and \textit{concave sharp edge} respectively \cite{abdelkader2020vorocrust, 2022MATFP}. Here $\phi$ is a user-defined variable and the user can also mark sharp edges manually.
For CAD models, the \textbf{external features} of medial axis consist of those convex sharp edges and their associated corners. The medial axis should pass through those surface points that are locally convex and non-smooth. As zero-radius spheres are placed on external features, these medial spheres can be either $T_1^2$ (on a convex edge) or $T_1^{\geq3}$ (on a convex corner). Please refer to Wang et al.~\shortcite{2022MATFP} for a more complete categorization.

\paragraph{Medial Mesh}
Following the convention in Q-MAT~\cite{li2015qmat}, we approximate the MAT of a 3D shape using a non-manifold simplicial complex, called \textit{medial mesh} $\mmesh$, consisting of three types of medial primitives: vertices, edges and triangles. Each vertex of $\mmesh$ represents a \textit{medial sphere} $\msphere=(\mcenter, r)$, with center $\mcenter \in \mathbb{R}^3$ and radius $r \in \mathbb{R}$. The union of enveloping volume of all the medial primitives can be used to reconstruct the surface. The enveloping volume of an edge of $\mmesh$ is called a \textit{medial cone}. As shown in Fig.~\ref{fig:pre_mm} (b), medial cone $\mathbf{e}_{ij}$ is a linear interpolation of two spheres $\mathbf{e}_{ij} = t\msphere_i+(1-t)\msphere_j$, $t\in [0,1]$. Similarly, the enveloping volume of a triangle $\mathbf{f}_{ijk}$  of $\mmesh$ is called \textit{medial slab}, shown in Fig.~\ref{fig:pre_mm} (c).

\subsection{Volumetric Restricted Power Diagram and its Dual}
\label{sec:pre_rpd_mm}

We follow a similar RPD-based strategy as MATFP~\cite{2022MATFP} to construct the medial mesh $\ma_s$ as the approximation of the 3D MAT $\ma$. Here, $\ma_s$ is generated from the \textit{restricted regular triangulation} (RRT), which is the dual of \textit{restricted power diagram} (RPD). The main difference is that MATFP computes the medial mesh $\ma_s$ by selecting a subset of simplices in \textit{regular triangulation} (RT) whose dual elements in \textit{power diagram} (PD) has non-empty intersections with the input shape $\model$, as MATFP relies on surface-RPD only. In this paper, we construct $\ma_s$ directly from the dual of volumetric RPD  (see Sec.~\ref{sec:tech_tet_sphere} for technical details). In this section, we discuss the formal definitions of RPD and its dual.

\begin{figure}[!h]
    \centering
    \begin{overpic}[width=0.7\linewidth]{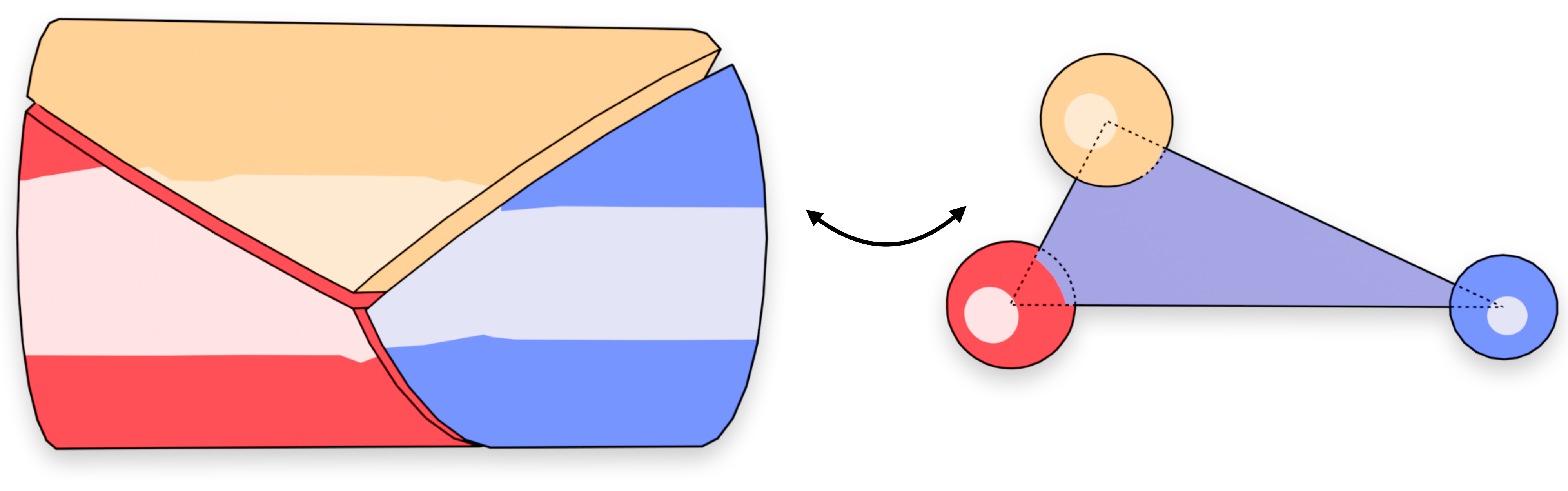}
    \put(22, -4){\textbf{(a)}}
    \put(73, -4){\textbf{(b)}}
    \put(76, 25){$\msphere_i$}
    \put(66, 5){$\msphere_j$}
    \put(95, 5){$\msphere_k$}
    \end{overpic}
    \caption{The duality between the volumetric RPD (a) of three medial spheres and the generated medial mesh (b)}
    \label{fig:pre_dual_mm_rpd}
\end{figure}
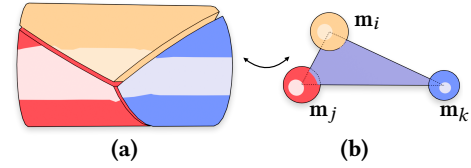

As a generalization of the Voronoi diagram, the power diagram (PD) \cite{aurenhammer1987power} is generated by a set of weighted points, and coincides with the Voronoi diagram in the case of equal weights. 
The power diagram of a set of medial spheres $\msphereset=\{\msphere_i\}_{i=1}^n$ is a partition of the domain $\domain \subset \rspace^d$ into a set of \textit{power cells}. Each power cell $\powercell_i$ consists of the points $\anypoint \in \domain$ closest to a particular sphere $\msphere_i$ as: 
\begin{equation}
    \powercell_i : \{ \anypoint \in\domain | \powerdist(\anypoint,\msphere_i) \leq \powerdist(\anypoint,\msphere_j), j \neq i \},
\end{equation}
where $\powerdist(\anypoint,\msphere_i)= ||\anypoint-\mcenter_i||^2 - r_i^2$ is the \textit{power distance} between the point $\anypoint$ and the medial sphere $\msphere_i=(\mcenter_i, r_i)$.

A power diagram restricted within a bounded shape $\model$ is called the \textit{restricted power diagram} (RPD) as:
\begin{equation}
    \rpd(\msphereset) = \bigcup_{\msphere_i\in\msphereset}\rpc(\msphere_i),
\end{equation}
where each sub-domain $\rpc(\msphere_i)$ is the restriction of the power cell $\powercell_i$ of the medial sphere $\msphere_i$ within $\model$:
\begin{equation}
    \rpc(\msphere_i) = \powercell_i \cap \model.
\end{equation}

Fig.~\ref{fig:pre_dual_mm_rpd} (a) shows the RPD of three spheres, where the input volume $\model$ is divided into three sub-domains clipped by three radical hyperplanes. We show its duality in Fig.~\ref{fig:pre_dual_mm_rpd} (b). The duality between the volumetric RPD and the medial mesh $\mmesh$  can be summarized as follows:

\begin{itemize}
    \item Each sub-domain $\rpc(\msphere_i)$ is called a \textit{restricted power cell} (RPC) of medial sphere $\msphere_i$, which is dual to a vertex on $\mmesh$.
    \item The face shared by two adjacent RPCs is called a \textit{restricted power face} (RPF), $\rpf(\msphere_i, \msphere_j)=\rpc(\msphere_i)\cap\rpc(\msphere_j)$, which is dual to an edge $\mathbf{e}_{ij}$ on $\mmesh$.
    \item The edge shared by three RPCs is called a \textit{restricted power edge} (RPE), $\rpe(\msphere_i, \msphere_j, \msphere_k)=\rpc(\msphere_i)\cap\rpc(\msphere_j)\cap\rpc(\msphere_k)$, which is dual to a triangle face $\mathbf{f}_{ijk}$ on $\mmesh$. 
    \item A vertex shared by four RPCs $\rpv(\msphere_i, \msphere_j, \msphere_k, \msphere_s)$ is called a \textit{restricted powper vertex} (RPV), which is dual to a tetrahedron on $\mmesh$, if it exists. Note that all tetrahedra in the medial mesh will be pruned using a geometry-guided thinning algorithm proposed by MATFP~\cite{2022MATFP} (see \ref{sec:tech_thinning} for details).
\end{itemize}

\subsection{Nerve Theorem and Homotopy Equivalence}
\label{sec:pre_nerve}


In this section, we present our assumptions on the input model, as well as the theoretical foundation of our RPD-based topology-preserving strategy, grounded on the \textit{Nerve Theorem}.

\begin{assumption}
\label{def:assumption}
The input for our method is a manifold tetrahedral mesh with a single connected component, no self-intersection, and \textbf{no cavity}. 
It should be noted that this “no cavity” assumption holds for all CAD models and organic models that we found online.
\end{assumption}

\begin{definition}[Nerve]
\label{def:nerve}
Let $X$ be a topological space, and $\mathcal{U}=\{U_i\}_{i\in I}$ be any covering of $X$ where $I$ is a set of indices, so $X=\bigcup_{i\in I} U_i$.
The \textbf{nerve} of $\mathcal{U}$, denoted $\mathcal{N}(\mathcal{U})$, is defined as a set of finite subsets of the index set $I$. It contains all finite subsets $J\subseteq I$ such that the intersection of the $U_i$ whose sub-indices are in $J$ is non-empty~\cite{carlsson2021topological}:
\begin{equation}
    \mathcal{N}(\mathcal{U}) = \Bigl\{ J \subseteq I: \bigcap_{j\in J} U_j \neq \emptyset,~J~\text{finite set} \Bigl\}.
\end{equation}
\end{definition}

\begin{theorem}[Nerve Theorem]
\label{def:nerve_theorem}
If for each finite subset $J\subseteq I$, the set $\bigcap_{j\in J} U_j$ is either empty or contractible, then $\mathcal{N}(\mathcal{U})$ is homotopy-equivalent to $X=\bigcup \mathcal{U}$, that is, $\mathcal{U}$ is a good cover~\cite{leray1950anneau}.  
\end{theorem}

In the context of this paper, the input shape $\model$ defines the topological space $X$, and RPD $\rpd(\msphereset)$ is a covering $\mathcal{U}$ of $X$ given the set of $n$ medial spheres $\msphereset=\{\msphere_i\}_{i=1}^n$. That is, the RPD $\rpd(\msphereset)$ is a covering of the shape $\model$.
The medial mesh $\mmesh$, which is computed from the dual of the RPD $\rpd(\msphereset)$, can be considered as the nerve of the RPD $\rpd(\msphereset)$. This is because every simplex of the mesh $\mmesh$, \ie vertex, edge, triangle, tetrahedron, is exactly dual to each RPC, intersecting face (RPF) of two adjacent RPCs, intersecting edge (RPE) of three adjacent RPCs, and intersecting vertex (RPV) of four adjacent RPCs, respectively.

According to the Nerve Theorem, if every restricted elements (\ie RPCs, RPFs, RPEs, RPVs) are contractible, then the medial mesh $\mmesh$ is homotopy-equivalent to the input shape $\model$.

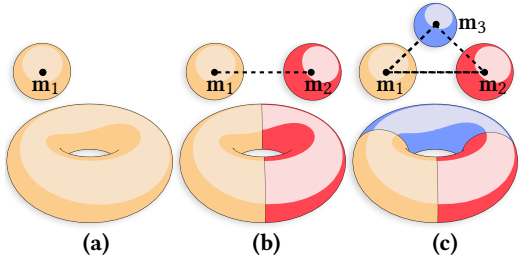
\begin{figure}[!h]
    \centering
    \begin{overpic}[width=0.8\linewidth]{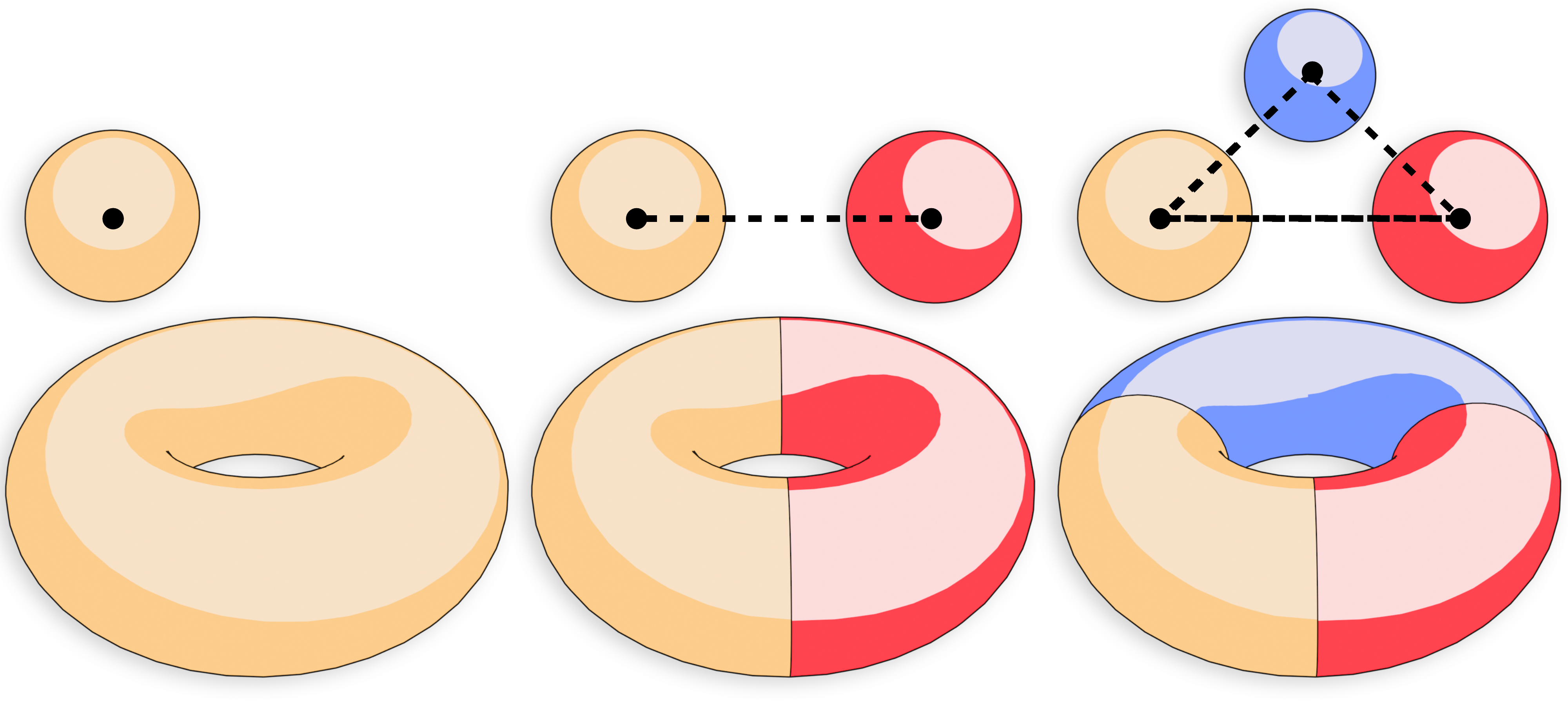}
    \put(15, -3){\textbf{(a)}}
    \put(5, 28){$\msphere_1$}
    \put(48, -3){\textbf{(b)}}
    \put(39, 28){$\msphere_1$}
    \put(58, 28){$\msphere_2$}
    \put(83, -3){\textbf{(c)}}
    \put(72, 28){$\msphere_1$}
    \put(92, 28){$\msphere_2$}
    \put(88, 40){$\msphere_3$}
    \end{overpic}
    \caption{The duality between the medial mesh (top) and volumetric RPD (bottom) for a 3D torus shape, where the homotopy equivalence does not hold for case (a) and case (b). An additional medial sphere $\msphere_3$ needs to be inserted in order to maintain the homotopy equivalence, as shown in (c). The dual edges are depicted as black dotted lines.}
    \label{fig:topo_fix}
\end{figure}

If we use $\rpset$ to represent any type of restricted element, including RPC $\rpc$, RPF $\rpf$, RPE $\rpe$, or RPV $\rpv$, the Euler-Poincar\'{e} formula shows:
\begin{equation}
    \euler(\rpset)=\beta_0(\rpset) - \beta_1(\rpset) + \beta_2(\rpset),
\end{equation}
where $\euler$ is the \textit{Euler characteristic} and $\beta_k$ is the $k$th Betti number representing the number of $k$-dimentional holes in $\rpset$. Since our input shape $\model$ contains no cavity, \ie $\beta_2(\model)=0$, all restricted elements also contains no cavity $\beta_2(\rpset)=0$. The formula can be rewritten as:
\begin{equation}
    \euler(\rpset)=\beta_0(\rpset) - \beta_1(\rpset).
\end{equation}
To test whether $\rpset$ is contractible, we need to check whether $\rpset$ is homotopy-equivalent to a one-point space. 
That is to say, the contractibility of $\rpset$ could be tested by two topological indicators:
(1) whether its number of \textit{connected components} $\beta_0(\rpset) = 1$; and if so, (2) whether its \textit{Euler characteristic} $\euler(\rpset) = 1$, which implies that its number of `circular' holes $\beta_1(\rpset) = 0$.

Let us take the shape torus in Fig~\ref{fig:topo_fix} as an example. When we use a single medial sphere $\msphere_1$ to represent the shape, its RPC $\rpc(\msphere_1)$ covers the whole torus but does not have the same topology as the medial sphere, as $\beta_0=1$ but $\euler=0$. The RPC $\rpc(\msphere_1)$ in Fig.~\ref{fig:topo_fix} (b) has $\beta_0=1$ and $\euler=1$ as expected, but its associated RPF $\rpf(\msphere_1, \msphere_2)$ has $\beta_0=2$. In Fig.~\ref{fig:topo_fix} (c) all restricted elements have $\beta_0=1$ and $\euler=1$ for all three medial spheres $\msphere_1$, $\msphere_2$ and $\msphere_3$.  Note that only RPCs and RPFs exist in this example (c) without any RPEs, hence only vertices and edges exist in the dual medial mesh. This contractibility testing process inspires our topology preservation algorithm proposed in Sec.~\ref{sec:fix_topo}.

%% file: 4_methods.tex
\section{Method}
\label{sec:method}
\begin{figure*}[!t]
    \centering
    \begin{overpic}[width=\linewidth]{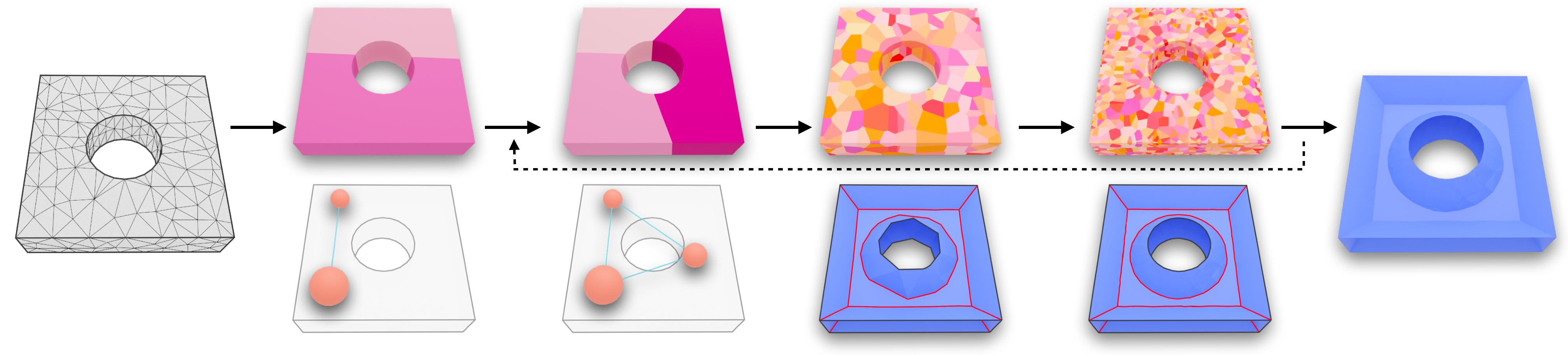}
    \put(15.5,15){\textbf{(a)}}
    \put(31.5,15){\textbf{(b)}}
    \put(49,15){\textbf{(c)}}
    \put(65.5,15){\textbf{(d)}}
    \put(82.5,15){\textbf{(e)}}
    \end{overpic}
    \caption{The overview of our computational pipeline. Given a tetrahedral mesh with surface features (sharp edges and corners) pre-detected as input, our method starts with an initial medial mesh of a small number of spheres ($\ie 2$ spheres) using the \textit{sphere-shrinking} algorithm~\cite{ma20123shrink}, shown in (a). Then the homotopy equivalence of the generated medial mesh w.r.t. the input shape is preserved by examining the topological equivalence of individual RPCs (Sec~\ref{sec:fix_topo}), shown in (b). We preserve the \textit{medial features} using the same method as MATFP~\cite{2022MATFP} (Sec~\ref{sec:fix_extf_intf}), shown in (c). Finally we preserve the geometric convergence based on a user-defined error threshold $\geoerror$ (Sec~\ref{sec:fix_geo}), shown in (d). We repeat this process until both topological preservation and geometric convergence are satisfied, and output the final result of generated medial mesh, shown in (e). }
    \label{fig:pipeline}
\end{figure*}
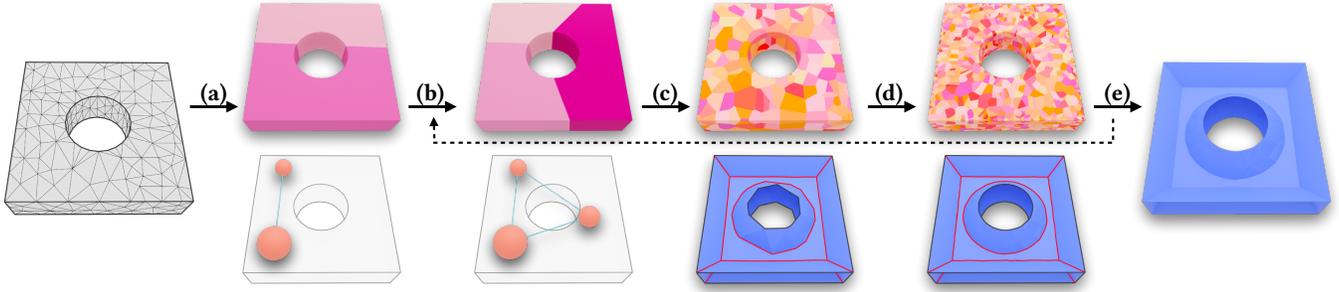

Given a manifold tetrahedral mesh (satisfying the Assumption in Sec.~\ref{sec:pre_nerve}) with its surface features (sharp edges and corners) pre-detected, our pipeline starts with an initial medial mesh of a small number ($\eg 50$) of randomly placed medial spheres computed with the \textit{sphere-shrinking} (Sec~\ref{sec:tech_sphere_gen}) algorithm~\cite{ma20123shrink}. 
The medial mesh is then refined iteratively through the following three steps. 
First, we preserve the homotopy equivalence of the generated medial mesh w.r.t. the input 3D shape by examining two topological indicators ($\beta_0$ and $\euler$) of all restricted elements for each medial sphere (Sec~\ref{sec:fix_topo}), as shown in Fig~\ref{fig:pipeline} (b).
Then, the \textit{medial features} are preserved by assessing if the RPCs of two adjacent medial spheres covers the same surface regions (Sec.~\ref{sec:fix_extf_intf}), as shown in Fig~\ref{fig:pipeline} (c). Finally, we ensure the geometric convergence by checking if the distance error between the shape boundary $\bmodel$ and the enveloping volume of medial mesh $\mmesh$ is smaller than a user-defined threshold $\geoerror$ (Sec.~\ref{sec:fix_geo}), as shown in Fig~\ref{fig:pipeline} (d). Each examination step is performed locally and new medial spheres are inserted for each preservation using two sphere generation strategies described in Sec.~\ref{sec:tech_sphere_gen}. For each iteration, we only update the RPD partially for cells that are related to the newly added medial spheres, rather than re-computing the whole RPD for all medial spheres. Please refer to both Sec.~\ref{sec:tech_detail} and  Supplementary Document for more implementation details.


\subsection{Topology Preservation}
\label{sec:fix_topo}

As discussed in Sec.~\ref{sec:pre_nerve}, the homotopy equivalence between the input shape $\model$ and the generated medial mesh $\mmesh$ is enforced by examining the following two topological indicators for each individual RPC and its associated RPFs and RPEs:
\begin{itemize}
    \item The number of \textit{connected components} (CC number) $\beta_0$ tells us the maximal subset of a topological space;
    \item The \textit{Euler characteristic} $\euler$ describes a topological space's structure regardless of the way it is bent. It can be calculated using following formula: $Euler = V - E + F - C$, where $V$ is the number of vertices, $E$ is the number of edges, $F$ is the number of faces, and $C$ is the number of volumetric cells.
\end{itemize}

To ensure homotopy equivalence, we are expecting each restricted element (\ie RPC, RPF, RPE) to have $CC~\beta_0=1$ and $Euler~\euler=1$.  
For each medial sphere $\msphere_i$, we perform a localized topological check for its corresponding restricted elements, and apply a straightforward refinement by adding new medial spheres if either their $CC\neq1$ or $Euler\neq1$. The details of our preservation strategy for those two topological indicators are given in Sec.~\ref{sec:cc_number} and Sec.~\ref{sec:euler_number}.

\subsubsection{CC Number $\beta_0$}
\label{sec:cc_number}

\begin{figure}[h!]
    \centering
    \begin{overpic}[width=0.6\linewidth]{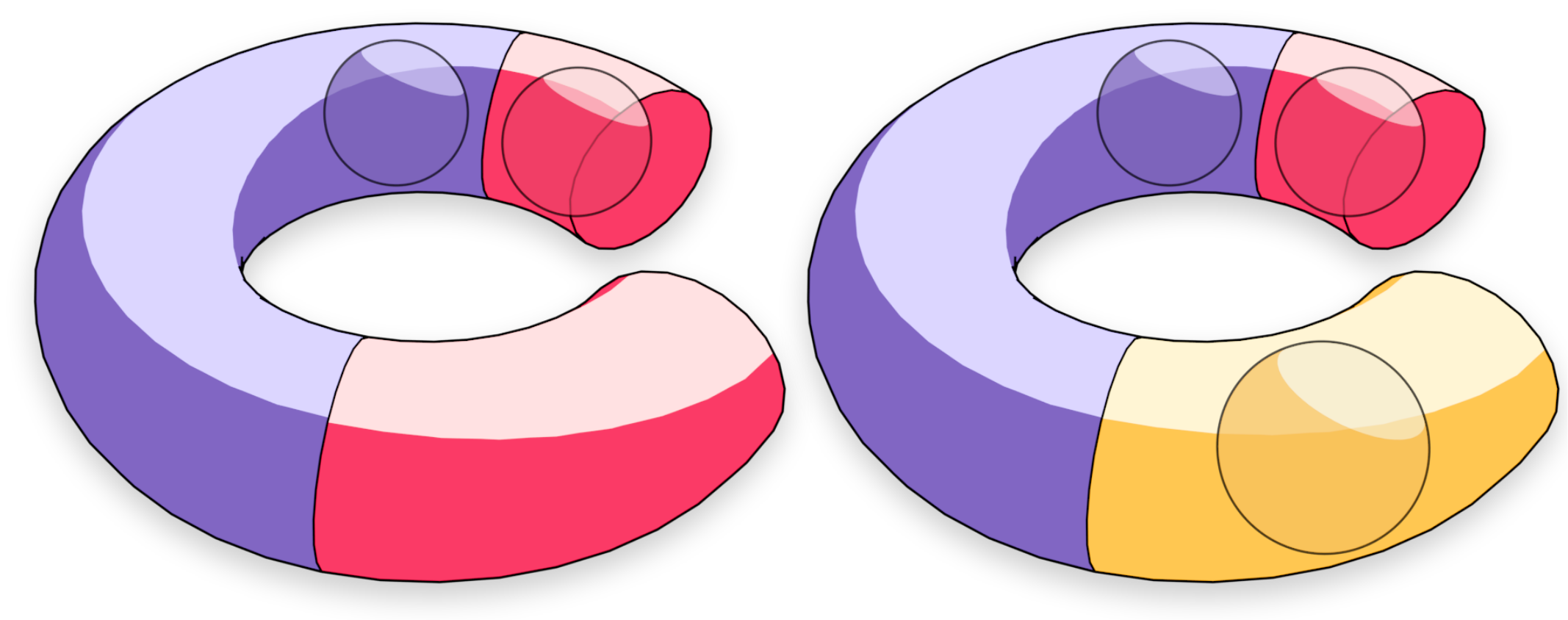}
    \put(22, -4){\textbf{(a)}}
    \put(73, -4){\textbf{(b)}}
    \put(22, 30){$\msphere_1$}
    \put(34, 29){$\msphere_2$}
    \put(72, 30){$\msphere_1$}
    \put(83, 29){$\msphere_2$}
    \put(81, 8){$\msphere_3$}
    \end{overpic}
    \caption{Illustration of solving the CC number inequivalence described in Sec~\ref{sec:cc_number}. The RPC of medial sphere $\msphere_2$ in (a) contains two connected components, which is not contractible. Hence we add a new sphere $\msphere_3$ to the other connected component of the RPC of $\msphere_2$ for maintaining the contractibility of each RPC.}
    \label{fig:topo_cc}
\end{figure}
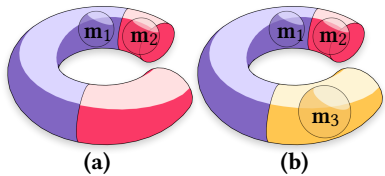

For a medial sphere $\msphere_i$ and its corresponding restricted elements $\rpc(\msphere_i)$, $\rpf(\msphere_i, \msphere_j)$ and $\rpe(\msphere_i, \msphere_j, \msphere_k)$, where $\msphere_j$ and $\msphere_k$ are the neighboring medial spheres of $\msphere_i$, we can trace their CC number $\beta_0$ using a simple traversal algorithm effortlessly. 
If $\beta_0>1$, we add a new medial sphere to the connected component other than the one of the current medial sphere $\msphere_i$. In Fig.~\ref{fig:topo_fix} (b), the RPF $\rpf(\msphere_1,\msphere_2)$ has $\beta_0(\rpf)=2$, while in Fig.~\ref{fig:topo_cc} (a), the RPC $\rpc(\msphere_2)$ has $\beta_0(\rpc)=2$. To fix these issues, 
we randomly choose a surface point on one of the CC of the RPF, or choose a surface point on the other CC of the RPC (other than where the medial sphere resides), and use it as the \textit{pin point} for the \textit{sphere-shrinking} algorithm (see Sec.~{\ref{sec:tech_sphere_gen}}) to add a new medial sphere $\msphere_3$, as shown in Fig.~\ref{fig:topo_fix} (c) and Fig~\ref{fig:topo_cc} (b).

\subsubsection{Euler Characteristic~$\euler$}
\label{sec:euler_number}

One can compute the Euler characteristic $\euler$ of each restricted element (RPC, RPF and RPE) after obtaining the explicit RPD representation through the clipping process (Sec.~\ref{sec:tech_tet_sphere}). However, current GPU-based implementations either directly evaluate integrals over every cells without computing the combinatorial data structure of power diagram~\cite{basselin2021RPD}, or store the dual form of cells using a simple triangle mesh ~\cite{liu2020RVD, ray2018meshless}, but requires costly post-processing steps to access the exact combinatorial data structure of the cells.
This is due to the fact that these methods are well-designed for the high parallelism of GPU, and are optimized for applications that only requires integrals over the cells, \eg fluid dynamics simulations. Therefore, in this paper we propose a \textit{Fractional Euler Characteristic} strategy to collect Euler characteristics on-the-fly, by taking full advantage of the existing GPU-based volumetric RPD computation pipeline. 

Similar to the \textit{Tet-Cell} strategy proposed by Liu et al.~\shortcite{liu2020RVD} (see Sec.~\ref{sec:tech_tet_sphere}), our pipeline takes the tetrahedral mesh as input, where the Euler characteristic is inherited during the clipping process. For each tetrahedron of the related medial sphere $\msphere_i$, we clip the RPC $\rpc(\msphere_i)$ as the intersection of half-spaces bounded by the bisectors of $\msphere_i$ and its power neighbors~\cite{ray2018meshless}.  Implementation details regarding finding tet-sphere relations and sphere neighbors are provided in the Supplementary Document.

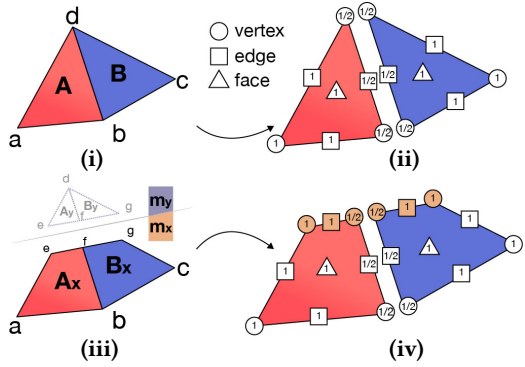
\begin{figure}[h!]
    \centering
    \begin{overpic}[width=0.8\linewidth]{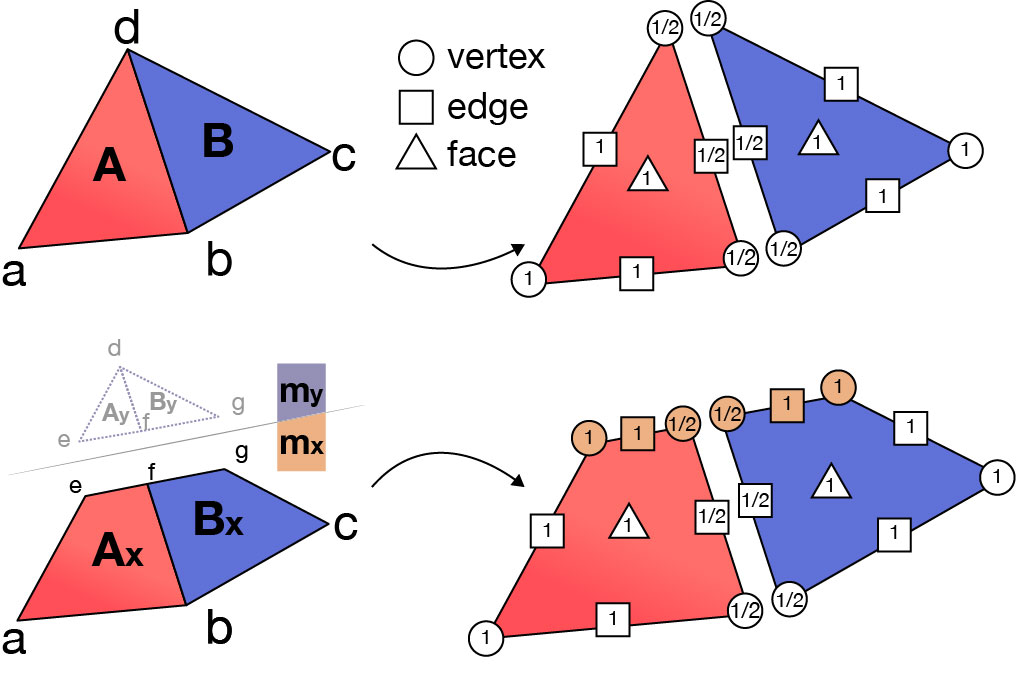}
    \put(14, 35){\textbf{(i)}}
    \put(74, 35){\textbf{(ii)}}
    \put(14, -2){\textbf{(iii)}}
    \put(74, -2){\textbf{(iv)}}
    \end{overpic}
    \caption{2D illustration of our GPU-based Euler characteristic computation on-the-fly during the clipping process of RPD. The number shown in each vertex, edge, and face represents its \textit{Fractional Euler Characteristic}, as described in Sec~\ref{sec:euler_number}.}
    \label{fig:topo_euler}
\end{figure}

\textbf{Fractional Euler Characteristics}. 
Fig.~\ref{fig:topo_euler} illustrates our proposed \textit{fractional Euler characteristics} in a 2D setting, demonstrating how these fractional numbers are inherited throughout the clipping process during RPD computation. In this example, the target Euler characteristic is $1$ for the 2D shape, \ie $\euler=$ $4$ (vertices) - $5$ (edges) + $2$ (faces) - $0$ (cells) $=1$. It contains 2 triangles $\mathbf{A}$ and $\mathbf{B}$ in Fig.~\ref{fig:topo_euler} (i), and 
the fractional Euler characteristic of each element is shown in Fig.~\ref{fig:topo_euler} (ii). 
Note that both vertex $\mathbf{b}$ and $\mathbf{d}$ are shared between $\mathbf{A}$ and $\mathbf{B}$, so their fractional Euler characteristic within each triangle is only $\frac{1}{2}$. The same rationale applies to the fractional Euler characteristic $\frac{1}{2}$ for the edge $(\mathbf{b}, \mathbf{d})$ within each triangle. As an initialization step for our GPU-based tet-cell clipping, whenever a tetrahedral mesh is inputted to the GPU, these fractional Euler characteristics are included alongside the mesh, derived from the combinatorial structure of the tetrahedral mesh.

\textit{Inheritance of Fractional Euler Characteristics during RPD Clipping}. During the parallel execution of clipping process \cite{ray2018meshless}, the half-space $\halfspace(\msphere_x, \msphere_y)$ of two medial spheres $\msphere_x$ and $\msphere_y$ cuts each triangle into two parts, thus new vertices and edges emerge. The fractional Euler characteristics of new vertices are inherited from their pre-clipped edges, and those of the new edges are inherited from its pre-clipped faces. For the example of Fig.~\ref{fig:topo_euler}, the new vertex $\mathbf{f}$ inherits the fractional Euler characteristic $\frac{1}{2}$ from its pre-clipped edge $(\mathbf{b}, \mathbf{d})$, and the new edge $(\mathbf{e}, \mathbf{f})$ inherits the fractional Euler characteristic $1$ from its pre-clipped face $\mathbf{A}$. Note that during the clipping process, although the new generated vertices, edges, and faces could be shared between different cells, we no longer divide their fractional Euler characteristics. For example, in Fig.~\ref{fig:topo_euler} (iii), although the new edge $(\mathbf{e}, \mathbf{f})$ is shared between two cells $\mathbf{A}_x$ and $\mathbf{A}_y$, we only set its fractional Euler characteristic as $1$ for cell $\mathbf{A}_x$, and $1$ for cell $\mathbf{A}_y$. This is because the use of fractional Euler characteristics is to facilitate the easy counting of Euler characteristic of each restricted power element associated with a particular sphere. For the example of Fig.~\ref{fig:topo_euler} (iv), we are only interested in the Euler characteristics for the restricted elements related to sphere $\mathbf{m_x}$, thus the fractional Euler characteristics of those newly generated vertices and edges will not be shared with the other spheres.

Through such inheritance, the fractional Euler characteristic of all restricted elements (RPCs, RPFs, RPEs) for sphere $\msphere_x$ can be obtained on-the-fly during the parallel clipping process. In Fig.~\ref{fig:topo_euler} (iv), the final RPFs of sphere $\mathbf{m}_x$ consists of two convex hulls $\mathbf{A}_x$ and $\mathbf{B}_x$, and its Euler characteristic can be computed by $\euler(\mathbf{A}_x)+\euler(\mathbf{B}_x)=1$, where: 
\begin{equation}
\begin{aligned}
\euler(\mathbf{A}_x)=(1+1+\frac{1}{2}+\frac{1}{2})-(1+1+1+\frac{1}{2})+(1)=\frac{1}{2},\\
\euler(\mathbf{B}_x)=(1+1+\frac{1}{2}+\frac{1}{2})-(1+1+1+\frac{1}{2})+(1)=\frac{1}{2}.
\end{aligned}
\end{equation}

For each medial sphere, we collect the fractional Euler characteristics for all of its restricted elements (RPCs, RPFs, RPEs) during the runtime of clipping process, then check if the target Euler characteristic for each element is expected or not, \eg in Fig.~\ref{fig:topo_fix} (a), the sphere $\msphere_1$ has its RPC $\rpc(\msphere_1)$ with $\euler(\rpc(\msphere_1))=0$. To fix this issue, we search all surface triangles among the RPC of sphere $\msphere_1$ and target the furthest one as the \textit{pin point} and uses the \textit{sphere-shrinking} algorithm (see Sec.~{\ref{sec:pre}}) to add a new sphere $\msphere_2$. After that, the RPF between $\msphere_1$ and $\msphere_2$ has CC number $\beta_0=2$, which will trigger the insertion of a new sphere $\msphere_3$, with the same process described in Sec.~\ref{sec:cc_number}.

\subsection{Medial Feature Preservation}
\label{sec:fix_extf_intf}

\begin{figure}[!h]
    \centering
    \begin{overpic}[width=0.8\linewidth]{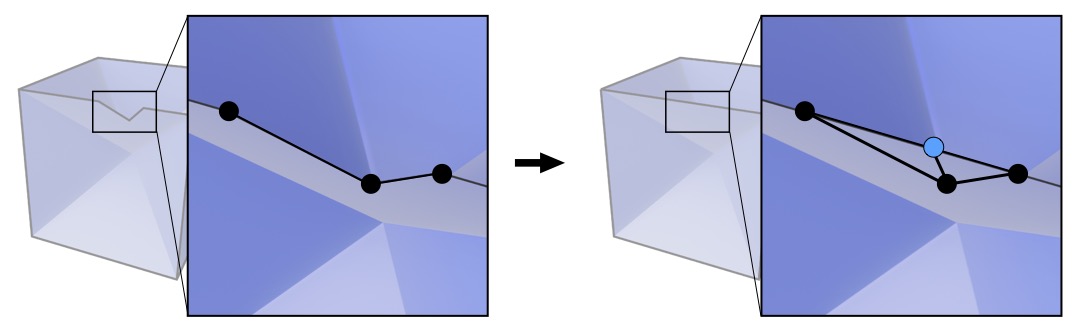}
    \put(10, 0){\textbf{(a)}}
    \put(63, 0){\textbf{(b)}}
    \put(18, 22){$\msphere_a$}
    \put(32, 10){$\msphere_b$}
    \put(39, 16){$\msphere_c$}
    \put(71, 22){$\msphere_a$}
    \put(85, 10){$\msphere_b$}
    \put(92, 16){$\msphere_c$}
    \put(85, 19){$\msphere_x$}
    \end{overpic}
    \caption{Illustration of external feature preservation. (a) A non-feature medial sphere $\msphere_b$ destroys the connectivity of two zero-radius feature spheres $\msphere_a\msphere_c$ in the medial mesh. (b) Adding a new zero-radius medial sphere $\msphere_x$ could preserve the  connectivity of external feature.}
    \label{fig:fix_extf}
\end{figure}
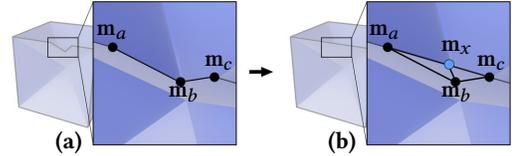

\textit{External features}, such as convex sharp edges and  corners where more than two convex sharp edges coincide, are commonly seen in CAD models and are pre-detected as input for our algorithm. We adopt the similar feature preservation strategies as MATFP~\cite{2022MATFP}, as shown in Fig~\ref{fig:fix_extf}. 
Since the generated medial mesh should pass through points where the surface is locally convex and non-smooth, we place zero-radius $T_1^2$ medial spheres on convex edges and a single zero-radius $T_1^u$ ($u \geq 3$) medial sphere at each corner.
Unlike MATFP which starts with a dense number of initial spheres, our method begins with a smaller set of non-feature medial spheres. Hence, instead of removing redundant non-feature spheres that `invade' the RPCs of two neighboring feature spheres on a sharp edge, we examine the individual RPC and add new zero-radius feature spheres iteratively if any sharp edge belongs to the cell of a non-feature sphere. As a result, all convex sharp edges reside in cells corresponding to some zero-radius feature spheres, which preserves the convex sharp edges in the generated medial mesh as a dual of volumetric RPD. 
For \textit{corners}, we adopt the same corner preservation strategy as MATFP~\shortcite{2022MATFP}, where we approximate the medial mesh structure within a small, selected region around the corner. This scheme operates by recursively tracing the medial axis sheets, beginning from those adjacent to convex sharp edges, until their intersecting seams are identified.
On a \textit{concave edge}, the medial spheres tangential to this feature are not zero-radius. To achieve a smooth transition on the medial axis, similar to MATFP~\shortcite{2022MATFP}, we densely sample medial spheres using the \textit{sphere-shrinking} algorithm~\shortcite{ma20123shrink} (see Sec.~\ref{sec:tech_sphere_gen}), with tangent surface points selected on the concave edge acting as the `pin'. For more details on the operations used to preserve \textit{corners} and \textit{concave features}, please refer to MATFP~\cite{2022MATFP}.

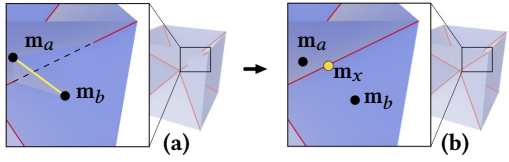
\begin{figure}[!h]
    \centering
    \begin{overpic}[width=0.8\linewidth]{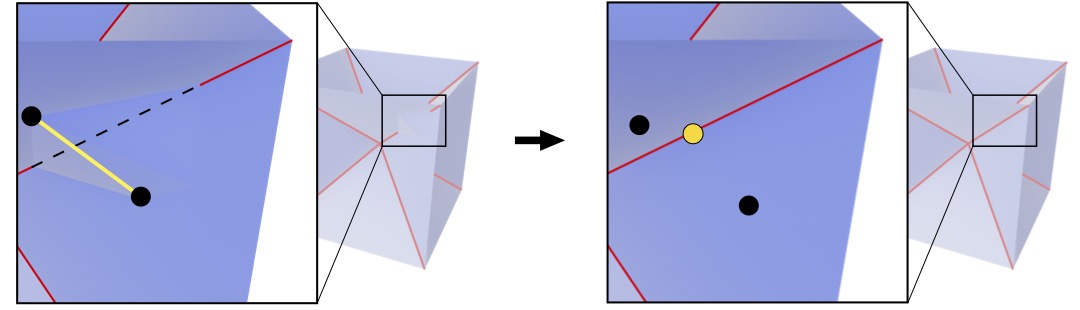}
    \put(32, 0){\textbf{(a)}}
    \put(86, 0){\textbf{(b)}}
    \put(5, 20){$\msphere_a$}
    \put(15, 10){$\msphere_b$}
    \put(58, 20){$\msphere_a$}
    \put(71, 9){$\msphere_b$}
    \put(65, 14){$\msphere_x$}
    \end{overpic}
    \caption{Illustration of internal feature preservation. (a) An ill-posed connection (yellow line) of two $T_2$ spheres $\msphere_a\msphere_b$ on two different medial sheets. (b) The internal medial feature (red line) is preserved after inserting a $T_3$ sphere $\msphere_x$.}
    \label{fig:fix_intf}
\end{figure}

\textit{Internal features} are preserved by inserting new internal feature spheres after detecting the deficiency, similar to MATFP~\cite{2022MATFP}. We show an illustration in Fig.~\ref{fig:fix_intf}. Here we maintain a queue of all medial edges in the medial mesh, and check whether two connected medial spheres of each edge belong to the same medial sheets as they touch the same surface regions using the surface part of their RPCs.


\subsection{Geometric Preservation}
\label{sec:fix_geo}

The medial mesh generated through the above steps is topologically correct and captures medial features. However, its reconstruction may not geometrically converge to the input shape, since we adaptively insert spheres from a low number. We propose an error-bounding strategy so that the geometric error from the input shape to the enveloping volume of medial mesh is bounded by a user-controlled threshold $\geoerror$.

\begin{figure}[!h]
    \centering
    \begin{overpic}[width=\linewidth]{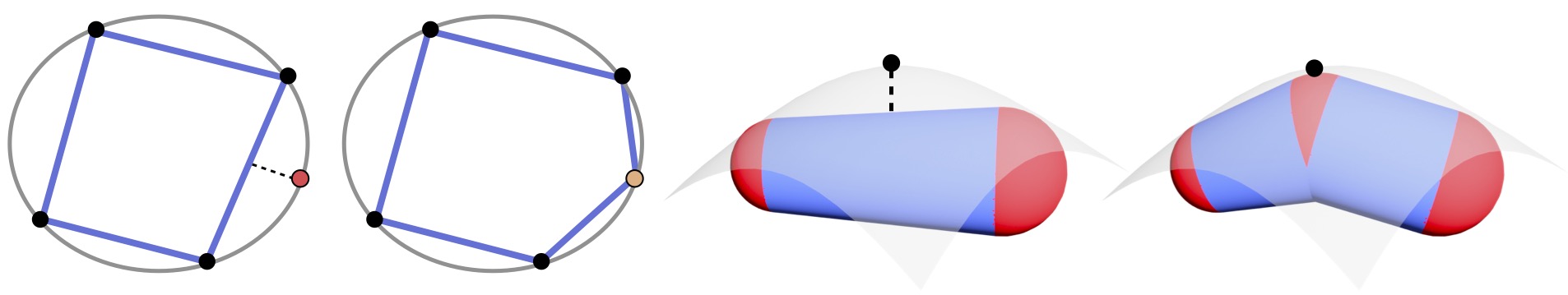}
    \put(20, -2){\textbf{(a)}}
    \put(70, -2){\textbf{(b)}}
    \put(17, 9){$\mathbf{d}$}
    \put(45, 2){$\msphere_1$}
    \put(64, 2){$\msphere_2$}
    \put(53, 12){$\mathbf{d}$}
    \put(56, 16){$\anypoint$}
    \put(73, 2){$\msphere_1$}
    \put(93, 2){$\msphere_2$}
    \put(82, 10){$\msphere_3$}
    \put(83, 16){$\anypoint$}
    \end{overpic}
    \caption{ Illustration of geometric preservation. (a) We compute the distance $d$ from the sampled points (in red) on external features to the nearest feature enveloping cone and add zero-radius medial sphere (in yellow) if the distance is too large. For other cases (b), we randomly sample surface points and compute their distance to the nearest enveloping primitive (\eg medial cone) of the medial mesh, and insert a new $T_2$ medial sphere (\eg $\msphere_3$) if the distance is too large.}
    \label{fig:fix_geo}
\end{figure}
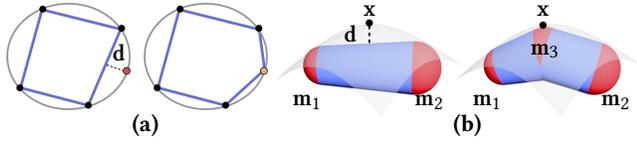

For preserving the geometry of external features, we sample points directly on those pre-detected features, such as Fig~\ref{fig:fix_geo} (a) in red, and compute their distance $d$ to the nearest enveloping cone. When the distance is larger than our threshold, which often happens for curved feature lines, we insert a new zero-radius feature sphere (Fig~\ref{fig:fix_geo} (a) in yellow). 

For other cases, we use the distance from any surface sample $\anypoint$ to its closest enveloping element $\melement_t$ as the metric. Each $\melement_t$ can be a medial sphere, a medial cone, or a medial slab (see Sec.~\ref{sec:pre_rpd_mm}). For each surface sample, we traverse all possible enveloping elements defined by the medial mesh and compute its distance to the closest enveloping element on GPU. Fig~\ref{fig:fix_geo} (b) shows an example of the distance from the surface sample $\anypoint$ to a medial cone defined by two medial spheres $\msphere_1$ and $\msphere_2$.
If the ratio of the distance $d$ over the diagonal of the bounding box is larger than a user-defined threshold $\geoerror$, we insert a new non-feature sphere with the surface sample as the pin point using the sphere-shrinking algorithm \cite{ma20123shrink} (see Sec.~\ref{sec:tech_sphere_gen}), shown in Fig.~\ref{fig:fix_geo} (b) as $\msphere_3$. As more non-feature spheres are inserted, the enveloping volume of the medial mesh is converging to the input shape geometrically. We show an ablation study on the impact of the value of $\geoerror$ in Sec~\ref{sec:ablation_bound}.

%% file: 5_tech_detail.tex
\section{Technical Details}
\label{sec:tech_detail}

\subsection{Generation of Medial Spheres}
\label{sec:tech_sphere_gen}

We use two strategies for computing medial spheres that are tangent to at least two surface points. These two strategies enable us to maintain the centeredness property of our generated medial mesh.

\begin{figure}[!h]
    \centering
    \begin{overpic}[width=0.9\linewidth]{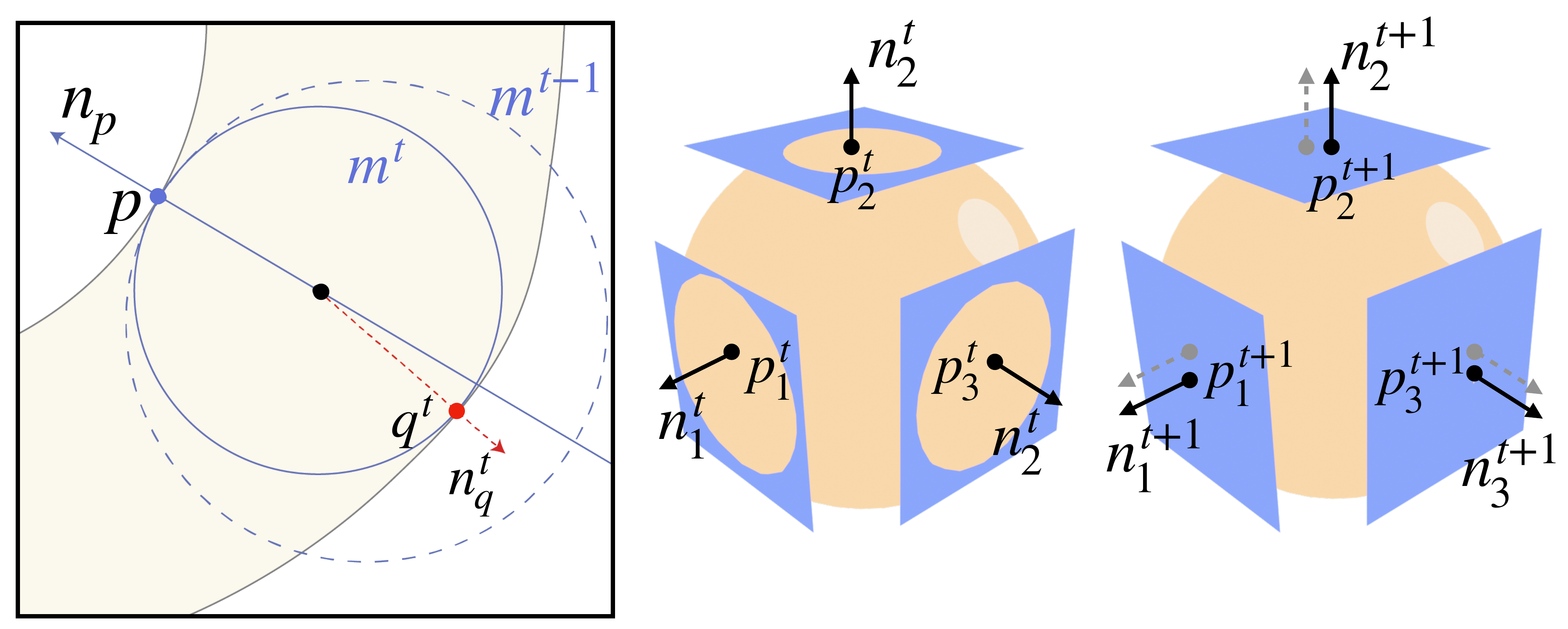}
    \put(20, -3){\textbf{(a)}}
    \put(68, -3){\textbf{(b)}}
    \end{overpic}
    \caption{Illustration of two medial sphere generation algorithms. (a) Given 'pin' point $\mathbf{p}$ on the model boundary $\bmodel$, the sphere-shrinking algorithm~\cite{ma20123shrink} decreases the sphere radius iteratively until the sphere $\msphere^{t}$ is a maximal empty ball (that is the interior of $\msphere^{t}$ contains no point of $\bmodel$), while the another tangent point touches $\mathbf{q}^{t}$. (b) The sphere-optimization algorithm \cite{2022MATFP} iteratively updates medial spheres in two alternating steps. The first step locks the aggregated tangent pairs $\{(\mathbf{p}_k, \mathbf{n}_k)\}_{k=1}^N$ and updates the sphere center and radius. The second step updates each tangent pair by fixing the previously updated medial sphere.}
    \label{fig:tech_sphere_gen}
\end{figure}

\begin{itemize}
    \item \textit{Sphere-shrinking} algorithm ~\cite{ma20123shrink} for computing $T_2$ medial spheres that are on \textbf{sheets} of the medial structure, and tangent to two different places on the surface, as shown in Fig.~\ref{fig:tech_sphere_gen} (a);
    \item \textit{Sphere-optimization} algorithm ~\cite{2022MATFP} for computing $T_N$ ($N\geq3$) medial spheres that are on \textbf{seams} or \textbf{junctions} of the medial structure, and tangent to at least three different places on the surface, as shown in Fig.~\ref{fig:tech_sphere_gen} (b).
\end{itemize}

\subsection{Computation of Restricted Power Cells}
\label{sec:tech_tet_sphere}

Similar to a Voronoi cell, a power cell is a convex polyhedron that is the intersection of half-spaces. The only difference is that these half-spaces are not generated by bisectors, but by radical hyperplanes.

We use the data structure introduced by Ray et al.~\shortcite{ray2018meshless} to represent the convex polyhedrons, which is highly compact and well-suited for GPU implementation. Each half-space of the polyhedron is represented by a float4 storing 4 coefficients $(a,b,c,d)$ of the plane equation $ax + by + cz + d > 0$. Each vertex of the polyhedron is the intersection of three half-spaces, therefore, is represented by a triplet of integers in a clockwise order, storing indices of three adjacent half-spaces.

\begin{figure}[!h]
    \centering
    \begin{overpic}[width=0.9\linewidth]{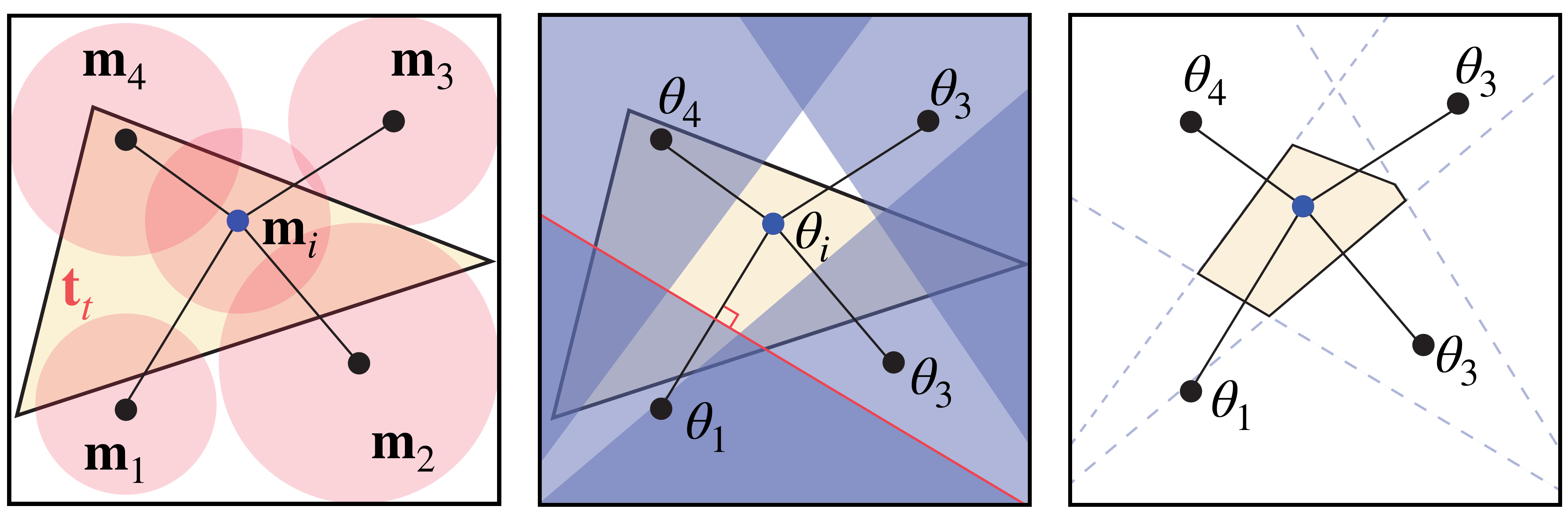}
    \put(16, -3){\textbf{(a)}}
    \put(49, -3){\textbf{(b)}}
    \put(82, -3){\textbf{(c)}}
    \end{overpic}
    \caption{Illustration of `Tet-Cell' strategy in 2D using triangles. Here we show the medial sphere as $\msphere_i=(\mcenter_i, r_i)$, where $\mcenter_i$ is the center of $\msphere_i$ and $r_i$ is its radius. (a) For each triangle-sphere pair $(\mathbf{t}_t, \msphere_i)$, we traverse all neighboring spheres of $\msphere_i$. The searching of sphere neighbors and tet-sphere relations are further discussed in Supplementary Document. (b) Each pair of two spheres $(\msphere_i, \msphere_k)$ defines a half-space that may potentially `clip' the triangle $\mathbf{t}_t$. (c) The part of the power cell inside the triangle (\ie tet in 3D) $\mathbf{t}_t$ for sphere $\msphere_i$ can be generated after the clipping process.}
    \label{fig:tech_tet_sphere}
\end{figure}

To restrict the power cells in an input domain, we extend the `Tet-Cell' strategy proposed by Liu et al.~\shortcite{liu2020RVD}. Given an input shape represented as a tetrahedral mesh~\cite{hu2020ftetwild}, we compute the intersection between the tetrahedral mesh and power cells. An illustration of `Tet-Cell' strategy in 2D is shown in Fig.~\ref{fig:tech_tet_sphere}. Each step discussed in Sec.~\ref{sec:method} will adaptively insert new medial spheres when topological or geometrical deficiency is detected. We thus only update the RPD partially with cells relating to the newly added medial spheres, instead of recomputing the whole RPD. Please refer to our implementation details in the Supplementary Document for further discussions.

\subsection{Thinning of Medial Mesh}
\label{sec:tech_thinning}

\begin{figure}[!h]
    \centering
    \begin{overpic}[width=\linewidth]{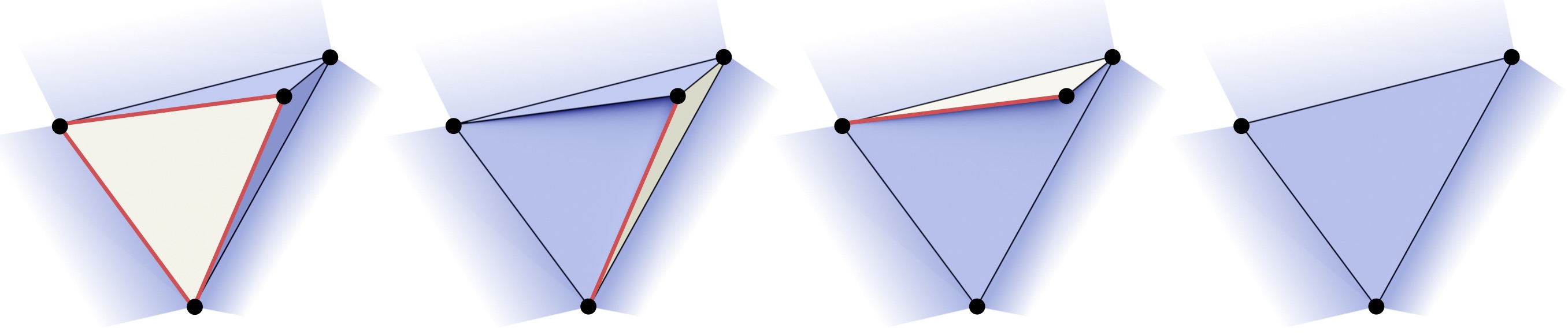}
    \put(10, -3){\textbf{(a)}}
    \put(35, -3){\textbf{(b)}}
    \put(60, -3){\textbf{(c)}}
    \put(86, -3){\textbf{(d)}}
    \put(10, 10){$f_1$}
    \put(18, 8){$t_1$}
    \put(38, 10){$e_2$}
    \put(43, 11){$f_2$}
    \put(59, 12){$e_3$}
    \put(63, 16){$f_3$}
    \end{overpic}
    \caption{The thinning process for one flat tetrahedron $t_1$ in a 3D medial mesh. The simple tet-face pair $(t_1, f_1)$ in (a) is selected to prune with the least importance. Then we continue pruning those face-edge simple pairs, \ie $(f_2, e_2)$ in (b) and $(f_3, e_3)$ in (c). The pruned result is shown in (d).}
    \label{fig:tech_thinning}
\end{figure}
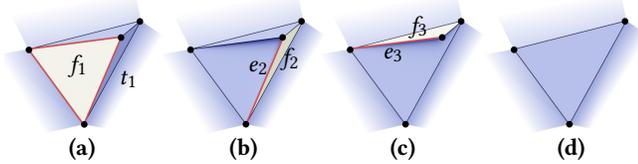

The medial mesh $\ma_s$ constructed as the dual of the RPD inevitably contains some fat but solid tetrahedrons. We apply the \textit{geometry-guided thinning algorithm} \cite{2022MATFP} that prunes simple pairs of simplices in the medial mesh. A \textit{simple pair} $(x, y)$ \cite{liu2010simple} is a pair of simplices such that $y$ is on the boundary of $x$, and there is no other cell in the complex with $y$ on its boundary. Ju et al. \shortcite{ju2007editing} have shown that the removal of simple pairs will not impact the topology of the complex. We first rank all possible tet-face pairs by the importance factor $\alpha_{ijk}$ of a given medial triangle $f_{ijk}$ in $\ma_s$ in ascending order. The $\alpha_{ijk}$ is defined as the ratio of the length of \textit{restricted power edge} (RPE) $\rpe(\msphere_i, \msphere_j, \msphere_k)$ over the average diameter of three medial spheres $\msphere_i$, $\msphere_j$ and $\msphere_k$. Note that the RPE $\rpe$ is dual to the medial triangle $f_{ijk}$ on $\ma_s$, discussed in Sec.~\ref{sec:pre_rpd_mm}. Then we remove tet-face pairs iteratively with the least importance until all tetrahedra are pruned. We continue pruning face-edge simple pairs that belong to the original tetrahedra until a target importance factor $\sigma$ is reached as a stop sign. To avoid over-prunning for models whose medial mesh boundaries are not external features, we set $\sigma=0.3$ for CAD models and $\sigma=0.1$ for organic models in our experiments. Fig~\ref{fig:tech_thinning} shows an illustration of the thinning process with one tetrahedron. 

%% file: 6_results.tex
\section{Experiments}
\label{sec:exp}

In this section, we show quantitative and qualitative evaluations of the proposed method. We implemented our algorithm in C++ and CUDA, using Geogram~\cite{levy2015geogram} for linear algebra routines. We ran our experiments on a computer with a 3.60GHz Intel(R) Core(TM) i7-9700K CPU, NVIDIA GeForce RTX 2080 Ti GPU, and 32 GB memory. 
We ran our method on the first $100$ models in the ABC dataset~\cite{koch2019abc} under the \textit{10k/test} folder using $2048$ as the number of mesh vertices, same as MATFP~\cite{2022MATFP}. We also test our method on $14$ organic models with various topology. All model sizes are normalized to the $[0, 1000]^3$ range. We use fTetwild~\cite{hu2020ftetwild} for computing the initial tetrahedral mesh from triangle mesh with parameters $l=0.5$. Our \href{https://github.com/ningnawang/MATTopo}{\violet{code \ExternalLink}} is available at \href{https://ningnawang.github.io/projects/2024_mattopo/}{\violet{our project website \ExternalLink}}, and the generated medial mesh can be viewed using the tool \href{https://github.com/songshibo/blender-mat-addon}{\violet{blender-mat-addon \ExternalLink}}~\cite{blender-mat-addon}.

\paragraph{Evaluation Metrics.}
We use the \textit{Euler characteristics} $\euler$ as the topology measures for the generated medial mesh, and show the ground truth Euler characteristic of the input shape as `GT $\euler$'.
We compute the Euler characteristic using $\euler = V - E + F - C$, where $V$ is the number of vertices, $E$ is the number of edges, $F$ is the number of faces, and $C$ is the number of volumetric cells. In our experiment, we consider only tetrahedrons as the 3D-complex cells for $C$. 
We use the two-sided Hausdorff distance error $\hd$ to measure the the surface reconstruction accuracy using the generated medial meshes. $\hd^1$ is the one-sided Hausdorff distance from the original surface to the surface reconstructed from MAT, and $\hd^2$ is the distance in reverse side. All Hausdorff distances are evaluated as percentages of the distance over the diagonal lengths of the models' bounding box. The $\hd^{max}$ is the maximum of $\hd^1$ and $\hd^2$. We show $\#s$ as the number of medial spheres for the medial meshes generated from each method, and use $\#t$ as the number of tets in the input tetrahedral mesh. 

\subsection{Comparison with MATFP Method}

\begin{figure}[h!]
    \centering
    \begin{overpic}[width=0.9\linewidth]{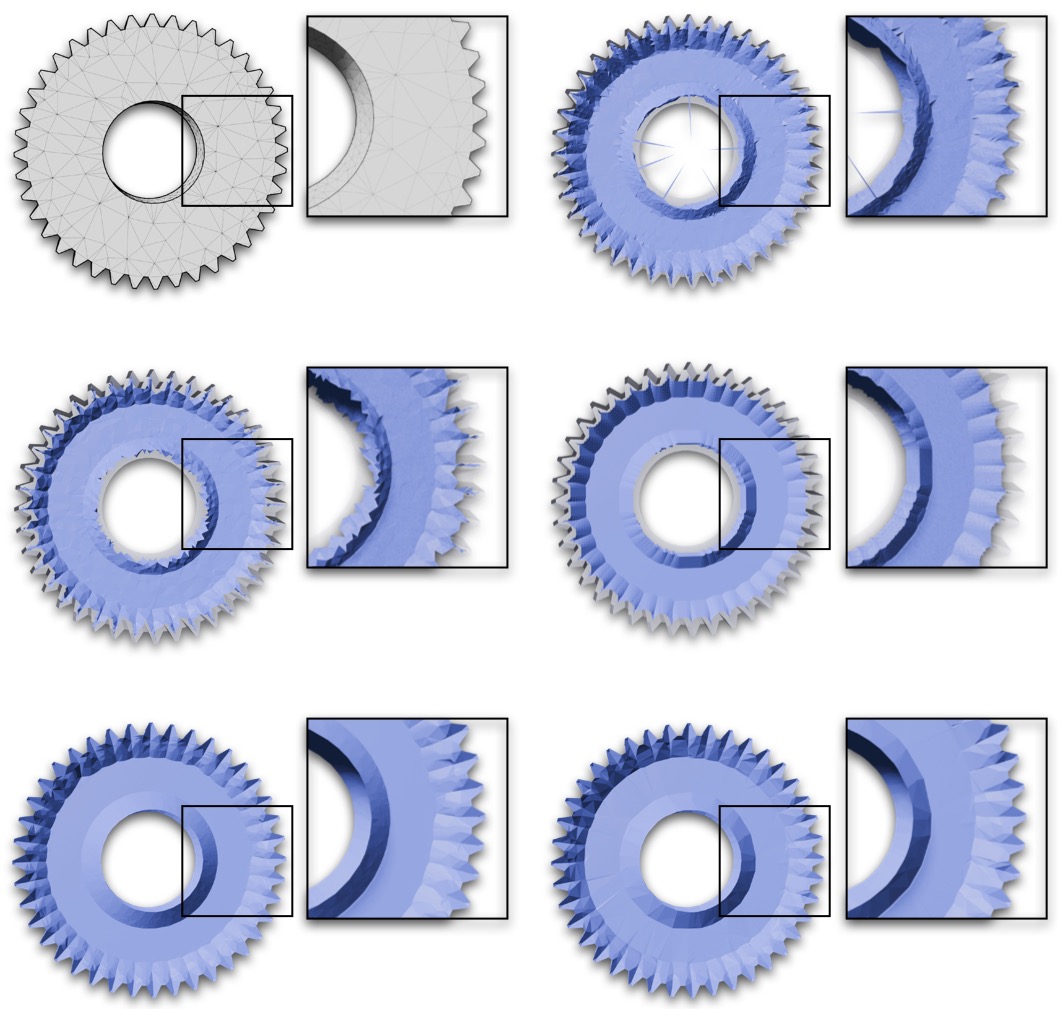}
    \put(6, 64){\textbf{(a)} Input}
    \put(33, 70){{\footnotesize GT $\euler$=0}}
    \put(60, 64){\textbf{(b)} PC}
    \put(83, 71){{\footnotesize $\#s=18k$}}
    \put(83, 68){{\footnotesize \red{$\mathbf{\euler=32}$}}}
    \put(6, 30){\textbf{(c)} SAT}
    \put(33, 38){{\footnotesize $\#s=25k$}}
    \put(33, 35){{\footnotesize \red{$\mathbf{\euler=8}$}}}
    \put(61, 30){\textbf{(d)} VC}
    \put(83, 38){{\footnotesize $\#s=30k$}}
    \put(83, 35){{\footnotesize $\mathbf{\euler=0}$}}
    \put(5, -3){\textbf{(e)} MATFP}
    \put(33, 5){{\footnotesize $\#s=17k$}}
    \put(33, 2){{\footnotesize \red{$\mathbf{\euler=2}$}}}
    \put(59, -3){\textbf{(f)} Ours}
    \put(83, 5){{\footnotesize $\#s=3.5k$}}
    \put(83, 2){{\footnotesize $\mathbf{\euler=0}$}}
    \end{overpic}
    \caption{Qualitative comparison of topology preservation between our method and MATFP~\cite{2022MATFP},  PC~\cite{amenta2001power},  SAT~\cite{miklos2010sat}, and VC~\cite{yan2018voxel}. The ground truth \textit{Euler characteristic} $\euler$ is shown in (a), and $\#s$ represents the number of medial spheres generated.}
    \label{fig:comp_topo}
\end{figure}
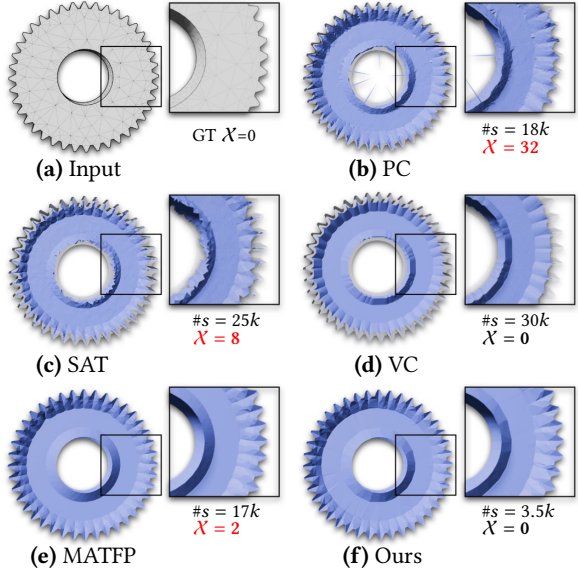

We compare our method with MATFP~\cite{2022MATFP} regarding the topology preservation. To our best knowledge, MATFP is the state-of-the-art method for computing medial axis of CAD models, which preserves both external and internal features. However, it also has a clear drawback with no guarantee of topology preservation for the generated medial mesh w.r.t. the input model. The key advantage of our method is its ability in preserving the topology, while maintaining the ability to capture external and internal features. We show the Euler characteristic $\euler$ in Table~\ref{tab:comp_euler} for those models that the output of MATFP deviates from the GT $\euler$. Our method, on the contrary, can preserve the topology while keeping a competitive reconstruction quality. 

Beside preserving topology, our method also generates almost $10$ times less number of medial spheres $\#s$ than MATFP in the final medial mesh, shown in  Table~\ref{tab:comp_euler} and Fig~\ref{fig:comp_topo}. MATFP favors preserving external or internal features 
by adding as many feature spheres as possible during each step. This will inevitable result in a large number of medial spheres, mostly redundant, in the generated medial mesh. Owing to the proposed adaptive refinement strategy, our method can add only few (even single) number of new medial spheres at each iteration and update the RPD partially using GPU. Each updated partial RPD will change the medial mesh connectivity within the corresponding local regions, as dual of volumetric RPD. These newly added medial spheres may already satisfy the criteria for local structures, thus avoiding other similar spheres (\ie spheres with similar centers and radii) to be inserted. 

Since we use the same feature preservation strategy as MATFP, our methods shows comparable if not better reconstruction quality, as shown in Table~\ref{tab:comp_euler} and Fig.~\ref{fig:comp_recon}. More visualizations are shown in Fig.~\ref{fig:mattopo_results}. Specifically, we use $\geoerror=0.6$ for models shown in Fig~\ref{fig:comp_recon}. For more detailed statistics, please refer to the Supplementary Document.

\begin{table}[!t]
\small
\caption{Quantitative comparison on topology preservation with MATFP~\cite{2022MATFP}. $\#s$ is the number of generated medial spheres. $\hd^{max}$ is the two-sided Hausdorff Distance between the original surface and reconstruction, as maximum of $\hd^1$ and $\hd^2$ described in Sec.~\ref{sec:exp}. We show the Euler characteristic as $\euler$ and ground truth as `GT $\euler$'. Comparing to MATFP which has no guarantee of homotopy equivalence, our method gives correct Euler characteristic with lower number of generated medial spheres and competitive reconstruction quality.}
\begin{center}
\scalebox{1}{
\begin{tabular}{c||c|c|c||c|c|c}
\hline
Model ID& \multicolumn{3}{c||}{MATFP} & \multicolumn{3}{c}{Ours} \\
(GT $\euler$) & $\#s$ & $\epsilon^{max}$ & $\euler$ & $\#s$ & $\epsilon^{max}$ & $\euler$ \\
\hline
549 (-6)&  21k&1.282&-5&     
$7.2k^{\star}$&\textbf{1.077}&\textbf{-6} \\

4123 (-3)&  17k&3.799&-1&    
$5.1k^{\star}$&\textbf{1.095}&\textbf{-3}  \\ 

5227 (-5)&  12k&\textbf{0.442}&5&      
$3.2k^{\star}$&1.419&\textbf{-5} \\

8315 (-4)&  11k&3.351&-2&     
$1.5k^{\star}$&\textbf{1.481}&\textbf{-4} \\

8964 (-72)&  25k&\textbf{0.251}&-37&  
$8.7k^{\star}$&0.753&\textbf{-72} \\

10836 (-1)&  3k&\textbf{1.067}&1&      
$2.7k^{\star}$&1.496&\textbf{-1}  \\ 

11299 (-24)& 19k&2.324&-13&   
$5.8k^{\star}$&\textbf{0.961}&\textbf{-24} \\

11790 (0)&   17k&1.79&2&     
$5.3k^{\star}$&\textbf{0.819}&\textbf{0}  \\

11835 (0)&    21k&2.252&-16&  
$5.4k^{\star}$&\textbf{1.015}&\textbf{0} \\

13026 (-4)&   26k&2.203&-2&    
$10k^{\star}$&\textbf{1.241}&\textbf{-4}  \\  

13607 (-13)&   19k&\textbf{0.956}&-19&    
$7.3k^{\star}$&1.486&\textbf{-13}  \\ 

14621 (-3)&   34k&2.812&-9&    
$12k^{\star}$&\textbf{0.982}&\textbf{-3}  \\ 

15094 (-8)&   35k&\textbf{1.045}&-4&    
$7.2k^{\star}$&1.462&\textbf{-8}  \\ 
\hline
\end{tabular}}
\end{center}
\label{tab:comp_euler}
\end{table}

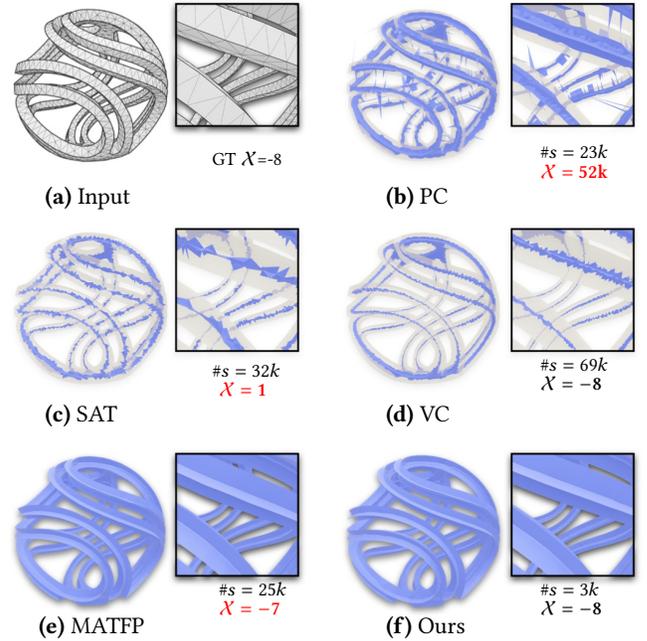
\begin{figure}[h!]
    \centering
    \begin{overpic}[width=\linewidth]{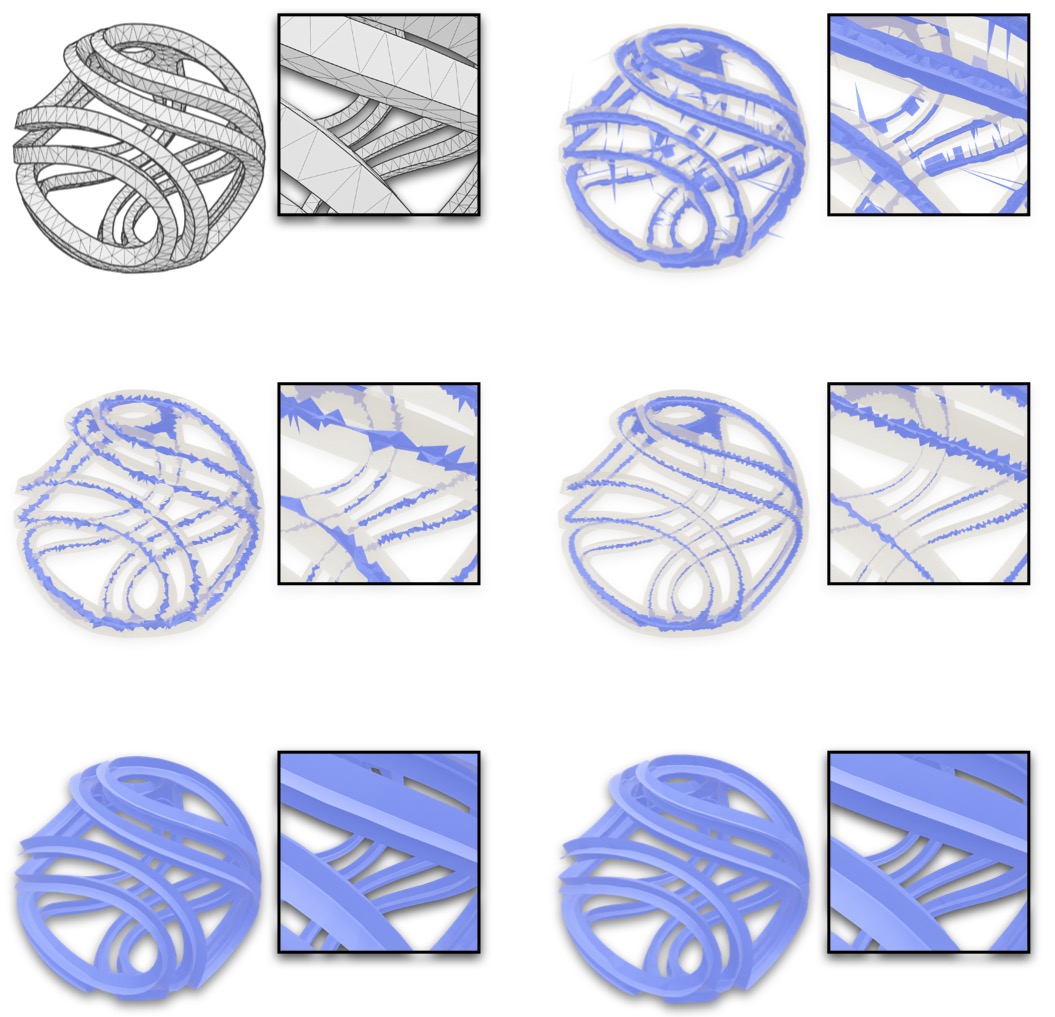}
    \put(6, 64){\textbf{(a)} Input}
    \put(32, 70){{\footnotesize GT $\euler$=-8}}
    \put(59, 64){\textbf{(b)} PC}
    \put(83, 71){{\footnotesize $\#s=23k$}}
    \put(83, 68){{\footnotesize \red{$\mathbf{\euler=52k}$}}}
    \put(6, 30){\textbf{(c)} SAT}
    \put(32, 37){{\footnotesize $\#s=32k$}}
    \put(33, 34){{\footnotesize \red{$\mathbf{\euler=1}$}}}
    \put(59, 30){\textbf{(d)} VC}
    \put(83, 38){{\footnotesize $\#s=69k$}}
    \put(83, 35){{\footnotesize $\mathbf{\euler=-8}$}}
    \put(5, -3){\textbf{(e)} MATFP}
    \put(33, 3){{\footnotesize $\#s=25k$}}
    \put(33, 0){{\footnotesize \red{$\mathbf{\euler=-7}$}}}
    \put(59, -3){\textbf{(f)} Ours}
    \put(83, 3){{\footnotesize $\#s=3k$}}
    \put(83, 0){{\footnotesize $\mathbf{\euler=-8}$}}
    \end{overpic}
    \caption{Comparison of the medial feature and topology preservation ability between our method and PC~\cite{amenta2001power}, SAT~\cite{miklos2010sat},  VC~\cite{yan2018voxel}, and  MATFP~\cite{2022MATFP}. Our method can not only preserve medial features, but also output the correct topology of the medial mesh.}
    \label{fig:comp_feature}
\end{figure}

\subsection{Comparison with PC, SAT, and VC Methods}

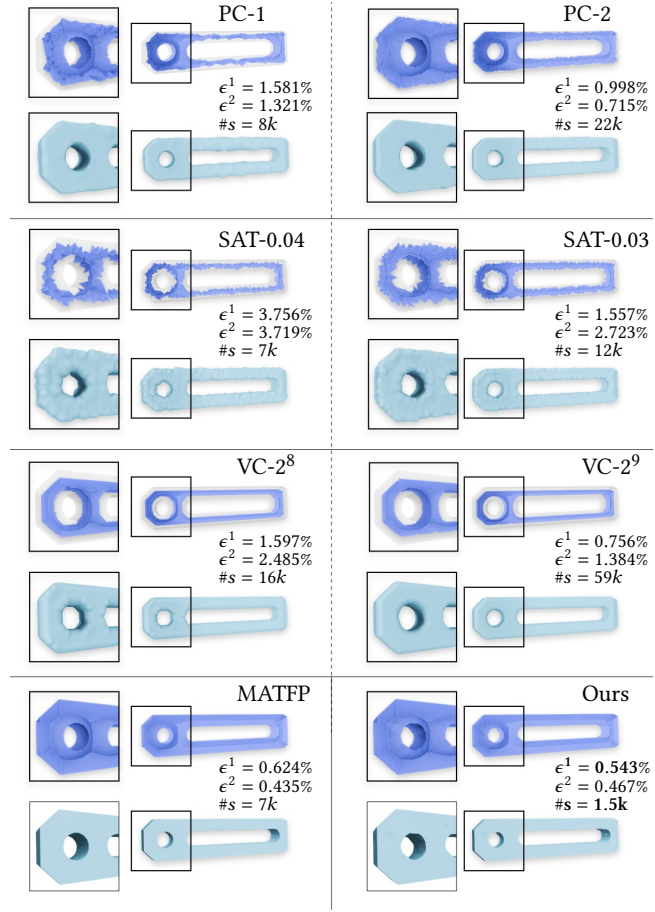
\begin{figure}[h!]
    \centering
    \begin{overpic}[width=\linewidth]{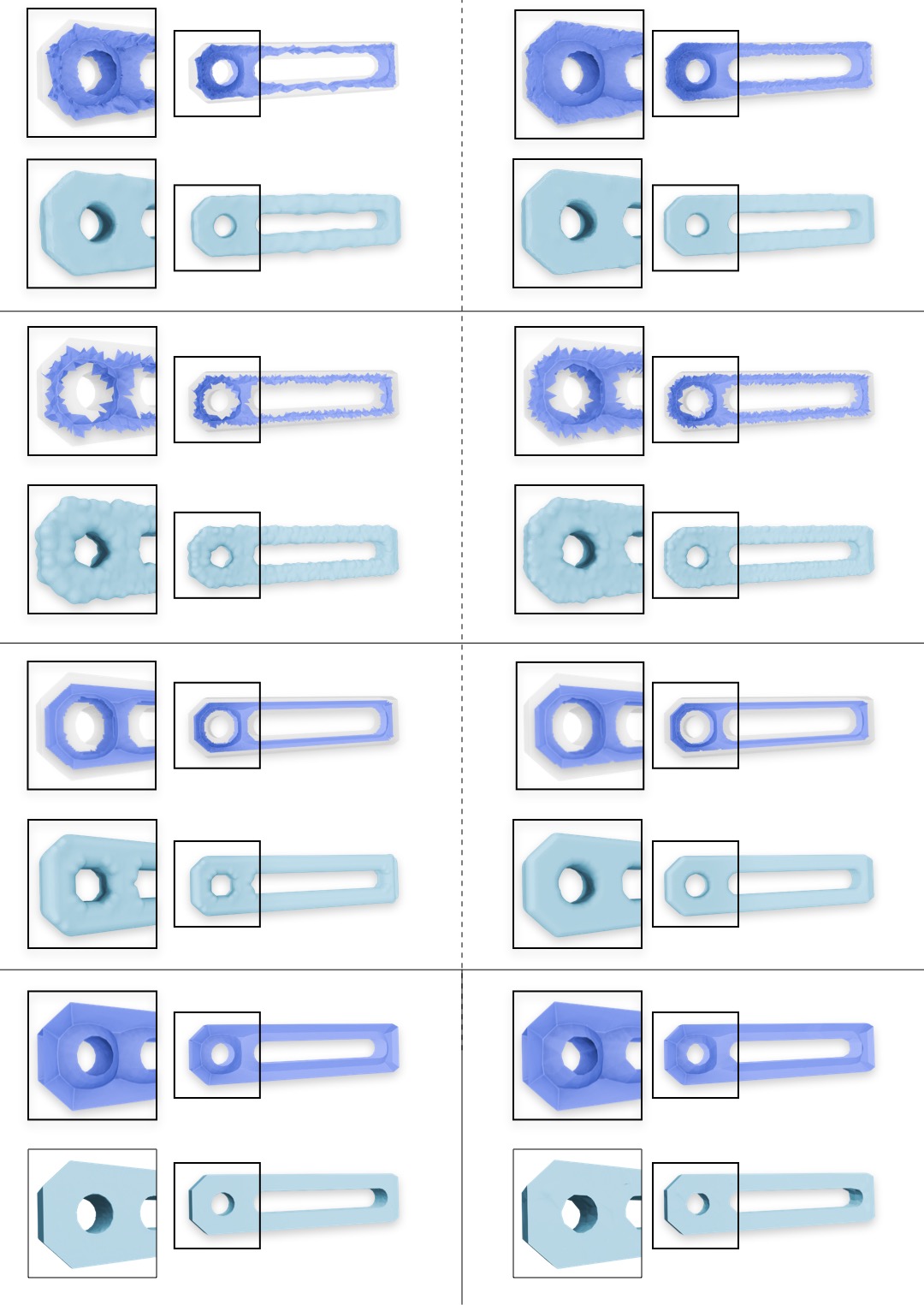}
    \put(23, 98){PC-1}
    \put(23, 90){{\footnotesize $\epsilon^1=1.581\%$}}
    \put(23, 88){{\footnotesize $\epsilon^2=1.321\%$}}
    \put(23, 86){{\footnotesize $\#s=8k$}}
    \put(61, 98){PC-2}
    \put(60, 90){{\footnotesize $\epsilon^1=0.998\%$}}
    \put(60, 88){{\footnotesize $\epsilon^2=0.715\%$}}
    \put(60, 86){{\footnotesize $\#s=22k$}}
    \put(23, 73){SAT-0.04}
    \put(23, 65){{\footnotesize $\epsilon^1=3.756\%$}}
    \put(23, 63){{\footnotesize $\epsilon^2=3.719\%$}}
    \put(23, 61){{\footnotesize $\#s=7k$}}
    \put(61, 73){SAT-0.03}
    \put(60, 65){{\footnotesize $\epsilon^1=1.557\%$}}
    \put(60, 63){{\footnotesize $\epsilon^2=2.723\%$}}
    \put(60, 61){{\footnotesize $\#s=12k$}}
    \put(25, 48){VC-$2^8$}
    \put(23, 40){{\footnotesize $\epsilon^1=1.597\%$}}
    \put(23, 38){{\footnotesize $\epsilon^2=2.485\%$}}
    \put(23, 36){{\footnotesize $\#s=16k$}}
    \put(63, 48){VC-$2^9$}
    \put(60, 40){{\footnotesize $\epsilon^1=0.756\%$}}
    \put(60, 38){{\footnotesize $\epsilon^2=1.384\%$}}
    \put(60, 36){{\footnotesize $\#s=59k$}}
    \put(25, 23){MATFP}
    \put(23, 15){{\footnotesize $\epsilon^1=0.624\%$}}
    \put(23, 13){{\footnotesize $\epsilon^2=0.435\%$}}
    \put(23, 11){{\footnotesize $\#s=7k$}}
    \put(63, 23){Ours}
    \put(60, 15){{\footnotesize $\mathbf{\epsilon^1=0.543\%}$}}
    \put(60, 13){{\footnotesize $\epsilon^2=0.467\%$}}
    \put(60, 11){{\footnotesize $\mathbf{\#s=1.5k}$}}
    \end{overpic}
    \caption{Qualitative comparison of the medial mesh and the reconstructed mesh among ours and  PC~\cite{amenta2001power}, SAT~\cite{miklos2010sat},  VC~\cite{yan2018voxel}, and  MATFP~\cite{2022MATFP}. Here $\epsilon^1$ and $\epsilon^2$ are the Hausdorff distance errors described in Sec~\ref{sec:exp}. 
    }
    \label{fig:comp_recon}
\end{figure}

We compare our method with three classical methods for approximating medial axis, including two point-cloud-based methods -- PC (Power Crust~\cite{amenta2001power}) and SAT (Scaled Axis Transform~\cite{miklos2010sat}), and one voxel-based method VC (Voxel Cores~\cite{yan2018voxel}), regarding feature and topology preservation quality (Fig~\ref{fig:comp_feature}), and surface reconstruction quality from the generated medial mesh (Fig~\ref{fig:comp_recon}).

We test the PC~\cite{amenta2001power} method using two different sampling densities, and experiment SAT~\cite{miklos2010sat} method using two values of sampling parameter: $\delta=0.04$ and $\delta=0.03$ in Fig~\ref{fig:comp_recon}, and set the scale parameter as default $1.0$ in all experiments. Similar to all other point-cloud-based methods, the quality of medial mesh generated would improve when the surface sampling density increases. However, these two methods generate medial meshes that are not thin with large number of flat tetrahedrons, and their Euler characteristic $\euler$ varies as shown in Fig~\ref{fig:comp_topo}, \ie $\euler=32$ and $\euler=8$ respectively. In addition, they cannot preserve any medial features, both externally and internally, as shown in Fig~\ref{fig:comp_feature}.

We also compare with VC~\cite{yan2018voxel} method regarding the reconstruction quality and the feature preservation quality of generated medial mesh. We use two voxel sizes $2^8$ and $2^9$ with pruning parameters $\lambda=0.03$, shown in Fig~\ref{fig:comp_recon}. Even though VC can output the topologically-correct medial mesh, it requires large number of medial spheres for outputting a smooth structure around the internal features, almost $10$ times more than our method (\ie $30k$ in Fig~\ref{fig:comp_topo}). This is due to the fact that VC controls the sampling globally based on the size of the voxels. Smaller voxel size would generates denser medial spheres. It cannot, however, directly control the sampling rate of medial spheres as needed. We also found that the medial mesh generated from VC shrinks more as the value of pruning parameter increases, which will result in a more rounded reconstruction result around the external features.

\subsection{Comparisons on Organic Models} 

We also test our method on $12$ organic models, and compare it with PC~\cite{amenta2001power}, SAT~\cite{miklos2010sat}, VC~\cite{yan2018voxel} and MATFP~\cite{2022MATFP}. We show the visual and quantitative comparison in Fig.~\ref{fig:comp_non_cad}. More statistics can be found in Table.~3 of the Supplementary Material.

We experiment SAT using the sampling parameter $\delta=0.03$ and set the scale parameter as default value $1.0$. We compare with the VC using the voxel size $2^8$ and the pruning parameter $\lambda=0.03$. For our method, we set $\geoerror=1.5$ for all organic models tested.

We have found that our method normally generates similar reconstruction accuracy with much fewer medial spheres (\ie PC $14k$, SAT $17k$, VC $36k$, MATFP $11k$, and ours $4k$ for the `Fertility' model). Our method can also maintain the topology and thinness property of the 3D medial axis w.r.t. the input shape, same as VC (\ie PC $35k$, SAT $7$, VC $1$, MATFP $4$, and ours $1$ for the Euler characteristic $\euler$ of `Rozy' model).

\subsection{Ablation Study}
\label{sec:ablation}

\paragraph{Geometric Error Bound $\geoerror$}
\label{sec:ablation_bound}

\begin{figure}[h!]
    \centering
    \begin{overpic}[width=\linewidth]{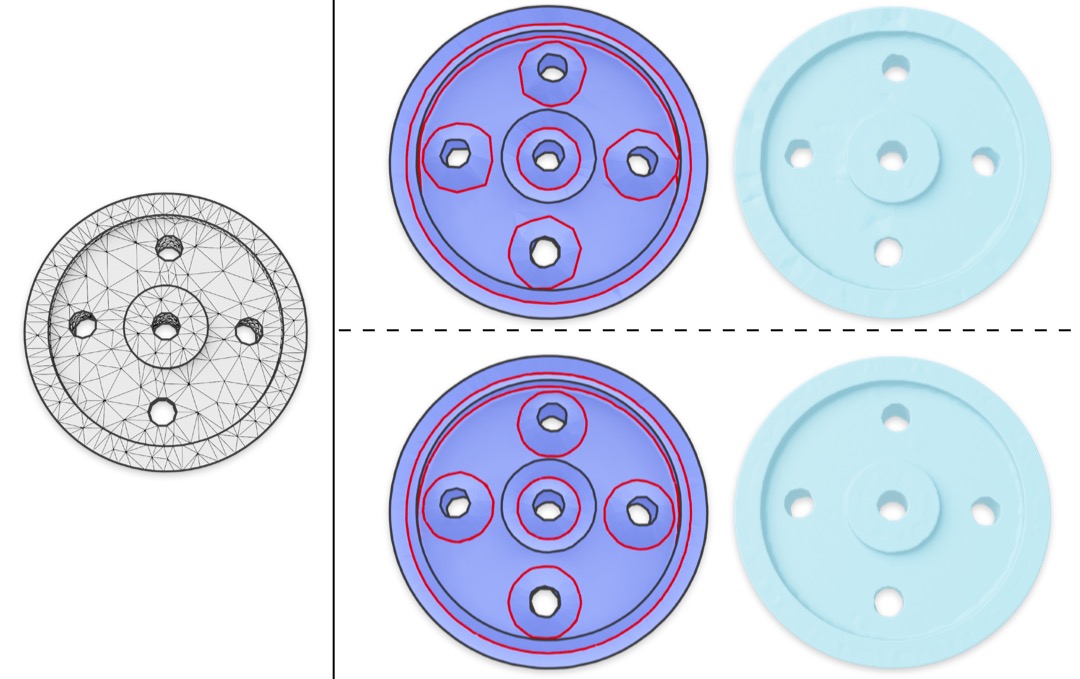}
    \put(6, 57){\textbf{(a) $\geoerror=1.5$} }
    \put(8, 53){{\footnotesize $\delta^1=1.48\%$}}
    \put(8, 50){{\footnotesize $\delta^2=1.13\%$}}
    \put(8, 47){{\footnotesize $\#s=1.5k$}}
    \put(6, 14){\textbf{(b) $\geoerror=0.6$} }
    \put(8, 10){{\footnotesize $\delta^1=0.47\%$}}
    \put(8, 7){{\footnotesize $\delta^2=0.58\%$}}
    \put(8, 4){{\footnotesize $\#s=6.2k$}}
    \end{overpic}
    \caption{Ablation study on geometric error bound $\geoerror$. We use two different error bounds: (a) $\geoerror=1.5$ and (b) $\geoerror=0.6$. Smaller error bound can generate a medial mesh with smoother features which results in a better reconstruction quality. }
    \label{fig:ablation_fix_geo}
\end{figure}

One important parameter used in our method is the user-defined geometric error bound $\geoerror$, which controls the one-sided reconstruction accuracy (from input surface to the reconstructed mesh). Here the error is measured by the distance from surface samples to the enveloping volume of the medial mesh. Similar to Hausdorff distance described in Sec~\ref{sec:exp}, this error is also scaled based on the diagonal length of the shape's bounding box. We show the effect of two different values of the parameter $\geoerror$ in Fig.~\ref{fig:ablation_fix_geo}. A smaller value of $\geoerror$ would generate a smoother medial mesh around non-feature regions, as more non-feature spheres are sampled. It can also reduce the non-smooth connections on medial features, either external or internal ones (see the black and red curves respectively in Fig.~\ref{fig:ablation_fix_geo}). As a result, a smaller error bound will inevitability output a reconstructed mesh with better quality.

\begin{figure}[h!]
    \centering
    \begin{overpic}[width=\linewidth]{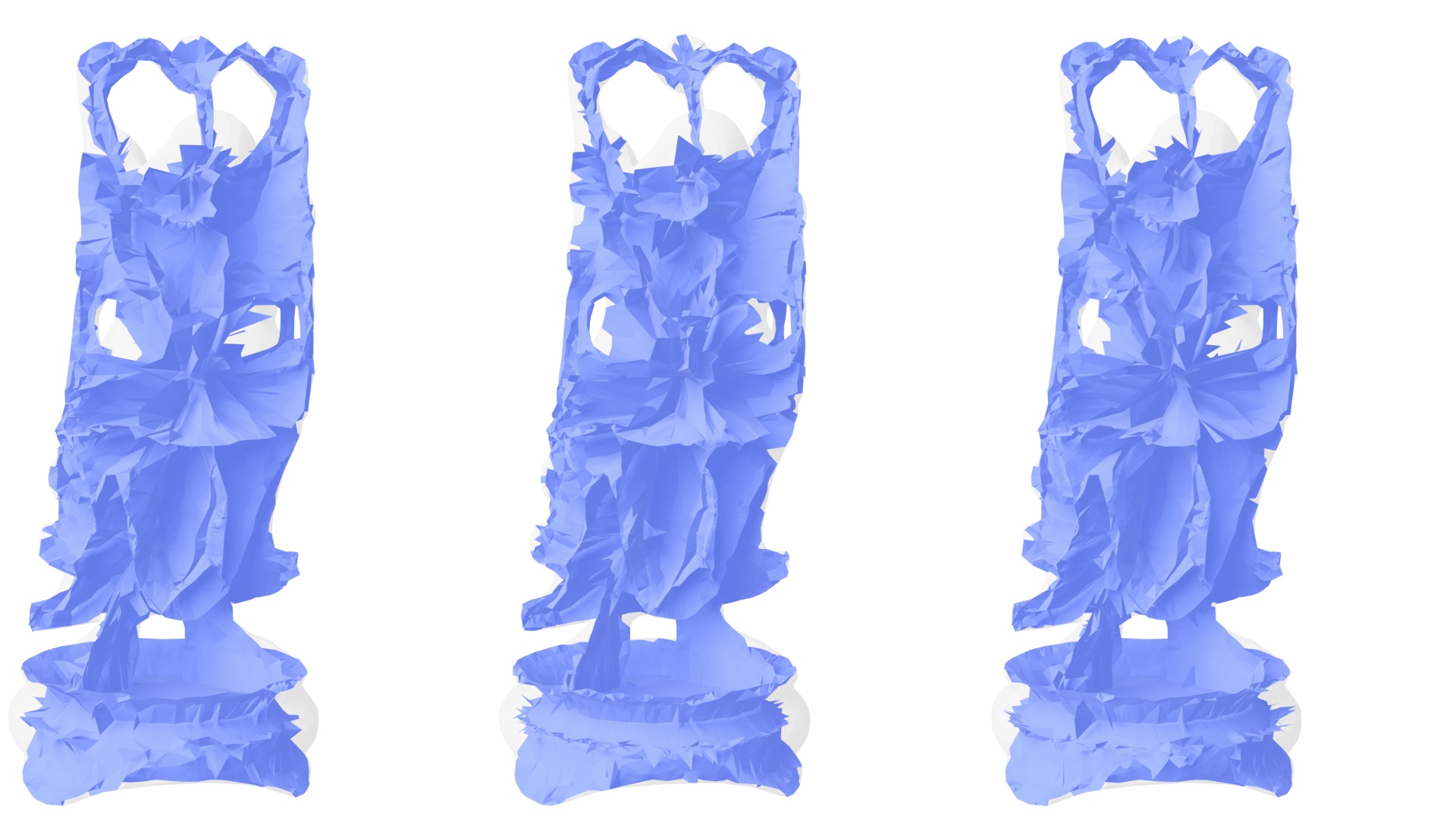}
    \put(9, -2){\textbf{(a)}}
    \put(44, -2){\textbf{(b)}}
    \put(78, -2){\textbf{(c)}}
    \put(22, 50){$\#t=7k$}
    \put(22, 45){$\#s_{init}=200$}
    \put(23, 40){$\euler=-7$}
    \put(23, 12){{\footnotesize $\delta^1=1.32\%$}}
    \put(23, 8){{\footnotesize $\delta^2=1.25\%$}}
    \put(23, 4){{\footnotesize $\#s=10k$}}
    \put(23, 0){{\footnotesize $T=639s$}}
    \put(56, 50){\underline{$\#t=10k$}}
    \put(55, 45){$\#s_{init}=200$}
    \put(56, 40){$\euler=-7$}
    \put(56, 12){{\footnotesize $\delta^1=1.16\%$}}
    \put(56, 8){{\footnotesize $\delta^2=1.02\%$}}
    \put(56, 4){{\footnotesize $\#s=10k$}}
    \put(56, 0){{\footnotesize $T=662s$}}
    \put(89, 50){$\#t=7k$}
    \put(89, 45){\underline{$\#s_{init}=1k$}}
    \put(90, 40){$\euler=-7$}
    \put(89, 12){{\footnotesize $\delta^1=1.42\%$}}
    \put(89, 8){{\footnotesize $\delta^2=1.14\%$}}
    \put(89, 4){{\footnotesize $\#s=9k$}}
    \put(89, 0){{\footnotesize $T=357s$}}
    \end{overpic}
    \caption{Ablation study on two initialization parameters: (1) $\#t$ as the number of tets in the input tetrahedral mesh; (2) $\#s_{init}$ as the number of initial spheres. The base case (a) uses $\#s_{init}=200$ and $\#t=7k$. While keeping the same and correct topology (GT genus=$8$, GT $\euler=-7$) and similar reconstruction ability, a higher number of $\#t$ would inevitably increase the running time (b), and a higher number of $\#s_{init}$ may reduce the processing time (c).}
    \label{fig:ablation_init}
\end{figure}
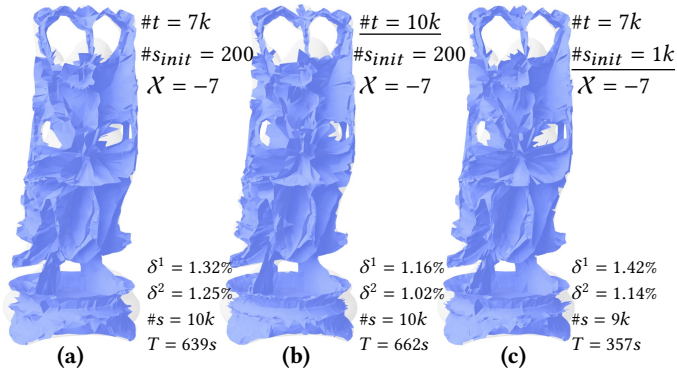

\paragraph{Initialization} 
There are two parameters that may impact our running time. One is $\#t$ as the number of tets in the input tetrahedral mesh; the other is $\#s_{init}$ as the number of initial spheres. We use the `Happy Buddha' model~\cite{dey2007computing} (genus=8) for testing with the error bound $\geoerror=1.5$ and show the results in Fig.~\ref{fig:ablation_init}. Our method can generate the medial mesh with the correct topology ($\euler$=-7) and similar reconstruction quality using different value of $\#t$ and $\#s_{init}$. Increasing $\#t$ (\ie smaller value of `l' parameter of fTetwild~\cite{hu2020ftetwild}) would inevitably make our RPD calculation slower as more `Tet-Cell' clipping steps need to be performed. A larger number of $\#s_{init}$ (\ie use $1k$ than $200$ in Fig.~\ref{fig:ablation_init}) would reduce the processing time.

%% file: 7_conclusion.tex
\begin{table}[!t]
\small
\caption{Statistic of our running time in seconds. $\#t$ is the number of tets in the given tetrahedral mesh. $\#s$ is the number of generated medial spheres. $\#\text{RPD}$ is the number of volumetric RPD calculated. $S_{topo}$ is the running time of topology preservation step (Sec.~\ref{sec:fix_topo}). $S_{extf}$ and $S_{intf}$ is the running time for preserving external features and internal features (Sec~\ref{sec:fix_extf_intf}). The model's ID\# correspond to those shown in Table~\ref{tab:comp_euler}. }
\begin{center}
\scalebox{0.9}{
\begin{tabular}{c||c|c|c|c|c|c|c|c}
\hline
Model ID& $\#t$ & $\#s$ & $\#\text{RPD}$ & $S_{topo}$ & $S_{extf}$ & $S_{intf}$ & $S_{geo}$ & Total (s)
\\
\hline
549 & 8k& 7.2k & 34 & 74 & 115 & 376 & 99 & 664\\
4123 & 1.2k & 5.1k & 35 & 15 & 42 & 109 & 32 & 198\\
5227 & 1k & 3.2k & 31 & 10 & 22 & 23 &  17 & 72 \\
8315 & 4k & 1.5k & 37 & 64 & 73 & 444 & 112 & 693 \\
8964 & 3.6k & 8.7k & 28 & 36 & 113 & 394 & 16 & 559 \\
10836 & 3.8k & 2.7k & 41 & 30 & 29 & 126 & 63 & 248\\
\hline
\end{tabular}}
\end{center}
\label{tab:runtime}
\end{table}

\section{Limitations and future work}
\label{sec:limitations}

Our method can produce an approximated MAT that is visually and quantitatively similar to those of MATFP~\cite{2022MATFP} geometry-wise, with less spheres and the additional benefit of topology preservation. 
Although both our method and MATFP rely on RPD, the former employs volumetric RPD, which necessitates cutting tets by half-spaces inside the volume, while the latter utilizes surface RPD, which only considers the intersection between triangles and half-spaces. Consequently, our method requires more computational time. As the runtime statistics shown in Table~\ref{tab:runtime}, our method takes minutes for each model even though each step's calculation of volumetric RPD on GPU only takes about 1-3 seconds. In the future, we will consider parallelising the feature preserving ($S_{extf}$ and $S_{intf}$) stage as much as possible to reduce the runtime.

Moreover, our current GPU-based implementation may fail to compute RPD if the given tetrahedral mesh has a very large number of tets. For example, the model $\#12280$ in Supplementary Document contains over 5.5 million tets even using the fTetwild~\cite{hu2020ftetwild} with the largest length parameter value $l=1$. We leave this computational issue for future exploration. 

%% file: 8_figureonly.tex
\begin{figure*}[h!]
    \centering
    \begin{overpic}[width=\linewidth]{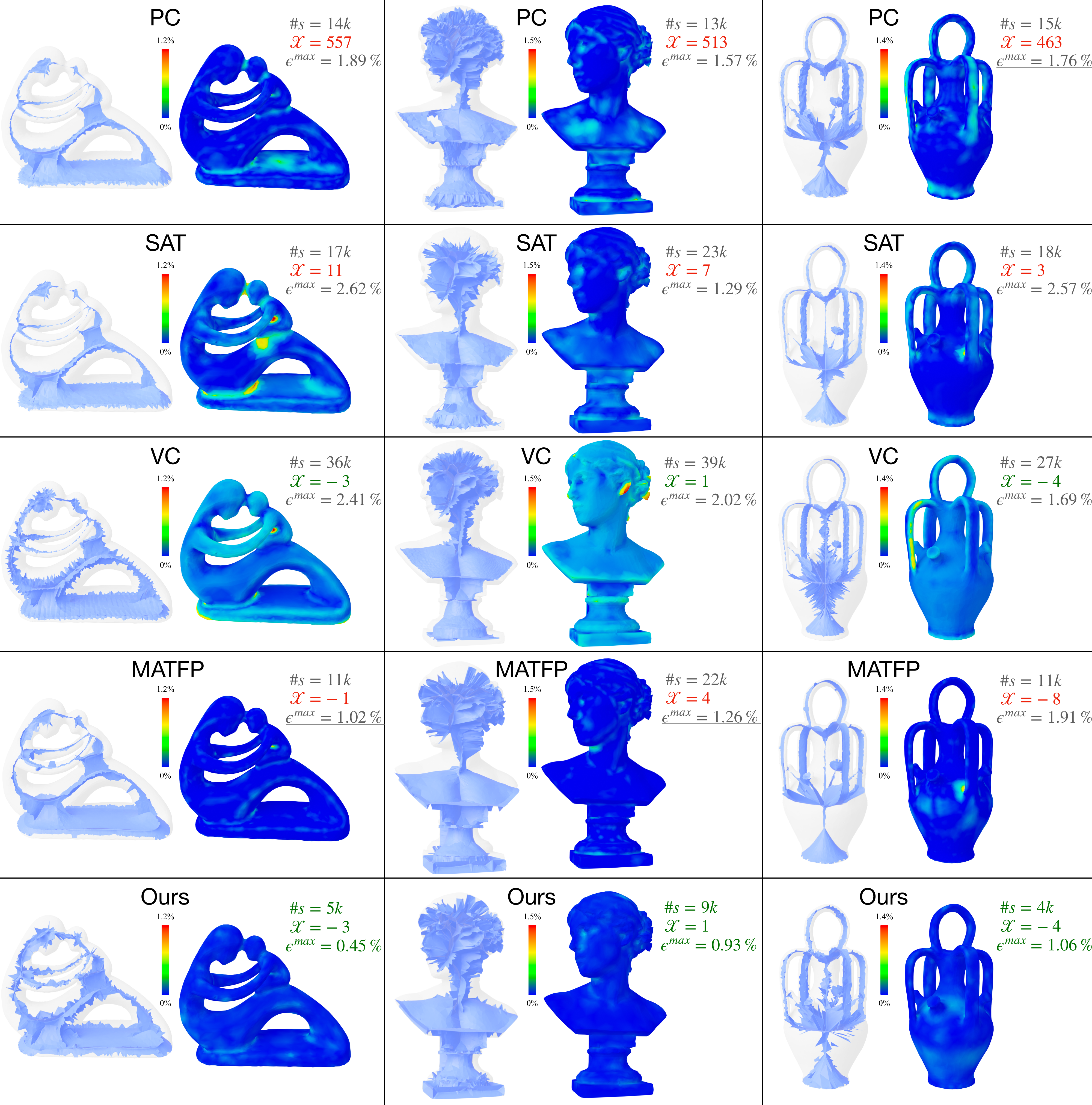}
    \end{overpic}
    \caption{Comparison with PC~\cite{amenta2001power}, SAT~\cite{miklos2010sat}, VC~\cite{yan2018voxel} and MATFP~\cite{2022MATFP} on three organic models. We show the generated medial mesh and the color-coded distribution of two-sided Hausdorff errors between the input surface and the reconstructed surface. Here, $s$ is the number of generated medial spheres. And $\euler$ is the Euler characteristic of the computed medial mesh, where a wrong number is shown in red. $\epsilon^{max}$ is the maximum of two-sided Hausdorff error, as the best result shown in green and the second best shown with underline.}
    \label{fig:comp_non_cad}
\end{figure*}

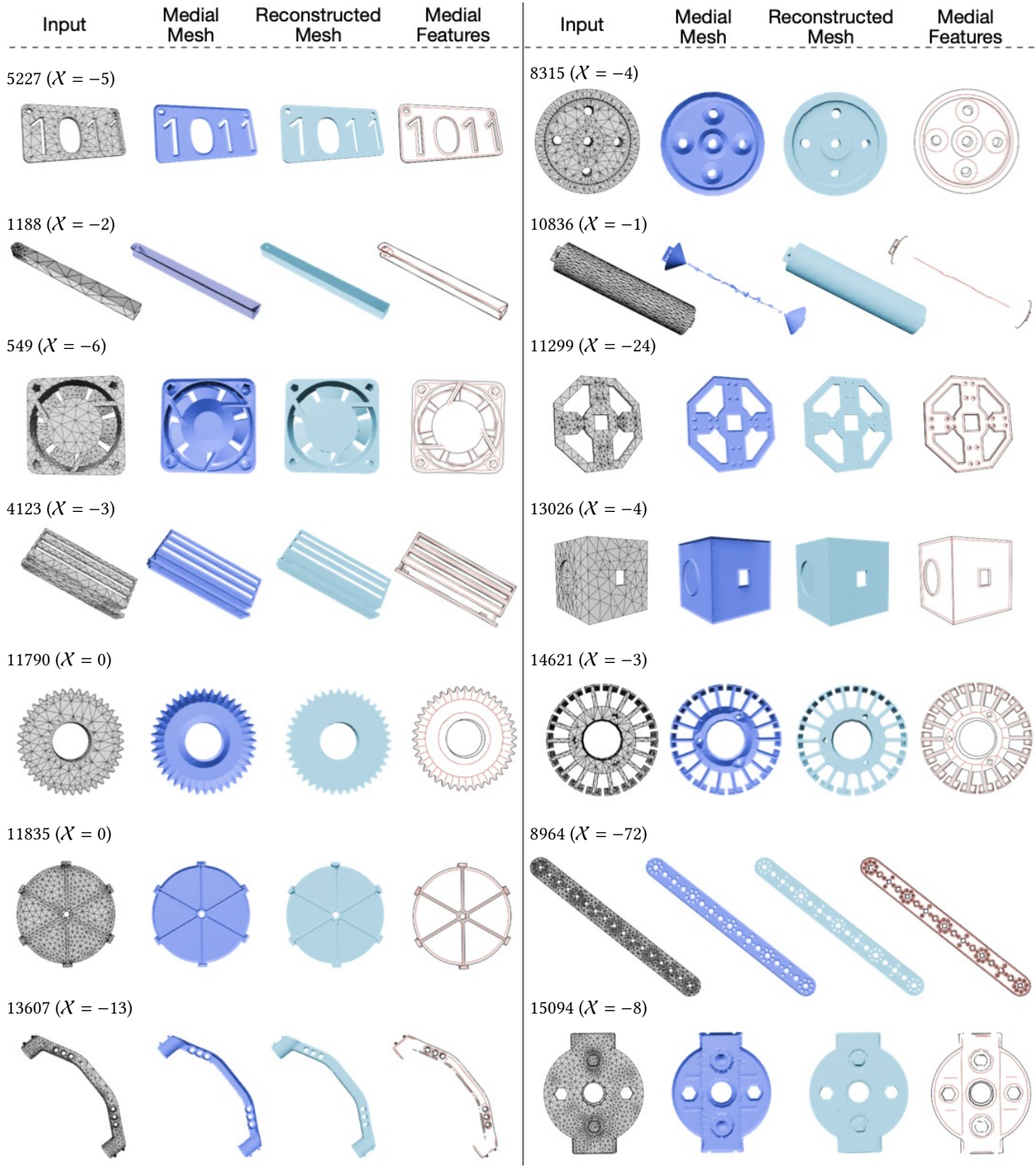
\begin{figure*}[h!]
    \centering
    \begin{overpic}[width=0.96\linewidth]{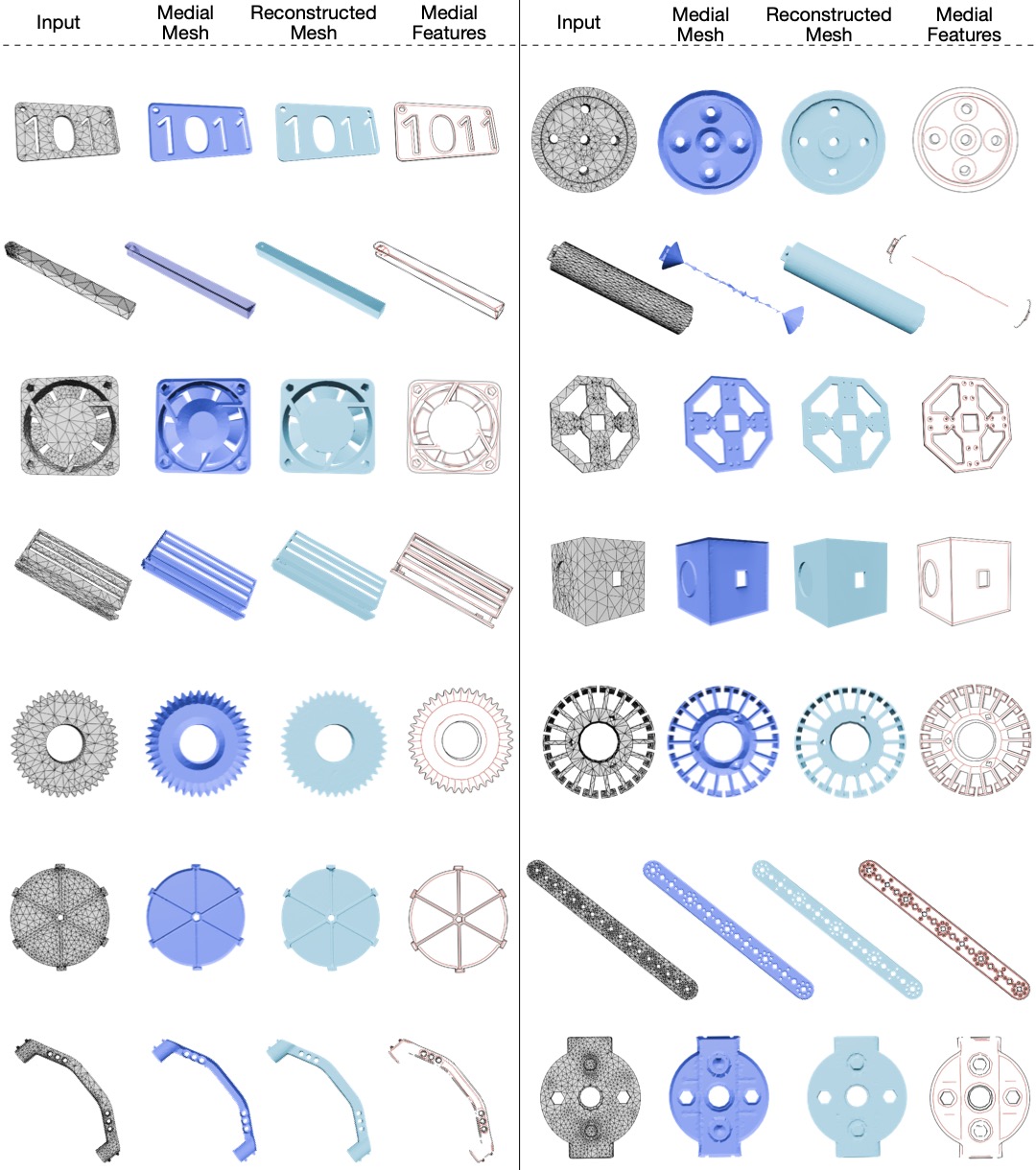}
    \put(0,93){5227 ($\euler=-5$)}
    \put(0,80.5){1188 ($\euler=-2$)}
    \put(0, 70){549 ($\euler=-6$)}
    \put(0, 56){4123 ($\euler=-3$)}
    \put(0, 43){11790 ($\euler=0$)}
    \put(0, 28){11835 ($\euler=0$)}
    \put(0, 13){13607 ($\euler=-13$)}
    \put(45,93.5){8315 ($\euler=-4$)}
    \put(45,80.5){10836 ($\euler=-1$)}
    \put(45, 70){11299 ($\euler=-24$)}
    \put(45, 56){13026 ($\euler=-4$)}
    \put(45, 43){14621 ($\euler=-3$)}
    \put(45, 28){8964 ($\euler=-72$)}
    \put(45, 13){15094 ($\euler=-8$)}
    \end{overpic}
    \caption{Visualization of models shown in Table~\ref{tab:comp_euler}, with their model ID and Euler characteristic $\euler$. From left to right are the input tetrahedral meshes, the generated medial meshes, the surfaces reconstructed from our medial meshes, and the extracted medial features. For the medial features, the black curves are the external features and the red curves are the internal features.}
    \label{fig:mattopo_results}
\end{figure*}

%% file: main.bbl

\begin{thebibliography}{62}


\ifx \showCODEN    \undefined \def \showCODEN     #1{\unskip}     \fi
\ifx \showDOI      \undefined \def \showDOI       #1{#1}\fi
\ifx \showISBNx    \undefined \def \showISBNx     #1{\unskip}     \fi
\ifx \showISBNxiii \undefined \def \showISBNxiii  #1{\unskip}     \fi
\ifx \showISSN     \undefined \def \showISSN      #1{\unskip}     \fi
\ifx \showLCCN     \undefined \def \showLCCN      #1{\unskip}     \fi
\ifx \shownote     \undefined \def \shownote      #1{#1}          \fi
\ifx \showarticletitle \undefined \def \showarticletitle #1{#1}   \fi
\ifx \showURL      \undefined \def \showURL       {\relax}        \fi
\providecommand\bibfield[2]{#2}
\providecommand\bibinfo[2]{#2}
\providecommand\natexlab[1]{#1}
\providecommand\showeprint[2][]{arXiv:#2}

\bibitem[\protect\citeauthoryear{Abdelkader, Bajaj, Ebeida, Mahmoud, Mitchell, Owens, and Rushdi}{Abdelkader et~al\mbox{.}}{2020}]%
        {abdelkader2020vorocrust}
\bibfield{author}{\bibinfo{person}{Ahmed Abdelkader}, \bibinfo{person}{Chandrajit~L Bajaj}, \bibinfo{person}{Mohamed~S Ebeida}, \bibinfo{person}{Ahmed~H Mahmoud}, \bibinfo{person}{Scott~A Mitchell}, \bibinfo{person}{John~D Owens}, {and} \bibinfo{person}{Ahmad~A Rushdi}.} \bibinfo{year}{2020}\natexlab{}.
\newblock \showarticletitle{VoroCrust: Voronoi meshing without clipping}.
\newblock \bibinfo{journal}{\emph{ACM Transactions on Graphics (TOG)}} \bibinfo{volume}{39}, \bibinfo{number}{3} (\bibinfo{year}{2020}), \bibinfo{pages}{1--16}.
\newblock


\bibitem[\protect\citeauthoryear{Amenta, Choi, and Kolluri}{Amenta et~al\mbox{.}}{2001}]%
        {amenta2001power}
\bibfield{author}{\bibinfo{person}{Nina Amenta}, \bibinfo{person}{Sunghee Choi}, {and} \bibinfo{person}{Ravi~Krishna Kolluri}.} \bibinfo{year}{2001}\natexlab{}.
\newblock \showarticletitle{The power crust}. In \bibinfo{booktitle}{\emph{Proceedings of the sixth ACM symposium on Solid modeling and applications}}. \bibinfo{pages}{249--266}.
\newblock


\bibitem[\protect\citeauthoryear{Aurenhammer}{Aurenhammer}{1987}]%
        {aurenhammer1987power}
\bibfield{author}{\bibinfo{person}{Franz Aurenhammer}.} \bibinfo{year}{1987}\natexlab{}.
\newblock \showarticletitle{Power diagrams: properties, algorithms and applications}.
\newblock \bibinfo{journal}{\emph{SIAM J. Comput.}} \bibinfo{volume}{16}, \bibinfo{number}{1} (\bibinfo{year}{1987}), \bibinfo{pages}{78--96}.
\newblock


\bibitem[\protect\citeauthoryear{Basselin, Alonso, Ray, Sokolov, Lefebvre, and L{\'e}vy}{Basselin et~al\mbox{.}}{2021}]%
        {basselin2021RPD}
\bibfield{author}{\bibinfo{person}{Justine Basselin}, \bibinfo{person}{Laurent Alonso}, \bibinfo{person}{Nicolas Ray}, \bibinfo{person}{Dmitry Sokolov}, \bibinfo{person}{Sylvain Lefebvre}, {and} \bibinfo{person}{Bruno L{\'e}vy}.} \bibinfo{year}{2021}\natexlab{}.
\newblock \showarticletitle{Restricted power diagrams on the GPU}. In \bibinfo{booktitle}{\emph{Computer Graphics Forum}}, Vol.~\bibinfo{volume}{40}. Wiley Online Library, \bibinfo{pages}{1--12}.
\newblock


\bibitem[\protect\citeauthoryear{Blum et~al\mbox{.}}{Blum et~al\mbox{.}}{1967}]%
        {blum1967transformation}
\bibfield{author}{\bibinfo{person}{Harry Blum} {et~al\mbox{.}}} \bibinfo{year}{1967}\natexlab{}.
\newblock \bibinfo{booktitle}{\emph{A transformation for extracting new descriptors of shape}}. Vol.~\bibinfo{volume}{43}.
\newblock \bibinfo{publisher}{MIT press Cambridge, MA}.
\newblock


\bibitem[\protect\citeauthoryear{Brandt and Algazi}{Brandt and Algazi}{1992}]%
        {brandt1992continuous}
\bibfield{author}{\bibinfo{person}{Jonathan~W Brandt} {and} \bibinfo{person}{V~Ralph Algazi}.} \bibinfo{year}{1992}\natexlab{}.
\newblock \showarticletitle{Continuous skeleton computation by Voronoi diagram}.
\newblock \bibinfo{journal}{\emph{CVGIP: Image understanding}} \bibinfo{volume}{55}, \bibinfo{number}{3} (\bibinfo{year}{1992}), \bibinfo{pages}{329--338}.
\newblock


\bibitem[\protect\citeauthoryear{Carlsson and Vejdemo-Johansson}{Carlsson and Vejdemo-Johansson}{2021}]%
        {carlsson2021topological}
\bibfield{author}{\bibinfo{person}{Gunnar Carlsson} {and} \bibinfo{person}{Mikael Vejdemo-Johansson}.} \bibinfo{year}{2021}\natexlab{}.
\newblock \bibinfo{booktitle}{\emph{Topological data analysis with applications}}.
\newblock \bibinfo{publisher}{Cambridge University Press}.
\newblock


\bibitem[\protect\citeauthoryear{Chazal and Lieutier}{Chazal and Lieutier}{2005}]%
        {chazal2005lambda}
\bibfield{author}{\bibinfo{person}{Fr{\'e}d{\'e}ric Chazal} {and} \bibinfo{person}{Andr{\'e} Lieutier}.} \bibinfo{year}{2005}\natexlab{}.
\newblock \showarticletitle{The “$\lambda$-medial axis”}.
\newblock \bibinfo{journal}{\emph{Graphical Models}} \bibinfo{volume}{67}, \bibinfo{number}{4} (\bibinfo{year}{2005}), \bibinfo{pages}{304--331}.
\newblock


\bibitem[\protect\citeauthoryear{Chazal and Lieutier}{Chazal and Lieutier}{2008}]%
        {chazal2008smooth}
\bibfield{author}{\bibinfo{person}{Fr{\'e}d{\'e}ric Chazal} {and} \bibinfo{person}{Andr{\'e} Lieutier}.} \bibinfo{year}{2008}\natexlab{}.
\newblock \showarticletitle{Smooth manifold reconstruction from noisy and non-uniform approximation with guarantees}.
\newblock \bibinfo{journal}{\emph{Computational Geometry}} \bibinfo{volume}{40}, \bibinfo{number}{2} (\bibinfo{year}{2008}), \bibinfo{pages}{156--170}.
\newblock


\bibitem[\protect\citeauthoryear{Culver, Keyser, and Manocha}{Culver et~al\mbox{.}}{2004}]%
        {culver2004exact}
\bibfield{author}{\bibinfo{person}{Tim Culver}, \bibinfo{person}{John Keyser}, {and} \bibinfo{person}{Dinesh Manocha}.} \bibinfo{year}{2004}\natexlab{}.
\newblock \showarticletitle{Exact computation of the medial axis of a polyhedron}.
\newblock \bibinfo{journal}{\emph{Computer Aided Geometric Design}} \bibinfo{volume}{21}, \bibinfo{number}{1} (\bibinfo{year}{2004}), \bibinfo{pages}{65--98}.
\newblock


\bibitem[\protect\citeauthoryear{Dey, Li, and Sun}{Dey et~al\mbox{.}}{2007}]%
        {dey2007computing}
\bibfield{author}{\bibinfo{person}{Tamal~K Dey}, \bibinfo{person}{Kuiyu Li}, {and} \bibinfo{person}{Jian Sun}.} \bibinfo{year}{2007}\natexlab{}.
\newblock \showarticletitle{On computing handle and tunnel loops}. In \bibinfo{booktitle}{\emph{2007 International Conference on Cyberworlds (CW'07)}}. IEEE, \bibinfo{pages}{357--366}.
\newblock


\bibitem[\protect\citeauthoryear{Dey and Zhao}{Dey and Zhao}{2002}]%
        {dey2002approximate}
\bibfield{author}{\bibinfo{person}{Tamal~K Dey} {and} \bibinfo{person}{Wulue Zhao}.} \bibinfo{year}{2002}\natexlab{}.
\newblock \showarticletitle{Approximate medial axis as a voronoi subcomplex}. In \bibinfo{booktitle}{\emph{Proceedings of the seventh ACM symposium on Solid modeling and applications}}. \bibinfo{pages}{356--366}.
\newblock


\bibitem[\protect\citeauthoryear{Dey and Zhao}{Dey and Zhao}{2004}]%
        {dey2004approximating}
\bibfield{author}{\bibinfo{person}{Tamal~K Dey} {and} \bibinfo{person}{Wulue Zhao}.} \bibinfo{year}{2004}\natexlab{}.
\newblock \showarticletitle{Approximating the medial axis from the Voronoi diagram with a convergence guarantee}.
\newblock \bibinfo{journal}{\emph{Algorithmica}} \bibinfo{volume}{38}, \bibinfo{number}{1} (\bibinfo{year}{2004}), \bibinfo{pages}{179--200}.
\newblock


\bibitem[\protect\citeauthoryear{Dou, Lin, Xu, Yang, Xin, Komura, and Wang}{Dou et~al\mbox{.}}{2022}]%
        {dou2021coverage}
\bibfield{author}{\bibinfo{person}{Zhiyang Dou}, \bibinfo{person}{Cheng Lin}, \bibinfo{person}{Rui Xu}, \bibinfo{person}{Lei Yang}, \bibinfo{person}{Shiqing Xin}, \bibinfo{person}{Taku Komura}, {and} \bibinfo{person}{Wenping Wang}.} \bibinfo{year}{2022}\natexlab{}.
\newblock \showarticletitle{Coverage Axis: Inner Point Selection for 3D Shape Skeletonization}. In \bibinfo{booktitle}{\emph{Computer Graphics Forum}}, Vol.~\bibinfo{volume}{41}. Wiley Online Library, \bibinfo{pages}{419--432}.
\newblock


\bibitem[\protect\citeauthoryear{Dou, Xin, Xu, Xu, Zhou, Chen, Wang, Zhao, and Tu}{Dou et~al\mbox{.}}{2020}]%
        {dou2020top}
\bibfield{author}{\bibinfo{person}{Zhiyang Dou}, \bibinfo{person}{Shiqing Xin}, \bibinfo{person}{Rui Xu}, \bibinfo{person}{Jian Xu}, \bibinfo{person}{Yuanfeng Zhou}, \bibinfo{person}{Shuangmin Chen}, \bibinfo{person}{Wenping Wang}, \bibinfo{person}{Xiuyang Zhao}, {and} \bibinfo{person}{Changhe Tu}.} \bibinfo{year}{2020}\natexlab{}.
\newblock \showarticletitle{Top-down shape abstraction based on greedy pole selection}.
\newblock \bibinfo{journal}{\emph{IEEE Transactions on Visualization and Computer Graphics}} \bibinfo{volume}{27}, \bibinfo{number}{10} (\bibinfo{year}{2020}), \bibinfo{pages}{3982--3993}.
\newblock


\bibitem[\protect\citeauthoryear{Fabri and Pion}{Fabri and Pion}{2009}]%
        {fabri2009cgal}
\bibfield{author}{\bibinfo{person}{Andreas Fabri} {and} \bibinfo{person}{Sylvain Pion}.} \bibinfo{year}{2009}\natexlab{}.
\newblock \showarticletitle{CGAL: The computational geometry algorithms library}. In \bibinfo{booktitle}{\emph{Proceedings of the 17th ACM SIGSPATIAL international conference on advances in geographic information systems}}. \bibinfo{pages}{538--539}.
\newblock


\bibitem[\protect\citeauthoryear{Fu, Xu, Xin, Chen, Tu, Yang, and Lu}{Fu et~al\mbox{.}}{2022}]%
        {fu2022easyvrmodeling}
\bibfield{author}{\bibinfo{person}{Zhiying Fu}, \bibinfo{person}{Rui Xu}, \bibinfo{person}{Shiqing Xin}, \bibinfo{person}{Shuangmin Chen}, \bibinfo{person}{Changhe Tu}, \bibinfo{person}{Chenglei Yang}, {and} \bibinfo{person}{Lin Lu}.} \bibinfo{year}{2022}\natexlab{}.
\newblock \showarticletitle{Easyvrmodeling: Easily create 3d models by an immersive vr system}.
\newblock \bibinfo{journal}{\emph{Proceedings of the ACM on Computer Graphics and Interactive Techniques}} \bibinfo{volume}{5}, \bibinfo{number}{1} (\bibinfo{year}{2022}), \bibinfo{pages}{1--14}.
\newblock


\bibitem[\protect\citeauthoryear{Ge, Yao, Yang, Wang, Chen, and Guo}{Ge et~al\mbox{.}}{2023}]%
        {ge2023point2mm}
\bibfield{author}{\bibinfo{person}{Mengyuan Ge}, \bibinfo{person}{Junfeng Yao}, \bibinfo{person}{Baorong Yang}, \bibinfo{person}{Ningna Wang}, \bibinfo{person}{Zhonggui Chen}, {and} \bibinfo{person}{Xiaohu Guo}.} \bibinfo{year}{2023}\natexlab{}.
\newblock \showarticletitle{Point2MM: Learning medial mesh from point clouds}.
\newblock \bibinfo{journal}{\emph{Computers \& Graphics}}  \bibinfo{volume}{115} (\bibinfo{year}{2023}), \bibinfo{pages}{511--521}.
\newblock


\bibitem[\protect\citeauthoryear{Hesselink and Roerdink}{Hesselink and Roerdink}{2008}]%
        {hesselink2008euclidean}
\bibfield{author}{\bibinfo{person}{Wim~H Hesselink} {and} \bibinfo{person}{Jos~BTM Roerdink}.} \bibinfo{year}{2008}\natexlab{}.
\newblock \showarticletitle{Euclidean skeletons of digital image and volume data in linear time by the integer medial axis transform}.
\newblock \bibinfo{journal}{\emph{IEEE Transactions on Pattern Analysis and Machine Intelligence}} \bibinfo{volume}{30}, \bibinfo{number}{12} (\bibinfo{year}{2008}), \bibinfo{pages}{2204--2217}.
\newblock


\bibitem[\protect\citeauthoryear{Hu, Chen, Yang, Wang, Guo, and Wang}{Hu et~al\mbox{.}}{2022}]%
        {hu2022immat}
\bibfield{author}{\bibinfo{person}{Jianwei Hu}, \bibinfo{person}{Gang Chen}, \bibinfo{person}{Baorong Yang}, \bibinfo{person}{Ningna Wang}, \bibinfo{person}{Xiaohu Guo}, {and} \bibinfo{person}{Bin Wang}.} \bibinfo{year}{2022}\natexlab{}.
\newblock \showarticletitle{IMMAT: Mesh reconstruction from single view images by medial axis transform prediction}.
\newblock \bibinfo{journal}{\emph{Computer-Aided Design}}  \bibinfo{volume}{150} (\bibinfo{year}{2022}), \bibinfo{pages}{103304}.
\newblock


\bibitem[\protect\citeauthoryear{Hu, Wang, Qian, Pan, Guo, Liu, and Wang}{Hu et~al\mbox{.}}{2019}]%
        {Hu2019MATNet}
\bibfield{author}{\bibinfo{person}{Jianwei Hu}, \bibinfo{person}{Bin Wang}, \bibinfo{person}{Lihui Qian}, \bibinfo{person}{Yiling Pan}, \bibinfo{person}{Xiaohu Guo}, \bibinfo{person}{Lingjie Liu}, {and} \bibinfo{person}{Wenping Wang}.} \bibinfo{year}{2019}\natexlab{}.
\newblock \showarticletitle{MAT-Net: Medial Axis Transform Network for 3D Object Recognition}. In \bibinfo{booktitle}{\emph{Proceedings of the 28th International Joint Conference on Artificial Intelligence}} \emph{(\bibinfo{series}{IJCAI'19})}. \bibinfo{pages}{774–781}.
\newblock


\bibitem[\protect\citeauthoryear{Hu, Wang, Yang, Chen, Guo, and Wang}{Hu et~al\mbox{.}}{2023}]%
        {hu2023s3ds}
\bibfield{author}{\bibinfo{person}{Jianwei Hu}, \bibinfo{person}{Ningna Wang}, \bibinfo{person}{Baorong Yang}, \bibinfo{person}{Gang Chen}, \bibinfo{person}{Xiaohu Guo}, {and} \bibinfo{person}{Bin Wang}.} \bibinfo{year}{2023}\natexlab{}.
\newblock \showarticletitle{S3DS: Self-supervised Learning of 3D Skeletons from Single View Images}. In \bibinfo{booktitle}{\emph{Proceedings of the 31st ACM International Conference on Multimedia}}. \bibinfo{pages}{6948--6958}.
\newblock


\bibitem[\protect\citeauthoryear{Hu, Schneider, Wang, Zorin, and Panozzo}{Hu et~al\mbox{.}}{2020}]%
        {hu2020ftetwild}
\bibfield{author}{\bibinfo{person}{Yixin Hu}, \bibinfo{person}{Teseo Schneider}, \bibinfo{person}{Bolun Wang}, \bibinfo{person}{Denis Zorin}, {and} \bibinfo{person}{Daniele Panozzo}.} \bibinfo{year}{2020}\natexlab{}.
\newblock \showarticletitle{Fast tetrahedral meshing in the wild}.
\newblock \bibinfo{journal}{\emph{ACM Transactions on Graphics (TOG)}} \bibinfo{volume}{39}, \bibinfo{number}{4} (\bibinfo{year}{2020}), \bibinfo{pages}{117--1}.
\newblock


\bibitem[\protect\citeauthoryear{Jalba, Kustra, and Telea}{Jalba et~al\mbox{.}}{2013}]%
        {kustra2013}
\bibfield{author}{\bibinfo{person}{Andrei~C. Jalba}, \bibinfo{person}{Jacek Kustra}, {and} \bibinfo{person}{Alexandru~C. Telea}.} \bibinfo{year}{2013}\natexlab{}.
\newblock \showarticletitle{Surface and Curve Skeletonization of Large 3D Models on the GPU}.
\newblock \bibinfo{journal}{\emph{IEEE Transactions on Pattern Analysis and Machine Intelligence}} \bibinfo{volume}{35}, \bibinfo{number}{6} (\bibinfo{year}{2013}), \bibinfo{pages}{1495--1508}.
\newblock
\urldef\tempurl%
\url{https://doi.org/10.1109/TPAMI.2012.212}
\showDOI{\tempurl}


\bibitem[\protect\citeauthoryear{Jalba, Sobiecki, and Telea}{Jalba et~al\mbox{.}}{2015}]%
        {jalba2015unified}
\bibfield{author}{\bibinfo{person}{Andrei~C Jalba}, \bibinfo{person}{Andre Sobiecki}, {and} \bibinfo{person}{Alexandru~C Telea}.} \bibinfo{year}{2015}\natexlab{}.
\newblock \showarticletitle{An unified multiscale framework for planar, surface, and curve skeletonization}.
\newblock \bibinfo{journal}{\emph{IEEE transactions on pattern analysis and machine intelligence}} \bibinfo{volume}{38}, \bibinfo{number}{1} (\bibinfo{year}{2015}), \bibinfo{pages}{30--45}.
\newblock


\bibitem[\protect\citeauthoryear{Ju, Zhou, and Hu}{Ju et~al\mbox{.}}{2007}]%
        {ju2007editing}
\bibfield{author}{\bibinfo{person}{Tao Ju}, \bibinfo{person}{Qian-Yi Zhou}, {and} \bibinfo{person}{Shi-Min Hu}.} \bibinfo{year}{2007}\natexlab{}.
\newblock \showarticletitle{Editing the topology of 3D models by sketching}.
\newblock \bibinfo{journal}{\emph{ACM Transactions on Graphics (TOG)}} \bibinfo{volume}{26}, \bibinfo{number}{3} (\bibinfo{year}{2007}), \bibinfo{pages}{42--es}.
\newblock


\bibitem[\protect\citeauthoryear{Koch, Matveev, Jiang, Williams, Artemov, Burnaev, Alexa, Zorin, and Panozzo}{Koch et~al\mbox{.}}{2019}]%
        {koch2019abc}
\bibfield{author}{\bibinfo{person}{Sebastian Koch}, \bibinfo{person}{Albert Matveev}, \bibinfo{person}{Zhongshi Jiang}, \bibinfo{person}{Francis Williams}, \bibinfo{person}{Alexey Artemov}, \bibinfo{person}{Evgeny Burnaev}, \bibinfo{person}{Marc Alexa}, \bibinfo{person}{Denis Zorin}, {and} \bibinfo{person}{Daniele Panozzo}.} \bibinfo{year}{2019}\natexlab{}.
\newblock \showarticletitle{Abc: A big cad model dataset for geometric deep learning}. In \bibinfo{booktitle}{\emph{Proceedings of the IEEE/CVF conference on computer vision and pattern recognition}}. \bibinfo{pages}{9601--9611}.
\newblock


\bibitem[\protect\citeauthoryear{Kustra, Jalba, and Telea}{Kustra et~al\mbox{.}}{2016}]%
        {kustra2015}
\bibfield{author}{\bibinfo{person}{Jacek Kustra}, \bibinfo{person}{Andrei Jalba}, {and} \bibinfo{person}{Alexandru Telea}.} \bibinfo{year}{2016}\natexlab{}.
\newblock \showarticletitle{Computing refined skeletal features from medial point clouds}.
\newblock \bibinfo{journal}{\emph{Pattern Recognition Letters}}  \bibinfo{volume}{76} (\bibinfo{year}{2016}), \bibinfo{pages}{13--21}.
\newblock
\showISSN{0167-8655}
\urldef\tempurl%
\url{https://doi.org/10.1016/j.patrec.2015.05.007}
\showDOI{\tempurl}
\newblock
\shownote{Special Issue on Skeletonization and its Application}.


\bibitem[\protect\citeauthoryear{Lan, Luo, Fratarcangeli, Xu, Wang, Guo, Yao, and Yang}{Lan et~al\mbox{.}}{2020}]%
        {Lan2020MedialElastics}
\bibfield{author}{\bibinfo{person}{Lei Lan}, \bibinfo{person}{Ran Luo}, \bibinfo{person}{Marco Fratarcangeli}, \bibinfo{person}{Weiwei Xu}, \bibinfo{person}{Huamin Wang}, \bibinfo{person}{Xiaohu Guo}, \bibinfo{person}{Junfeng Yao}, {and} \bibinfo{person}{Yin Yang}.} \bibinfo{year}{2020}\natexlab{}.
\newblock \showarticletitle{Medial Elastics: Efficient and Collision-Ready Deformation via Medial Axis Transform}.
\newblock \bibinfo{journal}{\emph{ACM Trans. Graph.}} \bibinfo{volume}{39}, \bibinfo{number}{3}, Article \bibinfo{articleno}{20} (\bibinfo{date}{apr} \bibinfo{year}{2020}).
\newblock


\bibitem[\protect\citeauthoryear{Leray}{Leray}{1950}]%
        {leray1950anneau}
\bibfield{author}{\bibinfo{person}{J. Leray}.} \bibinfo{year}{1950}\natexlab{}.
\newblock \bibinfo{booktitle}{\emph{L'anneau spectral et l'anneau filtr{\'e} d'homologie d'un espace localement compact et d'une application continue: (cours profess{\'e}s au Coll{\`e}ge de France en 1947-1948 et 1949-1950)}}.
\newblock \bibinfo{publisher}{Gauthier-Villars}.
\newblock
\urldef\tempurl%
\url{https://books.google.com/books?id=QYGKzwEACAAJ}
\showURL{%
\tempurl}


\bibitem[\protect\citeauthoryear{L{\'e}vy and Filbois}{L{\'e}vy and Filbois}{2015}]%
        {levy2015geogram}
\bibfield{author}{\bibinfo{person}{Bruno L{\'e}vy} {and} \bibinfo{person}{Alain Filbois}.} \bibinfo{year}{2015}\natexlab{}.
\newblock \showarticletitle{Geogram: a library for geometric algorithms}.
\newblock  (\bibinfo{year}{2015}).
\newblock


\bibitem[\protect\citeauthoryear{Li, Wang, Sun, Guo, Zhang, and Wang}{Li et~al\mbox{.}}{2015}]%
        {li2015qmat}
\bibfield{author}{\bibinfo{person}{Pan Li}, \bibinfo{person}{Bin Wang}, \bibinfo{person}{Feng Sun}, \bibinfo{person}{Xiaohu Guo}, \bibinfo{person}{Caiming Zhang}, {and} \bibinfo{person}{Wenping Wang}.} \bibinfo{year}{2015}\natexlab{}.
\newblock \showarticletitle{Q-mat: Computing medial axis transform by quadratic error minimization}.
\newblock \bibinfo{journal}{\emph{ACM Transactions on Graphics (TOG)}} \bibinfo{volume}{35}, \bibinfo{number}{1} (\bibinfo{year}{2015}), \bibinfo{pages}{1--16}.
\newblock


\bibitem[\protect\citeauthoryear{Lieutier}{Lieutier}{2004}]%
        {lieutier2004any}
\bibfield{author}{\bibinfo{person}{Andr{\'e} Lieutier}.} \bibinfo{year}{2004}\natexlab{}.
\newblock \showarticletitle{Any open bounded subset of Rn has the same homotopy type as its medial axis}.
\newblock \bibinfo{journal}{\emph{Computer-Aided Design}} \bibinfo{volume}{36}, \bibinfo{number}{11} (\bibinfo{year}{2004}), \bibinfo{pages}{1029--1046}.
\newblock


\bibitem[\protect\citeauthoryear{Lieutier and Wintraecken}{Lieutier and Wintraecken}{2023}]%
        {lieutier2023hausdorff}
\bibfield{author}{\bibinfo{person}{Andr{\'e} Lieutier} {and} \bibinfo{person}{Mathijs Wintraecken}.} \bibinfo{year}{2023}\natexlab{}.
\newblock \showarticletitle{Hausdorff and gromov-hausdorff stable subsets of the medial axis}. In \bibinfo{booktitle}{\emph{Proceedings of the 55th Annual ACM Symposium on Theory of Computing}}. \bibinfo{pages}{1768--1776}.
\newblock


\bibitem[\protect\citeauthoryear{Lin, Li, Liu, Chen, Choi, and Wang}{Lin et~al\mbox{.}}{2021}]%
        {lin2021point2skeleton}
\bibfield{author}{\bibinfo{person}{Cheng Lin}, \bibinfo{person}{Changjian Li}, \bibinfo{person}{Yuan Liu}, \bibinfo{person}{Nenglun Chen}, \bibinfo{person}{Yi-King Choi}, {and} \bibinfo{person}{Wenping Wang}.} \bibinfo{year}{2021}\natexlab{}.
\newblock \showarticletitle{Point2skeleton: Learning skeletal representations from point clouds}. In \bibinfo{booktitle}{\emph{Proceedings of the IEEE/CVF conference on computer vision and pattern recognition}}. \bibinfo{pages}{4277--4286}.
\newblock


\bibitem[\protect\citeauthoryear{Liu, Chambers, Letscher, and Ju}{Liu et~al\mbox{.}}{2010}]%
        {liu2010simple}
\bibfield{author}{\bibinfo{person}{Lu Liu}, \bibinfo{person}{Erin~W Chambers}, \bibinfo{person}{David Letscher}, {and} \bibinfo{person}{Tao Ju}.} \bibinfo{year}{2010}\natexlab{}.
\newblock \showarticletitle{A simple and robust thinning algorithm on cell complexes}. In \bibinfo{booktitle}{\emph{Computer Graphics Forum}}, Vol.~\bibinfo{volume}{29}. Wiley Online Library, \bibinfo{pages}{2253--2260}.
\newblock


\bibitem[\protect\citeauthoryear{Liu, Ma, Guo, and Yan}{Liu et~al\mbox{.}}{2020}]%
        {liu2020RVD}
\bibfield{author}{\bibinfo{person}{Xiaohan Liu}, \bibinfo{person}{Lei Ma}, \bibinfo{person}{Jianwei Guo}, {and} \bibinfo{person}{Dong-Ming Yan}.} \bibinfo{year}{2020}\natexlab{}.
\newblock \showarticletitle{Parallel computation of 3D clipped Voronoi diagrams}.
\newblock \bibinfo{journal}{\emph{IEEE Transactions on Visualization and Computer Graphics}} \bibinfo{volume}{28}, \bibinfo{number}{2} (\bibinfo{year}{2020}), \bibinfo{pages}{1363--1372}.
\newblock


\bibitem[\protect\citeauthoryear{Liu, Chen, Pan, Cohen-Or, Zhang, and Huang}{Liu et~al\mbox{.}}{2024}]%
        {BRepVP24}
\bibfield{author}{\bibinfo{person}{Yilin Liu}, \bibinfo{person}{Jiale Chen}, \bibinfo{person}{Shanshan Pan}, \bibinfo{person}{Daniel Cohen-Or}, \bibinfo{person}{Hao Zhang}, {and} \bibinfo{person}{Hui Huang}.} \bibinfo{year}{2024}\natexlab{}.
\newblock \showarticletitle{Split-and-Fit: Learning B-Reps via Structure-Aware Voronoi Partitioning}.
\newblock \bibinfo{journal}{\emph{ACM Transactions on Graphics (Proceedings of SIGGRAPH)}} \bibinfo{volume}{43}, \bibinfo{number}{4} (\bibinfo{year}{2024}), \bibinfo{pages}{108:1--108:13}.
\newblock


\bibitem[\protect\citeauthoryear{Ma, Bae, and Choi}{Ma et~al\mbox{.}}{2012}]%
        {ma20123shrink}
\bibfield{author}{\bibinfo{person}{Jaehwan Ma}, \bibinfo{person}{Sang~Won Bae}, {and} \bibinfo{person}{Sunghee Choi}.} \bibinfo{year}{2012}\natexlab{}.
\newblock \showarticletitle{3D medial axis point approximation using nearest neighbors and the normal field}.
\newblock \bibinfo{journal}{\emph{The Visual Computer}} \bibinfo{volume}{28}, \bibinfo{number}{1} (\bibinfo{year}{2012}), \bibinfo{pages}{7--19}.
\newblock


\bibitem[\protect\citeauthoryear{Miklos, Giesen, and Pauly}{Miklos et~al\mbox{.}}{2010}]%
        {miklos2010sat}
\bibfield{author}{\bibinfo{person}{Balint Miklos}, \bibinfo{person}{Joachim Giesen}, {and} \bibinfo{person}{Mark Pauly}.} \bibinfo{year}{2010}\natexlab{}.
\newblock \showarticletitle{Discrete scale axis representations for 3D geometry}.
\newblock In \bibinfo{booktitle}{\emph{ACM SIGGRAPH 2010 papers}}. \bibinfo{pages}{1--10}.
\newblock


\bibitem[\protect\citeauthoryear{Milenkovic}{Milenkovic}{1993}]%
        {milenkovic1993robust}
\bibfield{author}{\bibinfo{person}{Victor Milenkovic}.} \bibinfo{year}{1993}\natexlab{}.
\newblock \showarticletitle{Robust Construction of the Voronoi Diagram of a Polyhedron.}. In \bibinfo{booktitle}{\emph{CCCG}}, Vol.~\bibinfo{volume}{93}. Citeseer, \bibinfo{pages}{473--478}.
\newblock


\bibitem[\protect\citeauthoryear{Noma, Sell{\'a}n, Sharp, Singh, and Jacobson}{Noma et~al\mbox{.}}{2024}]%
        {noma2024surface}
\bibfield{author}{\bibinfo{person}{Yuta Noma}, \bibinfo{person}{Silvia Sell{\'a}n}, \bibinfo{person}{Nicholas Sharp}, \bibinfo{person}{Karan Singh}, {and} \bibinfo{person}{Alec Jacobson}.} \bibinfo{year}{2024}\natexlab{}.
\newblock \showarticletitle{Surface-Filling Curve Flows via Implicit Medial Axes}.
\newblock \bibinfo{journal}{\emph{ACM Transactions on Graphics (TOG)}} \bibinfo{volume}{43}, \bibinfo{number}{4} (\bibinfo{year}{2024}), \bibinfo{pages}{1--12}.
\newblock


\bibitem[\protect\citeauthoryear{Petrov, Goyal, Thamizharasan, Kim, Gadelha, Averkiou, Chaudhuri, and Kalogerakis}{Petrov et~al\mbox{.}}{2024}]%
        {petrov2024gem3d}
\bibfield{author}{\bibinfo{person}{Dmitry Petrov}, \bibinfo{person}{Pradyumn Goyal}, \bibinfo{person}{Vikas Thamizharasan}, \bibinfo{person}{Vladimir Kim}, \bibinfo{person}{Matheus Gadelha}, \bibinfo{person}{Melinos Averkiou}, \bibinfo{person}{Siddhartha Chaudhuri}, {and} \bibinfo{person}{Evangelos Kalogerakis}.} \bibinfo{year}{2024}\natexlab{}.
\newblock \showarticletitle{GEM3D: GEnerative Medial Abstractions for 3D Shape Synthesis}. In \bibinfo{booktitle}{\emph{ACM SIGGRAPH 2024 Conference Papers}}. \bibinfo{pages}{1--11}.
\newblock


\bibitem[\protect\citeauthoryear{Ray, Sokolov, Lefebvre, and L{\'e}vy}{Ray et~al\mbox{.}}{2018}]%
        {ray2018meshless}
\bibfield{author}{\bibinfo{person}{Nicolas Ray}, \bibinfo{person}{Dmitry Sokolov}, \bibinfo{person}{Sylvain Lefebvre}, {and} \bibinfo{person}{Bruno L{\'e}vy}.} \bibinfo{year}{2018}\natexlab{}.
\newblock \showarticletitle{Meshless Voronoi on the GPU}.
\newblock \bibinfo{journal}{\emph{ACM Transactions on Graphics (TOG)}} \bibinfo{volume}{37}, \bibinfo{number}{6} (\bibinfo{year}{2018}), \bibinfo{pages}{1--12}.
\newblock


\bibitem[\protect\citeauthoryear{Rumpf and Telea}{Rumpf and Telea}{2002}]%
        {rumpf2002continuous}
\bibfield{author}{\bibinfo{person}{Martin Rumpf} {and} \bibinfo{person}{Alexandru Telea}.} \bibinfo{year}{2002}\natexlab{}.
\newblock \showarticletitle{A continuous skeletonization method based on level sets}.
\newblock In \bibinfo{booktitle}{\emph{EPRINTS-BOOK-TITLE}}. \bibinfo{publisher}{University of Groningen, Johann Bernoulli Institute for Mathematics and Computer Science}.
\newblock


\bibitem[\protect\citeauthoryear{Saha, Borgefors, and di~Baja}{Saha et~al\mbox{.}}{2016}]%
        {saha2016survey}
\bibfield{author}{\bibinfo{person}{Punam~K Saha}, \bibinfo{person}{Gunilla Borgefors}, {and} \bibinfo{person}{Gabriella~Sanniti di Baja}.} \bibinfo{year}{2016}\natexlab{}.
\newblock \showarticletitle{A survey on skeletonization algorithms and their applications}.
\newblock \bibinfo{journal}{\emph{Pattern recognition letters}}  \bibinfo{volume}{76} (\bibinfo{year}{2016}), \bibinfo{pages}{3--12}.
\newblock


\bibitem[\protect\citeauthoryear{Sherbrooke, Patrikalakis, and Brisson}{Sherbrooke et~al\mbox{.}}{1996}]%
        {sherbrooke1996algorithm}
\bibfield{author}{\bibinfo{person}{Evan~C Sherbrooke}, \bibinfo{person}{Nicholas~M Patrikalakis}, {and} \bibinfo{person}{Erik Brisson}.} \bibinfo{year}{1996}\natexlab{}.
\newblock \showarticletitle{An algorithm for the medial axis transform of 3D polyhedral solids}.
\newblock \bibinfo{journal}{\emph{IEEE transactions on visualization and computer graphics}} \bibinfo{volume}{2}, \bibinfo{number}{1} (\bibinfo{year}{1996}), \bibinfo{pages}{44--61}.
\newblock


\bibitem[\protect\citeauthoryear{Siddiqi, Bouix, Tannenbaum, and Zucker}{Siddiqi et~al\mbox{.}}{2002}]%
        {siddiqi2002hamilton}
\bibfield{author}{\bibinfo{person}{Kaleem Siddiqi}, \bibinfo{person}{Sylvain Bouix}, \bibinfo{person}{Allen Tannenbaum}, {and} \bibinfo{person}{Steven~W Zucker}.} \bibinfo{year}{2002}\natexlab{}.
\newblock \showarticletitle{Hamilton-jacobi skeletons}.
\newblock \bibinfo{journal}{\emph{International Journal of Computer Vision}}  \bibinfo{volume}{48} (\bibinfo{year}{2002}), \bibinfo{pages}{215--231}.
\newblock


\bibitem[\protect\citeauthoryear{Siddiqi and Pizer}{Siddiqi and Pizer}{2008}]%
        {siddiqi2008medial}
\bibfield{author}{\bibinfo{person}{Kaleem Siddiqi} {and} \bibinfo{person}{Stephen Pizer}.} \bibinfo{year}{2008}\natexlab{}.
\newblock \bibinfo{booktitle}{\emph{Medial representations: mathematics, algorithms and applications}}. Vol.~\bibinfo{volume}{37}.
\newblock \bibinfo{publisher}{Springer Science \& Business Media}.
\newblock


\bibitem[\protect\citeauthoryear{Sobiecki, Jalba, and Telea}{Sobiecki et~al\mbox{.}}{2014}]%
        {sobiecki2014comparison}
\bibfield{author}{\bibinfo{person}{Andr{\'e} Sobiecki}, \bibinfo{person}{Andrei Jalba}, {and} \bibinfo{person}{Alexandru Telea}.} \bibinfo{year}{2014}\natexlab{}.
\newblock \showarticletitle{Comparison of curve and surface skeletonization methods for voxel shapes}.
\newblock \bibinfo{journal}{\emph{Pattern Recognition Letters}}  \bibinfo{volume}{47} (\bibinfo{year}{2014}), \bibinfo{pages}{147--156}.
\newblock


\bibitem[\protect\citeauthoryear{Song and Wang}{Song and Wang}{2023}]%
        {blender-mat-addon}
\bibfield{author}{\bibinfo{person}{Shibo Song} {and} \bibinfo{person}{Ningna Wang}.} \bibinfo{year}{2023}\natexlab{}.
\newblock \bibinfo{title}{blender-mat-addon}.
\newblock \bibinfo{howpublished}{\url{https://github.com/songshibo/blender-mat-addon}}.
\newblock


\bibitem[\protect\citeauthoryear{Tagliasacchi, Delame, Spagnuolo, Amenta, and Telea}{Tagliasacchi et~al\mbox{.}}{2016}]%
        {tagliasacchi20163d}
\bibfield{author}{\bibinfo{person}{Andrea Tagliasacchi}, \bibinfo{person}{Thomas Delame}, \bibinfo{person}{Michela Spagnuolo}, \bibinfo{person}{Nina Amenta}, {and} \bibinfo{person}{Alexandru Telea}.} \bibinfo{year}{2016}\natexlab{}.
\newblock \showarticletitle{3d skeletons: A state-of-the-art report}. In \bibinfo{booktitle}{\emph{Computer Graphics Forum}}, Vol.~\bibinfo{volume}{35}. Wiley Online Library, \bibinfo{pages}{573--597}.
\newblock


\bibitem[\protect\citeauthoryear{Wang, Wang, Wang, and Guo}{Wang et~al\mbox{.}}{2022}]%
        {2022MATFP}
\bibfield{author}{\bibinfo{person}{Ningna Wang}, \bibinfo{person}{Bin Wang}, \bibinfo{person}{Wenping Wang}, {and} \bibinfo{person}{Xiaohu Guo}.} \bibinfo{year}{2022}\natexlab{}.
\newblock \showarticletitle{Computing Medial Axis Transform with Feature Preservation via Restricted Power Diagram}.
\newblock \bibinfo{journal}{\emph{ACM Transactions on Graphics (Proceedings of SIGGRAPH Asia 2022)}} \bibinfo{volume}{41}, \bibinfo{number}{6} (\bibinfo{year}{2022}).
\newblock


\bibitem[\protect\citeauthoryear{Wang, Dou, Xu, Lin, Liu, Long, Xin, Liu, Komura, Yuan, et~al\mbox{.}}{Wang et~al\mbox{.}}{2024}]%
        {wang2024coverage}
\bibfield{author}{\bibinfo{person}{Zimeng Wang}, \bibinfo{person}{Zhiyang Dou}, \bibinfo{person}{Rui Xu}, \bibinfo{person}{Cheng Lin}, \bibinfo{person}{Yuan Liu}, \bibinfo{person}{Xiaoxiao Long}, \bibinfo{person}{Shiqing Xin}, \bibinfo{person}{Lingjie Liu}, \bibinfo{person}{Taku Komura}, \bibinfo{person}{Xiaoming Yuan}, {et~al\mbox{.}}} \bibinfo{year}{2024}\natexlab{}.
\newblock \showarticletitle{Coverage Axis++: Efficient Inner Point Selection for 3D Shape Skeletonization}.
\newblock \bibinfo{journal}{\emph{arXiv preprint arXiv:2401.12946}} (\bibinfo{year}{2024}).
\newblock


\bibitem[\protect\citeauthoryear{Xu, Dou, Wang, Xin, Chen, Jiang, Guo, Wang, and Tu}{Xu et~al\mbox{.}}{2023}]%
        {xu2023globally}
\bibfield{author}{\bibinfo{person}{Rui Xu}, \bibinfo{person}{Zhiyang Dou}, \bibinfo{person}{Ningna Wang}, \bibinfo{person}{Shiqing Xin}, \bibinfo{person}{Shuangmin Chen}, \bibinfo{person}{Mingyan Jiang}, \bibinfo{person}{Xiaohu Guo}, \bibinfo{person}{Wenping Wang}, {and} \bibinfo{person}{Changhe Tu}.} \bibinfo{year}{2023}\natexlab{}.
\newblock \showarticletitle{Globally consistent normal orientation for point clouds by regularizing the winding-number field}.
\newblock \bibinfo{journal}{\emph{ACM Transactions on Graphics (TOG)}} \bibinfo{volume}{42}, \bibinfo{number}{4} (\bibinfo{year}{2023}), \bibinfo{pages}{1--15}.
\newblock


\bibitem[\protect\citeauthoryear{Xu, Liu, Wang, Chen, Xin, Guo, Zhong, Komura, Wang, and Tu}{Xu et~al\mbox{.}}{2024}]%
        {xu2024cwf}
\bibfield{author}{\bibinfo{person}{Rui Xu}, \bibinfo{person}{Longdu Liu}, \bibinfo{person}{Ningna Wang}, \bibinfo{person}{Shuangmin Chen}, \bibinfo{person}{Shiqing Xin}, \bibinfo{person}{Xiaohu Guo}, \bibinfo{person}{Zichun Zhong}, \bibinfo{person}{Taku Komura}, \bibinfo{person}{Wenping Wang}, {and} \bibinfo{person}{Changhe Tu}.} \bibinfo{year}{2024}\natexlab{}.
\newblock \showarticletitle{CWF: Consolidating Weak Features in High-quality Mesh Simplification}.
\newblock \bibinfo{journal}{\emph{ACM Transactions on Graphics (TOG)}} \bibinfo{volume}{43}, \bibinfo{number}{4} (\bibinfo{year}{2024}).
\newblock
\showISSN{0730-0301}
\urldef\tempurl%
\url{https://doi.org/10.1145/3658159}
\showDOI{\tempurl}


\bibitem[\protect\citeauthoryear{Xu, Wang, Dou, Zong, Xin, Jiang, Ju, and Tu}{Xu et~al\mbox{.}}{2022}]%
        {xu2022rfeps}
\bibfield{author}{\bibinfo{person}{Rui Xu}, \bibinfo{person}{Zixiong Wang}, \bibinfo{person}{Zhiyang Dou}, \bibinfo{person}{Chen Zong}, \bibinfo{person}{Shiqing Xin}, \bibinfo{person}{Mingyan Jiang}, \bibinfo{person}{Tao Ju}, {and} \bibinfo{person}{Changhe Tu}.} \bibinfo{year}{2022}\natexlab{}.
\newblock \showarticletitle{RFEPS: Reconstructing feature-line equipped polygonal surface}.
\newblock \bibinfo{journal}{\emph{ACM Transactions on Graphics (TOG)}} \bibinfo{volume}{41}, \bibinfo{number}{6} (\bibinfo{year}{2022}), \bibinfo{pages}{1--15}.
\newblock


\bibitem[\protect\citeauthoryear{Yan, Wang, L{\'e}vy, and Liu}{Yan et~al\mbox{.}}{2010}]%
        {yan2010efficient}
\bibfield{author}{\bibinfo{person}{Dong-Ming Yan}, \bibinfo{person}{Wenping Wang}, \bibinfo{person}{Bruno L{\'e}vy}, {and} \bibinfo{person}{Yang Liu}.} \bibinfo{year}{2010}\natexlab{}.
\newblock \showarticletitle{Efficient computation of 3D clipped Voronoi diagram}. In \bibinfo{booktitle}{\emph{Advances in Geometric Modeling and Processing: 6th International Conference, GMP 2010, Castro Urdiales, Spain, June 16-18, 2010. Proceedings 6}}. Springer, \bibinfo{pages}{269--282}.
\newblock


\bibitem[\protect\citeauthoryear{Yan, Letscher, and Ju}{Yan et~al\mbox{.}}{2018}]%
        {yan2018voxel}
\bibfield{author}{\bibinfo{person}{Yajie Yan}, \bibinfo{person}{David Letscher}, {and} \bibinfo{person}{Tao Ju}.} \bibinfo{year}{2018}\natexlab{}.
\newblock \showarticletitle{Voxel cores: Efficient, robust, and provably good approximation of 3d medial axes}.
\newblock \bibinfo{journal}{\emph{ACM Transactions on Graphics (TOG)}} \bibinfo{volume}{37}, \bibinfo{number}{4} (\bibinfo{year}{2018}), \bibinfo{pages}{1--13}.
\newblock


\bibitem[\protect\citeauthoryear{Yan, Sykes, Chambers, Letscher, and Ju}{Yan et~al\mbox{.}}{2016}]%
        {yan2016erosion}
\bibfield{author}{\bibinfo{person}{Yajie Yan}, \bibinfo{person}{Kyle Sykes}, \bibinfo{person}{Erin Chambers}, \bibinfo{person}{David Letscher}, {and} \bibinfo{person}{Tao Ju}.} \bibinfo{year}{2016}\natexlab{}.
\newblock \showarticletitle{Erosion thickness on medial axes of 3D shapes}.
\newblock \bibinfo{journal}{\emph{ACM Transactions on Graphics (TOG)}} \bibinfo{volume}{35}, \bibinfo{number}{4} (\bibinfo{year}{2016}), \bibinfo{pages}{1--12}.
\newblock


\bibitem[\protect\citeauthoryear{Yang, Yao, and Guo}{Yang et~al\mbox{.}}{2018}]%
        {yang2018dmat}
\bibfield{author}{\bibinfo{person}{Baorong Yang}, \bibinfo{person}{Junfeng Yao}, {and} \bibinfo{person}{Xiaohu Guo}.} \bibinfo{year}{2018}\natexlab{}.
\newblock \showarticletitle{DMAT: Deformable medial axis transform for animated mesh approximation}. In \bibinfo{booktitle}{\emph{Computer Graphics Forum}}, Vol.~\bibinfo{volume}{37}. Wiley Online Library, \bibinfo{pages}{301--311}.
\newblock


\bibitem[\protect\citeauthoryear{Yang, Yao, Wang, Hu, Pan, Pan, Wang, and Guo}{Yang et~al\mbox{.}}{2020}]%
        {yang2020p2mat}
\bibfield{author}{\bibinfo{person}{Baorong Yang}, \bibinfo{person}{Junfeng Yao}, \bibinfo{person}{Bin Wang}, \bibinfo{person}{Jianwei Hu}, \bibinfo{person}{Yiling Pan}, \bibinfo{person}{Tianxiang Pan}, \bibinfo{person}{Wenping Wang}, {and} \bibinfo{person}{Xiaohu Guo}.} \bibinfo{year}{2020}\natexlab{}.
\newblock \showarticletitle{P2MAT-NET: Learning medial axis transform from sparse point clouds}.
\newblock \bibinfo{journal}{\emph{Computer Aided Geometric Design}}  \bibinfo{volume}{80} (\bibinfo{year}{2020}), \bibinfo{pages}{101874}.
\newblock


\end{thebibliography}
